\documentclass[%
 reprint,
 amsmath,amssymb,
 aps,
 prb,
]{revtex4-2}

\usepackage{graphicx}
\usepackage{dcolumn}
\usepackage{bm}
\usepackage[colorlinks = true,
            linkcolor = blue,
            urlcolor  = blue,
            citecolor = blue,
            anchorcolor = blue]{hyperref}
\usepackage{hyperref}

\begin{document}

\preprint{APS/123-QED}

\title{Pressure induced evolution of anisotropic superconductivity and Fermi surface nesting in a ternary boride}

\author{Subhajit Pramanick\textsuperscript{1}}
\email{subhajitbhu@kgpian.iitkgp.ac.in}
\author{Sudip Chakraborty\textsuperscript{2}}
\email{sudipchakraborty@hri.res.in}
\author{A. Taraphder\textsuperscript{1}}
\email{arghya@phy.iitkgp.ac.in}

\affiliation{\textsuperscript{1}Department of Physics, Indian Institute of Technology Kharagpur, Kharagpur 721302, India}
\affiliation{\textsuperscript{2}Materials Theory for Energy Scavenging Lab, Harish-Chandra Research Institute, A CI of Homi Bhabha National Institute, Chhatnag Road, Jhunsi, Prayagraj 211019, India}

\date{\today}

\begin{abstract}
Using Migdal-Eliashberg theory implemented in Electron Phonon Wannier (EPW) code, we have investigated anisotropic superconductivity in a ternary boride $\mathrm{Ta(MoB)_2}$. It is a single-gap, anisotropic, phonon-mediated superconductor having a critical temperature $\mathrm{T_c}\sim  \, 19.3$ K. A dominant contribution to superconductivity arises from the robust coupling between electronic states, primarily created by the $\mathrm{d_{xy}}$,$\mathrm{d_{x^2 - y^2}}$ orbitals of Mo atoms and the in-plane vibrations of Mo atoms. A weak Fermi surface nesting and a small electron-phonon coupling cannot induce charge density wave-like instabilities, as evidenced by the lack of a significant  peak in the real part of the total Lindhard susceptibility and the absence of phonon softening. Furthermore, we have studied its electronic and superconducting properties under hydrostatic pressure up to 76.69 GPa, owing to its low bulk modulus and metastability. The persistent reduction in the density of states at the Fermi level, Fermi surface nesting and the stiffening of phonon modes lead to a diminution of superconductivity under pressure up to 59.71 GPa. At 76.69 GPa, a modification in the topology of the Fermi surface, namely a Lifshitz transition, occurs resulting in a sudden enhancement of nesting. This enhanced nesting, in turn, induces an abrupt stabilisation of superconductivity at 76.69 GPa, resulting in a V-shaped response to pressure.

\end{abstract}

\vspace*{0.8cm}                              
\maketitle


\section{\label{sec:1}Introduction}

Discovery of new superconducting materials continues to be a formidable challenge in material science and condensed matter physics \cite{Yao2021}. Superconductivity in conventional superconductors is attributed to the condensation of Cooper pairs, which are formed by phonon-mediated attractive electron-electron interactions \cite{bardeen1957theory} overcoming the Coulomb repulsion between electrons. Following the ground breaking theory of Bardeen, Cooper, and Schrieffer (BCS) \cite{bardeen1957theory}, numerous techniques \cite{mcmillan1968transition,allen1975transition,migdal1958interaction,eliashberg1960interactions} for examining superconductivity have been developed. These range from semi-empirical ones like the McMillan formula \cite{mcmillan1968transition,allen1975transition} to advanced computational approaches like the Migdal-Eliashberg (ME) formalism \cite{migdal1958interaction,eliashberg1960interactions}, which uses first-principles Green's function methods. Recently, new approaches based on density-functional theory (DFT) \cite{hohenberg1964inhomogeneous} have evolved. Density-functional theory for superconductors (SCDFT) \cite{oliveira1988density}, which is becoming increasingly popular, is one prominent example. Among the phonon-mediated conventional superconductors, $\mathrm{MgB_2}$ exhibits one of the highest superconducting critical temperature ($\mathrm{T_c}$) \cite{nagamatsu2001superconductivity,kong2001electron,kortus2001superconductivity}, measured experimentally at 39 K \cite{nagamatsu2001superconductivity} and its superconducting properties are characterized by significant anisotropy and a two-gap nature \cite{liu2001beyond,iavarone2002two,szabo2001evidence,putti2011mgb2}, which arise from the presence of multiple bands across its Fermi surface. Prior research has reported strong coupling between $\mathrm{\sigma}$-bonding bands from B $\mathrm{p_{x,y}}$ orbitals and in-plane vibration within honeycombed B layer, i.e. $\mathrm{E_{2g}}$ phonon mode leads to the generation of strong electron pairs \cite{choi2002origin,margine2013anisotropic,souma2003origin}. However, weak coupling between $\mathrm{\pi}$ bonding and antibonding bands from B $\mathrm{p_z}$ orbital and the $\mathrm{E_{2g}}$ phonon mode, generates comparatively smaller gap \cite{choi2002origin,margine2013anisotropic,souma2003origin}. Extensive efforts have been made to explore novel superconducting phases derived from $\mathrm{MgB_2}$ \cite{akimitsu2005superconductivity,katsura2010possibility,yu2022superconductive,miao2016first,an2021superconductivity,wang2023covalent}. These methods involve chemical doping or substitution, resulting in the formation of compounds such as $\mathrm{Mg_{1-x}Li_xB_2}$ \cite{pallecchi2009investigation}, $\mathrm{Mg_{1-x}Zr_xB_2}$ \cite{feng2002enhanced}, and boride systems similar to $\mathrm{MgB_2}$, such as $\mathrm{MoB_2}$ \cite{pei2023pressure} and $\mathrm{WB_2}$ \cite{lim2022creating}. Yu et al. employed data-driven screening to identify binary boride superconductors that resembled $\mathrm{MgB_2}$ and identified previously unreported superconductors including $\mathrm{SrGa_2}$, $\mathrm{BaGa_2}$, $\mathrm{BaAu_2}$ and $\mathrm{LaCu_2}$ \cite{yu2022superconductive}. Superconductivity in binary borides has been validated through several experimental and computational studies \cite{katsura2010possibility,an2021superconductivity,wang2023covalent}, whereas superconductivity in ternary borides is still relatively less explored \cite{chen2024high}. 
\par
Tuning of the parameters of a lattice, such as its spacing, shape, or interaction strengths, can significantly influence its physical properties. Typically, such tuning is carried out by employing hydrostatic pressure \cite{wudil2022hydrostatic,zhang2021hydrostatic}, uniaxial strain \cite{hemme2023tuning,wang2023controlled}, biaxial strain \cite{frisenda2017biaxial}, doping \cite{gofryk2012electronic}, alloying \cite{bi2022giant} etc. Hydrostatic pressure is widely recognized to significantly modify the electronic bonding state, and consequently, the physical properties of materials by reducing interatomic distances. Several high-pressure studies were performed on $\mathrm{MgB_2}$ to examine the effect of pressure on $\mathrm{T_c}$ \cite{razavi2002effect,prassides2001compressibility,wang2009superconductivity,shao2004pressure,ma2009absence}. The first principle study conducted by Wang et. al. \cite{wang2009superconductivity} and Ma et. al. \cite{ma2009absence} demonstrated that the density of states of B $\mathrm{p_{x,y}}$ orbitals decreases with hydrostatic pressure, while $\mathrm{E_{2g}}$ phonon modes stiffen. This led to a gradual decrease in electron-phonon coupling strength and a consequent reduction of superconductivity. Additionally, a structural phase transition occurred at pressures exceeding 200 GPa, changing the $\mathrm{AlB_2}$-type hexagonal symmetry of the material to $\mathrm{KHg_2}$-type orthorhombic symmetry, ultimately transforming it into a weak metal \cite{wang2009superconductivity,ma2009absence}. Pressure can also induce other electronic phase changes, such as the transition from a semiconductor to a superconductor \cite{hu2013pressure}, the occurrence of V-shaped \cite{tafti2015universal,yue2018electron} and dome-shaped superconductivity \cite{zhao2021inverted,pan2015pressure}, the transition from a topological insulator to a topological superconductor, \cite{zhu2013superconductivity,matsubayashi2014superconductivity} etc. In addition to the effects of pressure, doping, alloying and other factors, researchers have also investigated superconductivity in appropriate heterostructures \cite{mohanta2015multiband, meng2024layer,patel2024electron}. For instance, Patel et al. have recently discovered a two-gap superconductivity in $\mathrm{NbSe_2}$/$\mathrm{MoS_2}$ AB-heterostacking using first principle investigation \cite{patel2024electron}. Additionally, they observed a decrease in superconductivity as the thickness of $\mathrm{NbSe_2}$ increased and also detected the potential occurrence of Ising superconductivity \cite{patel2024electron}.

\par

Transition-metal borides represent a novel category of materials characterized by superconductivity, refractory characteristics, unique magnetic properties and exceptional stability under extreme conditions \cite{zhao2022synthesis,jatmika2020superconducting,wang2021nd2fe14b,kanoun2012structure}. Ternary transition-metal borides (TTMBs) exhibit unique boron substructures and diverse characteristics, in contrast to binary borides, resulting in remarkable designability. Specifically, TTMBs have drawn much interest as the potential candidates for multifunctional integration materials since they contain more than one metallic phases \cite{zhao2022synthesis}. Notable documentation on TTMBs include: a modified sol-gel method for synthesizing $\mathrm{Nd_2Fe_{14}B}$ magnetic powders proposed by Wang et al. \cite{wang2021nd2fe14b}, the identification of $\mathrm{YRh_4B_4}$ as a conventional s-wave superconductor with a $\mathrm{T_c}$ of 10.6 K by Jatmika et al. \cite{jatmika2020superconducting}, the ultraincompressibility and exceptional hardness of $\mathrm{Os_{1-x}Ru_{x}B_2}$, which may serve as viable candidates for cutting tools or wear-resistant coatings, reported by Kanoun et al. \cite{kanoun2012structure}, the ferromagnetic metastable properties of $\mathrm{Fe(MoB)_2}$ demonstrated by Zhao et al. \cite{zhao2022synthesis} etc. Therefore, innovative functional materials can be developed by adjusting the bimetallic composition of TTMBs. Recently Chen et al. has predicted the critical temperature of various binary and ternary boride superconductors, including $\mathrm{ZrB_{12}}$, $\mathrm{WB}$, $\mathrm{Mo_2B}$, $\mathrm{MoIrB_4}$, $\mathrm{Ta(MoB)_2}$ and others using high-throughput screening \cite{chen2024high}. Among them, $\mathrm{Ta(MoB)_2}$ is one of the previously synthesized TTMBs \cite{sobolev1968phase} that has no experimental data so far on its superconducting properties. Chen et al. have theoretically anticipated its critical temperature to be 12 K \cite{chen2024high}. However, they employed the Allen-Dynes modified McMillan formula \cite{mcmillan1968transition,allen1975transition} to compute superconductivity, which fails to consider anisotropy. In this work, we explore anisotropic superconductivity of the bulk ternary boride $\mathrm{Ta(MoB)_2}$ utilizing Migdal-Eliashberg formalism \cite{migdal1958interaction,eliashberg1960interactions}. Considering the presence of anisotropy in superconductivity, we simulated the superconducting property of $\mathrm{Ta(MoB)_2}$ at hydrostatic pressure up to 76.69 GPa and observed a nesting-induced v-shaped response.

\section{\label{sec:2}Computational Methodology}

\begin{figure*}[t]
\includegraphics[width=0.30\linewidth]{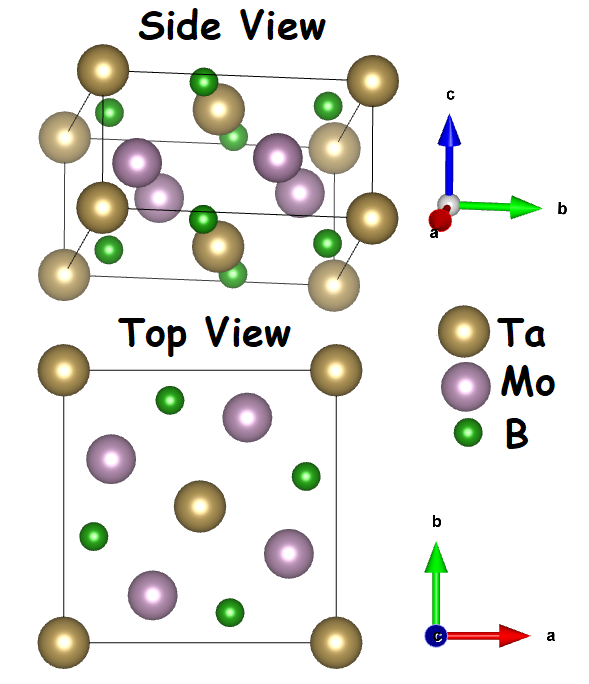}
\includegraphics[width= 0.46\linewidth]{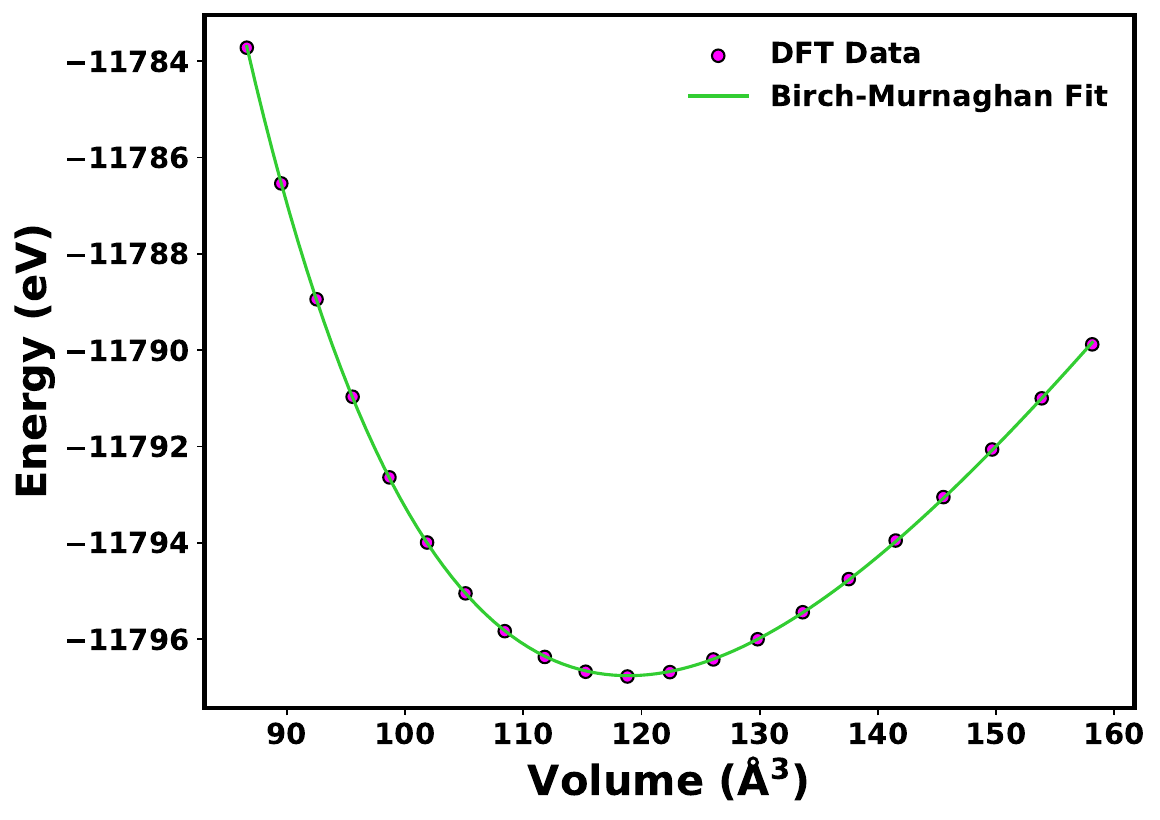}
\vspace*{-0.25cm}
    \begin{center}
        \hspace*{-1.7cm}
        \textbf{\small{(a)}}
        \hspace*{8cm}
        \textbf{\small{(b)}}
    \end{center}
\vspace{-12pt}
\caption{\label{fig:1} (a) Crystal structure of $\mathrm{Ta(MoB)_2}$ at ambient pressure, (b) Birch-Murnaghan equation fitting for finding bulk modulus and its pressure derivative}
\end{figure*}
We performed an electronic structure calculation by employing Vienna ab-initio simulation package (VASP) \cite{kresse1993ab,kresse1996efficiency}, where the Projector-Augmented Wave(PAW) formalism \cite{blochl1994projector,kresse1999ultrasoft} has been implemented. For this purpose, the generalized gradient approximation(GGA) as proposed by Perdew-Burke-Ernzerhof (PBE)  \cite{perdew1996generalized} has been used for the exchange-correlation functional. We have considered the conjugate gradient algorithm to achieve robust ionic relaxation until the force on each atom falls below -0.001 eV/\text{\AA} along with 415 eV energy cut off and an energy tolerance $\mathrm{10^{-8}}$ eV. We have sampled the Brillouin zone with 12 $\times$ 12 $\times$ 24 Monkhorst-Pack k-points. One of the goals of this investigation is to unravel the pressure-driven structural evolution of $\mathrm{Ta(MoB)_2}$ crystal structure. We therefore determine the dynamical stability at each pressure point, which is based on phonon dispersion calculations within the framework of Density Functional Perturbation Theory (DFPT), as implemented in Quantum Espresso package \cite{giannozzi2009quantum,giannozzi2017advanced,giannozzi2020quantum}. During this calculation, we have used ultra-soft pseudo-potentials from Garrity-Bennett-Rabe-Vanderbilt (GBRV) library \cite{garrity2014pseudopotentials} with PBE functional and fully relativistic pseudo-potentials for performing electronic structure calculations, considering spin-orbit coupling. We have set the convergence criteria for plane wave energy cut off at 80 Ry, charge density cut off at 640 Ry and gaussian smearing having a width of 0.015 Ry. It is worth mentioning that, we obtained phonon dispersions through Fourier interpolation of the dynamical matrices, which have been calculated using a 12 $\times$ 12 $\times$ 24 k-point mesh and a 3 $\times$ 3 $\times$ 6 q-point mesh. The corresponding electron-phonon coupling and superconductivity have been computed using Allen-Dynes modified McMillan formula \cite{mcmillan1968transition,allen1975transition} and Migdal-Eliashberg formalism \cite{migdal1958interaction,eliashberg1960interactions}. The superconducting critical temperature is estimated using the following Allen-Dynes modified McMillan formula within the isotropic regime \cite{mcmillan1968transition,allen1975transition}:

\begin{eqnarray}
\mathrm{T_c = \frac{{\omega}_{log}}{1.20} \hspace{0.1cm} exp\left[\frac{-1.04 \hspace{0.1cm} (1 + \lambda)}{\lambda - {\mu}_c^{\ast} (1+0.62\lambda)}\right]} 
\label{eq:1}
\end{eqnarray}
where, $\mathrm{\lambda}$ is the isotropic electron-phonon coupling (EPC) strength, $\mathrm{\alpha^2 F(\omega)}$ is the isotropic Eliashberg spectral function, $\mathrm{{\mu}_c^{\ast}}$ is the effective Coulomb potential and the corresponding logarithmic average phonon frequency $\mathrm{{\omega}_{log}}$ is:
\begin{eqnarray}
\mathrm{{\omega}_{log} = exp\left[\frac{\int d\omega \frac{\alpha^2 F(\omega)}{\omega}ln\omega}{\int d\omega \frac{\alpha^2 F(\omega)}{\omega}}\right]}    
\label{eq:2}
\end{eqnarray}
The isotropic Eliashberg electron-phonon spectral function is determined from\cite{ponce2016epw,margine2013anisotropic,allen1983theory}:
\begin{eqnarray}
\mathrm{\alpha^2 F(\omega) = \frac{1}{2\pi N_F} \sum_{\textbf{q},\nu} \frac{\gamma_{\textbf{q}\nu}}{\omega_{\textbf{q}\nu}} \delta (\omega - \omega_{\textbf{q}\nu})}   
\label{eq:3}
\end{eqnarray}
where, $\mathrm{\gamma_{\textbf{q}\nu}}$ is the linewidth of phonon corresponding to branch index $\nu$ and momentum $\textbf{q}$, $\mathrm{N_F}$ is the electron density of states at Fermi level and $\mathrm{\omega_{\textbf{q}\nu}}$ is the phonon frequency of the corresponding mode and momentum. The inverse of phonon linewidth is an indicator of phonon life-time, which is directly associated with electron-phonon coupling strength. Here  \cite{ponce2016epw,margine2013anisotropic,allen1983theory}:
\begin{eqnarray}
\mathrm{\gamma_{\textbf{q}\nu} =} && \mathrm{2\pi \omega_{\textbf{q}\nu}\sum_{m,n} \int_{BZ} \frac{d\textbf{k}}{\Omega_{BZ}}|{g_{mn\nu}(\textbf{k},\textbf{q})|^2 }}  \nonumber\\
&& \mathrm{\times \delta (\epsilon_{n\textbf{k}} - \epsilon_F) \delta (\epsilon_{m\textbf{k}+\textbf{q}} - \epsilon_F)}
\label{eq:4}
\end{eqnarray}
and the cumulative EPC strength $\mathrm{\lambda}$ is \cite{ponce2016epw,margine2013anisotropic}:
\begin{eqnarray}
\mathrm{\lambda = 2 \int_{0}^{\omega} \frac{\alpha^2 F(\omega)}{\omega} d\omega}  
\label{eq:5}
\end{eqnarray}
In equation (4), the integration is over the entire first Brillouin zone. Here, $\mathrm{\epsilon_{n\textbf{k}},\, \epsilon_{m\textbf{k}+\textbf{q}}}$ are the eigen values of Kohn-Sham orbitals at bands n, m and wave vectors \textbf{k},\,  \textbf{k}+\textbf{q}; $\mathrm{g_{mn\nu}(\textbf{k}, \, \textbf{q})}$ is the electron-phonon coupling (EPC) matrix element and $\mathrm{\epsilon_F}$ is the Fermi energy. The anisotropic superconductivity is determined by solving the following Migdal-Eliashberg equations self-consistently \cite{ponce2016epw,margine2013anisotropic,allen1983theory}:
\begin{eqnarray}
\mathrm{Z(\textbf{k}_1, i\omega_{n_{1}}) =} && \mathrm{1 + \frac{\pi T}{N_F \omega_{n_{1}}} \sum_{\textbf{k}_2, n_2} \frac{\omega_{n_{2}}}{\sqrt{\omega_{n_{2}}^2 + \Delta^2(\textbf{k}_2, i\omega_{n_{2}})}}}  \nonumber\\ 
&& \mathrm{\times \lambda(\textbf{k}_1, \textbf{k}_2, n_1 - n_2) \delta(\epsilon_{\textbf{k}_{2}})}  
\label{eq:6}
\end{eqnarray}
\begin{eqnarray}
\mathrm{Z(\textbf{k}_1,} &&\mathrm{i\omega_{n_{1}}) \Delta (\textbf{k}_1, i\omega_{n_{1}}) = \frac{\pi T}{N_F} \sum_{\textbf{k}_2, n_2} \frac{\Delta (\textbf{k}_2, i\omega_{n_{2}})}{\sqrt{\omega_{n_{2}}^2 + \Delta^2(\textbf{k}_2, i\omega_{n_{2}})}}}\nonumber\\ 
&& \mathrm{\times [\lambda (\textbf{k}_1, \textbf{k}_2, n_1 - n_2) - N_F V(\textbf{k}_1 - \textbf{k}_2)] \delta (\epsilon_{\textbf{k}_{2}})}
\label{eq:7}
\end{eqnarray}
where, Z is renormalization function, $\mathrm{\Delta}$ is superconducting gap,  $\mathrm{i\omega_n = i (2n + 1) \pi T}$ (n = integer) are fermionic Matsubara frequencies and  $\mathrm{V(\textbf{k}_1 - \textbf{k}_2)}$ is screened Coulomb interaction. For solving these two equations self-consistently, we have employed Electron Phonon Wannier (EPW) code \cite{ponce2016epw,giustino2007electron,margine2013anisotropic}. The upper limit of frequency integration in Eliashberg equations has been kept as 0.4 eV, which is five times larger than the maximum frequency in phonon dispersion. For interpolation from Wannier basis to Bloch basis using EPW code \cite{ponce2016epw,giustino2007electron,margine2013anisotropic}, a fine 15 $\times$ 15 $\times$ 30 k and q grid have been used for 0.10 eV smearing width. However, Fermi nesting has been calculated using $\mathrm{75\times75\times150}$ fine k-grid and a 0.02 eV smearing width for the energy-conserving delta function. To solve the Eliashberg equations, the effective Coulomb potential $\mathrm{\mu^\ast_c}$ and a convergence threshold have been set to 0.1 and $\mathrm{10^{-4}}$ eV, respectively. VESTA \cite{momma2011vesta} and XCrySDen \cite{kokalj1999xcrysden} are used for the visualization of crystal structure and Fermi surface of the systems under the influence of pressure, respectively.

\section{\label{sec:3}Results and Discussions}

\subsection{\label{sec:3.1}Crystal Structure Analysis}

\begin{figure*}[t]
\includegraphics[width=0.475\linewidth]{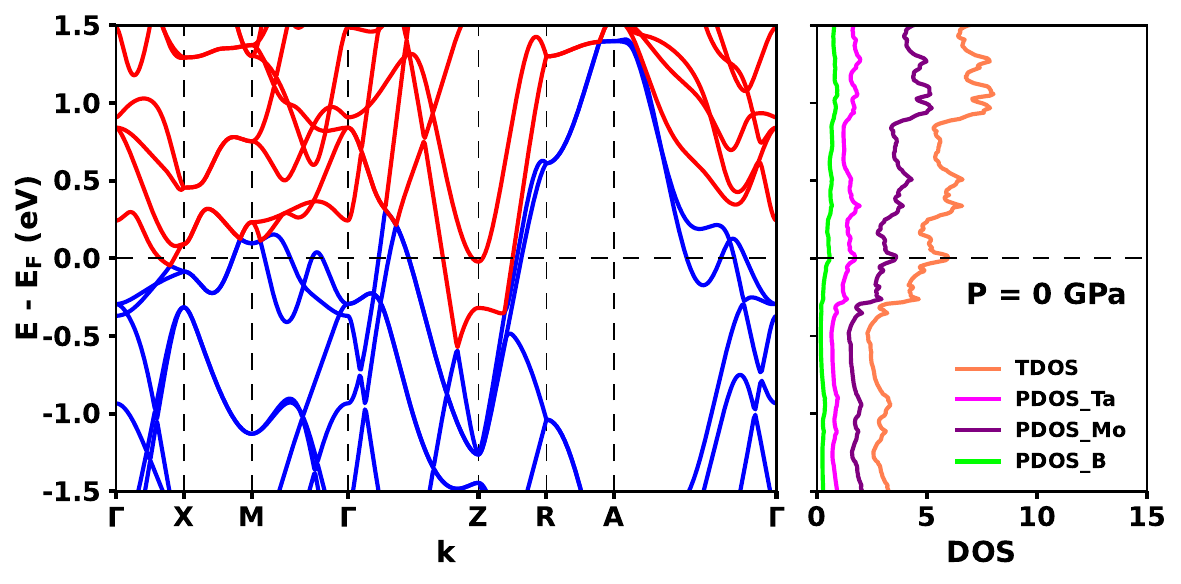}
\includegraphics[width=0.515\linewidth]{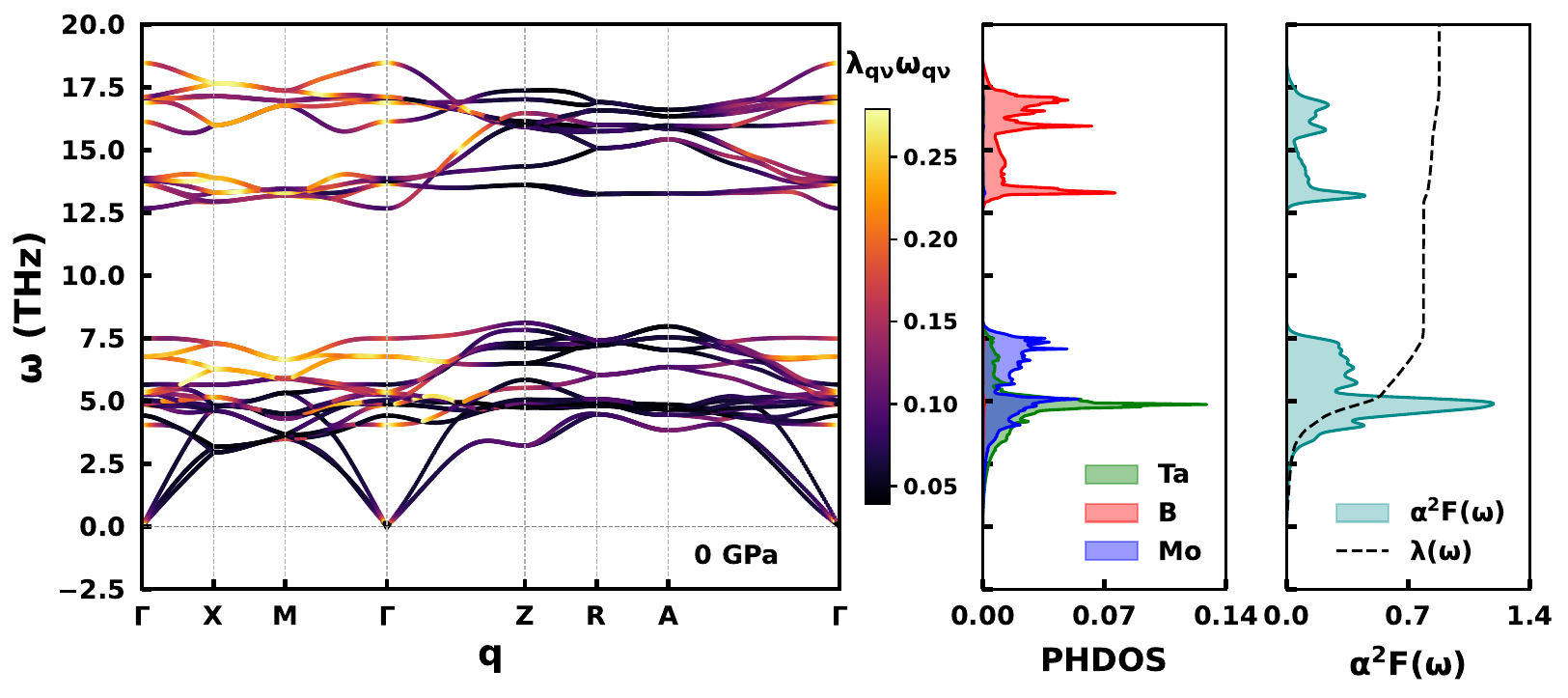}
\vspace*{-0.65cm}
    \begin{center}
        \hspace*{-1.7cm}
        \textbf{\small{(a)}}
        \hspace*{8cm}
        \textbf{\small{(b)}}
    \end{center}
\vspace*{-0.32cm}
\includegraphics[width= 0.25\linewidth]{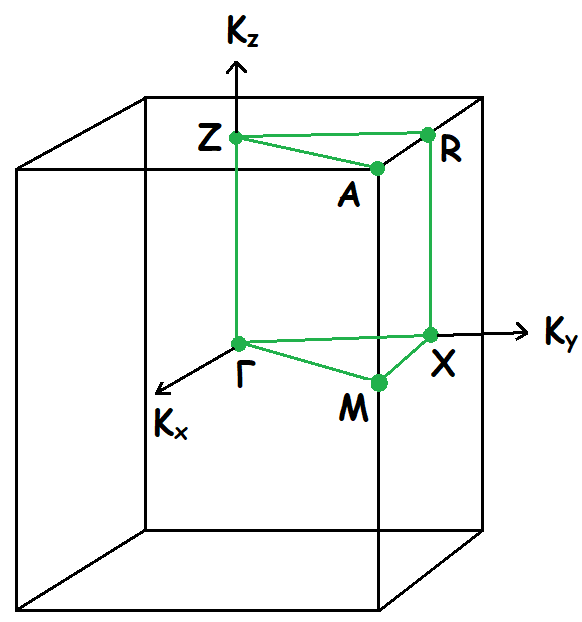}
\includegraphics[width= 0.14\linewidth]{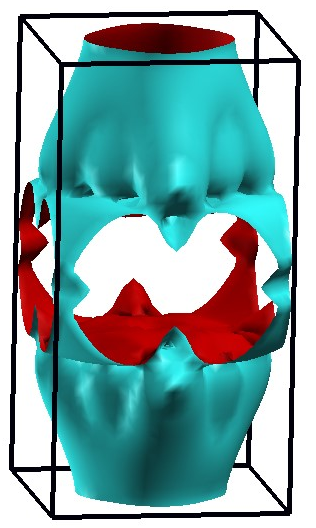}
\includegraphics[width= 0.14\linewidth]{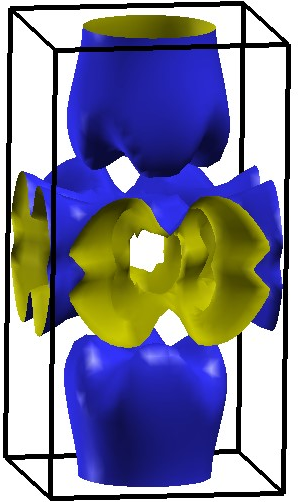}
\includegraphics[width= 0.14\linewidth]{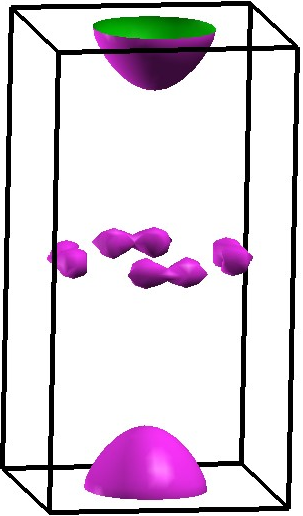}
\includegraphics[width= 0.143\linewidth]{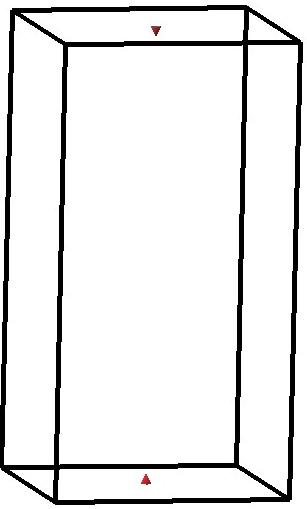}
\includegraphics[width= 0.14\linewidth]{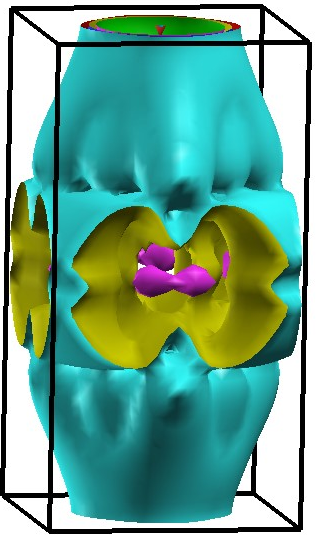}
\vspace*{-0.05cm}
    \begin{center}
        \hspace*{0.48cm}
        \textbf{\small{(c)}}
        \hspace*{3.4cm}
        \textbf{\small{(d)}}
        \hspace*{1.8cm}
        \textbf{\small{(e)}}
        \hspace*{1.8cm}
        \textbf{\small{(f)}}
        \hspace*{1.8cm}
        \textbf{\small{(g)}}
        \hspace*{1.8cm}
        \textbf{\small{(h)}}
    \end{center}
\vspace{-11pt}
\caption{\label{fig:2} (a) Electronic band structure and desnity of states at ambient pressure, (b) phonon band structure and color plot of $\mathrm{\lambda_{q\nu} \omega_{q\nu}}$ on the phonon bands at ambient pressure. Right panel describes corresponding phonon density of states and isotropic Eliashberg spectral function, cumulative electron-phonon coupling strength, (c) First Brillouin zone of $\mathrm{Ta(MoB)_2}$ (tetragonal lattice having space group p4/mbm). There are four bands that cross the Fermi level at ambient pressure. Fermi surface plot of $\mathrm{Ta(MoB)_2}$ at ambient pressure corresponding to (d),(e),(f) and (g) individual bands (two different colours have been used for two opposite faces), (h) merged bands}
\end{figure*}

At ambient pressure, $\mathrm{Ta(MoB)_2}$ crystallizes in the tetragonal phase having space group p4/mbm \cite{jain2013commentary,sobolev1968phase}, where tantalum (Ta) is bonded to eight molybdenum (Mo) and four boron (B) atoms, resulting in a combination of distorted cuboctahedra consisting of $\mathrm{TaB_4Mo_8}$ units that share both face and corner positions. The bond length of all Ta-Mo bonds measure exactly 2.77 Å. Each Ta-B bond has a length of 2.49 Å. Molybdenum (Mo) exhibits a 10-coordinate geometry, being surrounded by four equivalent Ta atoms and six equivalent B atoms. Among the Mo-B bonds, there are four shorter ones at 2.39 Å and two longer ones at 2.47 Å. Boron (B) is in a nine-coordinate configuration, bonded to two identical Ta atoms, six identical Mo atoms, and one B atom. The bond length between the two B atoms is 1.89 Å. The thermodynamic stability of $\mathrm{Ta(MoB)_2}$ under different hydrostatic pressures have been examined based on the calculation of hull energies \cite{chen2024high,twyman2022environmental} and formation energies \cite{peterson2021materials,curtarolo2005accuracy} shown in Table \hyperref[tab:1]{I}. Hull energy quantifies how far a material is from its decomposition into the most stable competing phase. A zero hull energy indicates the most stable phase, whereas higher values suggest diminished stability \cite{chen2024high,twyman2022environmental}. Compounds with hull energies below approximately 0.1 eV/atom are frequently considered metastable, as this corresponds to the thermal fluctuation of roughly 1100 K, a standard synthesis temperature for borides \cite{chen2024high,zheng2023superconductivity}. The formation energies \cite{peterson2021materials,curtarolo2005accuracy} have been computed using the following relation: 
\begin{eqnarray}
\mathrm{\Delta E_F = E_{Ta(MoB)_2} - 2E_{Ta} - 4E_{Mo} - 4E_{B}}  
\label{eq:8}
\end{eqnarray}

\noindent where, $\mathrm{E_{Ta(MoB)_2}}$ is the total energy of $\mathrm{Ta(MoB)_2}$ and $\mathrm{E_{Ta}}$, $\mathrm{E_{Mo}}$, $\mathrm{E_{B}}$ are the total energy of Ta, Mo and B atoms respectively in their corresponding stable most bulk configuration \cite{arakcheeva2003commensurate,ross1963high,decker1959crystal}. The method of calculating $\mathrm{E_{form_1}}$, $\mathrm{E_{form_2}}$ and hull energy in Table \hyperref[tab:1]{I} has been described in our supplementary material \cite{supplementary}. At ambient pressure, the hull energy of $\mathrm{Ta(MoB)_2}$ is +0.058 eV/atom indicating it as a metastable material. The values of hull energy in Table \hyperref[tab:1]{I} also suggest its metastability with pressure upto 76.69 GPa. This also ensures phase separation

\begin{table}[b]
\renewcommand{\arraystretch}{1.05}  
\caption{\label{tab:1}%
Formation and hull energy at different pressures
}
\begin{ruledtabular}
\begin{tabular}{cccc}
\textrm{Pressure} & \multicolumn{2}{c}{\textrm{Formation Energy}} & \textrm{Hull Energy} \\
\textrm{(GPa)} & \multicolumn{2}{c}{\textrm{(eV/atom)}} & \textrm{(eV/atom)} \\
\cline{2-3}
& $\mathrm{E_{form_1}}$ & $\mathrm{E_{form_2}}$ & \\
\colrule
0.00 & -0.480 & -0.480 & +0.058 \\
9.11 & -0.470 & -0.484 & +0.057 \\
19.49 & -0.440 & -0.493 & +0.057 \\
31.28 & -0.386 & -0.508 & +0.057 \\
44.62 & -0.308 & -0.527 & +0.058 \\
59.71 & -0.202 & -0.551 & +0.059 \\
76.69 & -0.067 & -0.580 & +0.061 \\
\end{tabular}
\end{ruledtabular}
\end{table}

\noindent of ternary and binary borides is very likely under pressure what Quan et al. \cite{quan2021mob} has obtained for binary borides. The bulk modulus, which is an important parameter for understanding how a material can respond under uniform compression, can be determined from the third order Birch-Murnaghan isothermal equation as follows \cite{birch1947finite}:

\vspace{-10pt}
\begin{eqnarray}
\mathrm{P(V) =} && \mathrm{\frac{3B_0}{V}\left[\left(\frac{V_0}{V}\right)^\frac{7}{3} - \left(\frac{V_0}{V}\right)^\frac{5}{3}\right]}  \nonumber\\ 
&& \mathrm{\times \left [1 + \frac{3}{4}(B_0^\prime - 4) \Biggl\{ \left(\frac{V_0}{V}\right)^\frac{2}{3} -1 \Biggr\} \right]}
\label{eq:9}
\end{eqnarray}
where $\mathrm{P}$ is the pressure, $\mathrm{V_0}$ is the volume at ambient pressure, $\mathrm{V}$ is the volume as a function of pressure, $\mathrm{B_0}$ is the bulk modulus and $\mathrm{B_0^\prime}$ is the pressure derivative of the bulk modulus. This equation of state in terms of total energy ($\mathrm{E}$) and volume ($\mathrm{V}$) can be expressed as \cite{birch1947finite}:
\begin{eqnarray}
\mathrm{E(V) =} && \mathrm{E_0 + \frac{9V_0B_0}{16} \left [ \left ( \frac{V_0}{V}\right)^\frac{2}{3} - 1\right]^3 B_0^\prime}  \nonumber\\ 
&& \hspace*{-0.5cm} \mathrm{+ \frac{9V_0B_0}{16}\left [ \left ( \frac{V_0}{V}\right)^\frac{2}{3} - 1\right]^2 \left [ 6 - 4 \left ( \frac{V_0}{V}\right)^\frac{2}{3}\right ]}
\label{eq:10}
\end{eqnarray}
Figure \hyperref[fig:1]{1(b)} displays the plot of the equation of state derived from the Birch-Murnaghan fit. The Bulk modulus and its pressure derivative have been determined to be 1.742 GPa and 4.130, respectively. Such a low bulk modulus indicates that the material is more sensitive to pressure, which motivates us further to investigate electron-phonon coupling under the influence of external pressure. 

\subsection{\label{sec:3.2}Electronic Structure and Dynamical Stability of $\mathbf{Ta(MoB)_2}$ at ambient Pressure}
The density of states at the Fermi level is a preliminary factor in determining the presence of conventional superconductivity in a material \cite{mcmillan1968transition}, which underscores the importance of electronic structures before delving into superconductivity. Furthermore, the examination of phonon-mediated superconducting properties requires a careful consideration of the phonon dispersions of any material under perturbations \cite{baroni2001phonons}. At ambient pressure, the left panel of Figure \hyperref[fig:2]{2(a)} displays the conduction and valence bands, represented by red and blue colors, respectively, while four significant bands intersecting the Fermi level. This band structure contains several electron and hole pockets. Electron pockets are found around X and Z points, while hole pockets are located around M point, as well as along the path from M to Z and from R to $\mathrm{\Gamma}$ points. The right panel of the Figure \hyperref[fig:2]{2(a)} displays the total and projected density of states, indicating that the Mo atoms contribute more significantly to the density of states at the Fermi level than the B and Ta atoms. Figure 2(a) and 2(b) of our supplementary material \cite{supplementary} contains the fat-band analysis of $\mathrm{Ta(MoB)_2}$, which clearly demonstrates that the d-orbitals of Mo atoms make a significant contribution to the density of states at the Fermi level. The four bands intersecting the Fermi level predominantly display the characteristics of the d orbitals of Mo atoms. The electronic band structure of $\mathrm{Ta(MoB)_2}$ after incorporating spin-orbit coupling (SOC) at ambient pressure is shown in Figure 2(c) of supplementary material \cite{supplementary}, where we can observe that the fourth band crossing the Fermi level at Z point previously, is moving above it due to the shift of the Fermi level. Having inversion as well as time reversal symmetry (as it is non-magnetic), no band splitting in bands of $\mathrm{Ta(MoB)_2}$ is observed in presence of SOC and each bands keep double degeneracy for up and down spin states. The degeneracy of the d orbitals of Mo atoms has been removed due to atomic SOC around the $\mathrm{\Gamma}$ and $\mathrm{X}$ points, as well as along the $\mathrm{\Gamma - Z}$ and $\mathrm{R - \Gamma}$ path near the Fermi level. The Fermi window for the calculation of electron-phonon coupling (EPC) is having a width of 0.4 eV, while there is no substantial changes within this narrow range. Consequently, in $\mathrm{Ta(MoB)_2}$, we believe that SOC cannot significantly alter the characteristics of superconductivity, and hence we exclude SOC in superconductivity calculations for each pressure. The absence of negative phonon frequencies in the phonon dispersion, as shown in Figure \hyperref[fig:2]{2(b)}, signifies the absence of imaginary phonon modes, hence confirming the dynamical stability of the system under ambient pressure condition \cite{baroni2001phonons}. The phonon density of states in the right panel of Figure \hyperref[fig:2]{2(b)} demonstrates that Ta and Mo atoms primarily contribute to the low-frequency acoustic branch, whereas B atoms primarily contribute to the high-frequency optical branch. We have also analyzed the colour plot of $\mathrm{\lambda_{q\nu} \omega_{q\nu}}$ on the phonon bands along high symmetry points. The primary contribution to $\mathrm{\lambda_{q\nu} \omega_{q\nu}}$ originates from some modes of the acoustic and optical branch along $\mathrm{\Gamma - X - M - \Gamma}$ path. Few modes along $\mathrm{\Gamma - Z}$ path also contribute to it slightly. The first Brillouin zone of $\mathrm{Ta(MoB)_2}$ has been illustrated in Figure \hyperref[fig:2]{2(c)}, highlighting its high symmetry points. The Fermi surface plots corresponding to the four bands that intersect the Fermi level have been displayed in Figure \hyperref[fig:2]{2 (d-g)}. Along the $\mathrm{\Gamma - Z - R}$ path around Z point the third and fourth bands (Figure \hyperref[fig:2]{2(f-g)}) exhibit two electron-like open hemispherical shaped Fermi sheets.

\subsection{\label{sec:3.3}Fermi Surface Nesting of $\mathbf{Ta(MoB)_2}$ at ambient pressure}

\begin{figure*}[t]
\hspace{-0.5cm}
\includegraphics[width=0.275\linewidth]{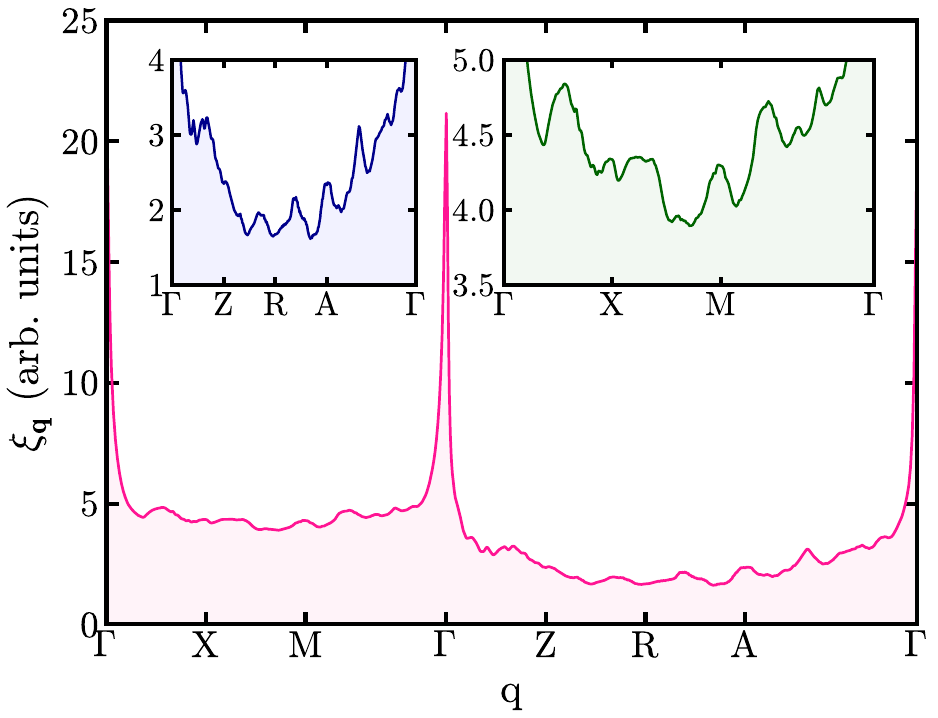}
\hspace{0.8cm}
\includegraphics[width=0.295\linewidth]{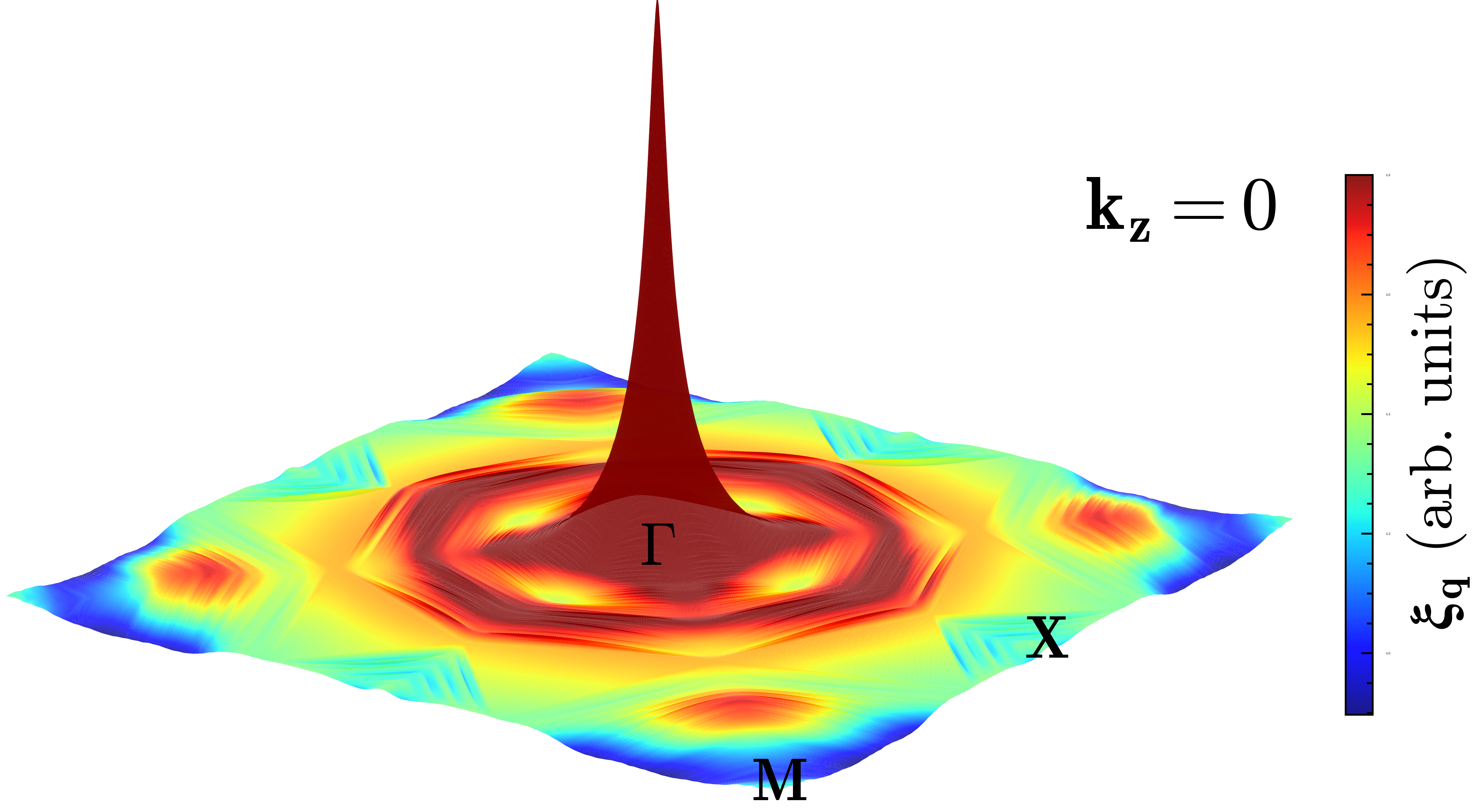}
\hspace{1.0cm}
\includegraphics[width=0.295\linewidth]{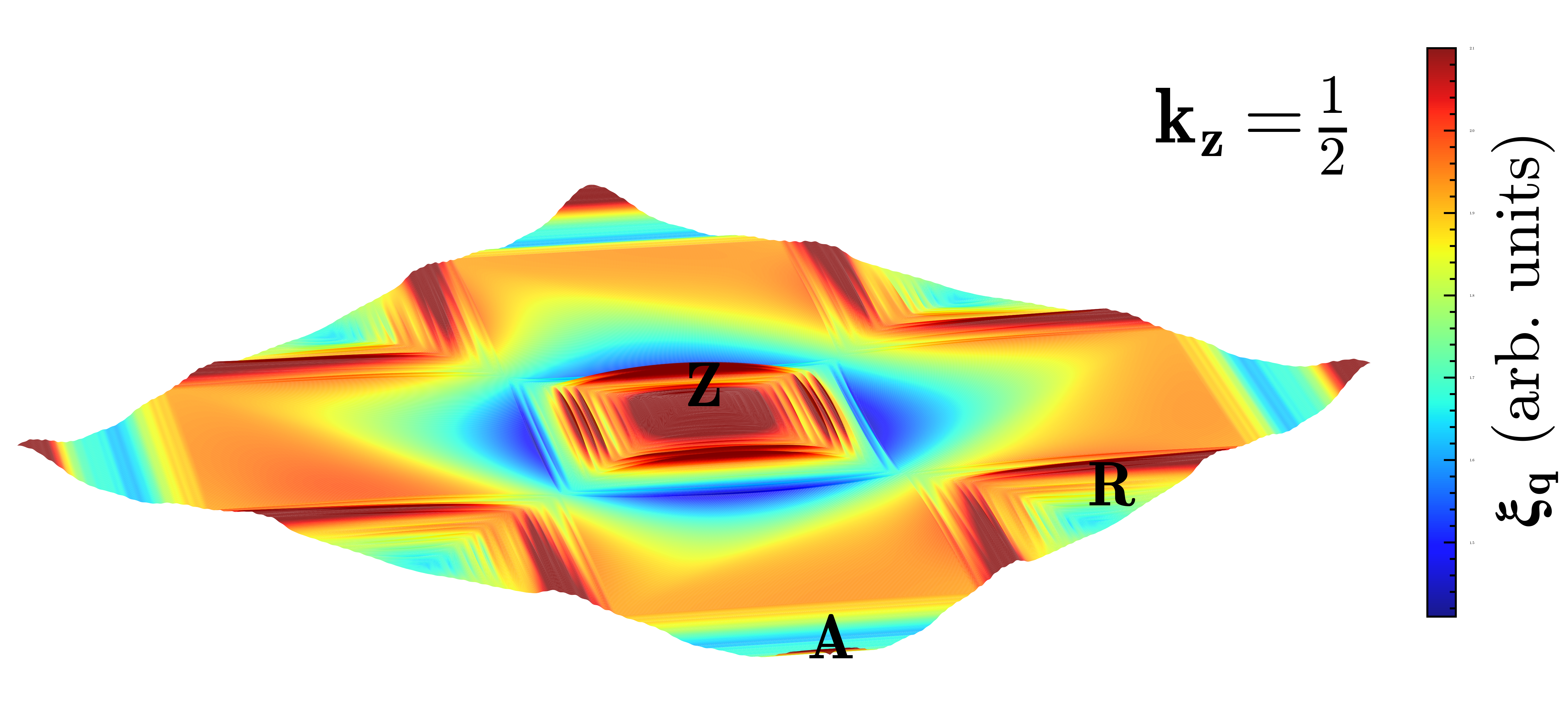}
\begin{center}
        \hspace*{0.01cm}
        \textbf{\small{(a)}}
        \hspace*{4.9cm}
        \textbf{\small{(b)}}
        \hspace*{6.0cm}
        \textbf{\small{(c)}}
    \end{center}
\includegraphics[width=0.275\linewidth]{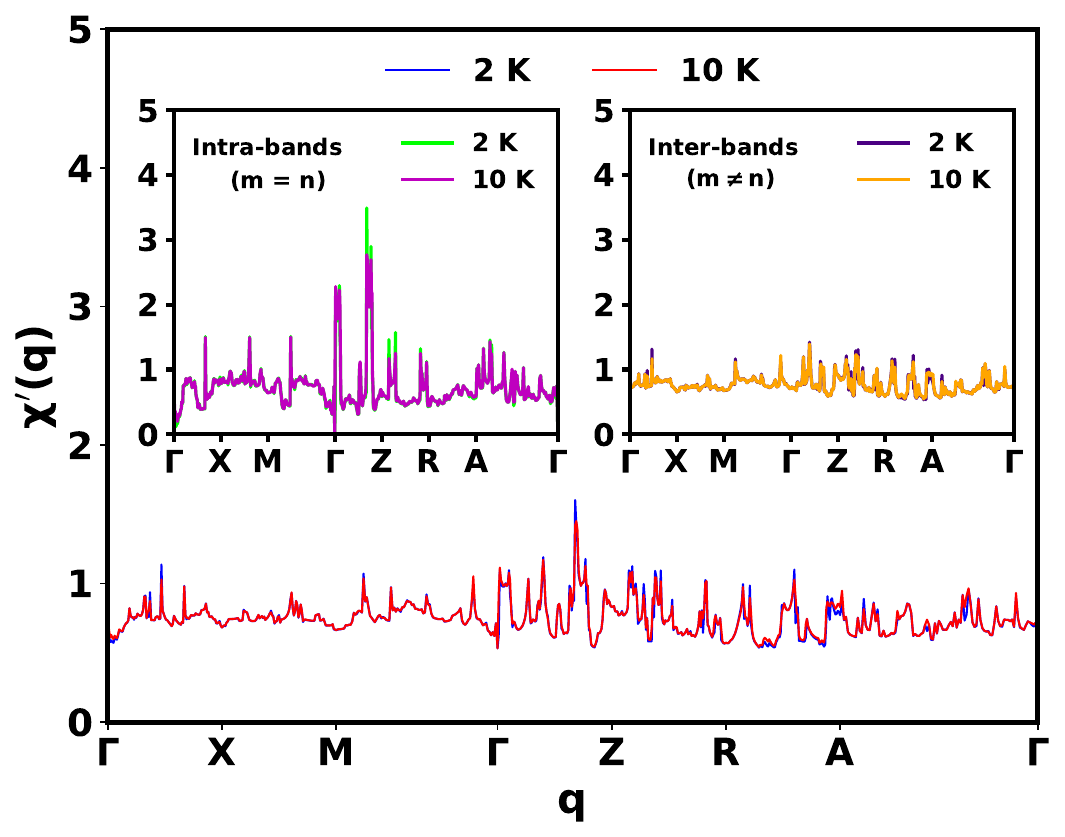}
\includegraphics[width=0.355\linewidth]{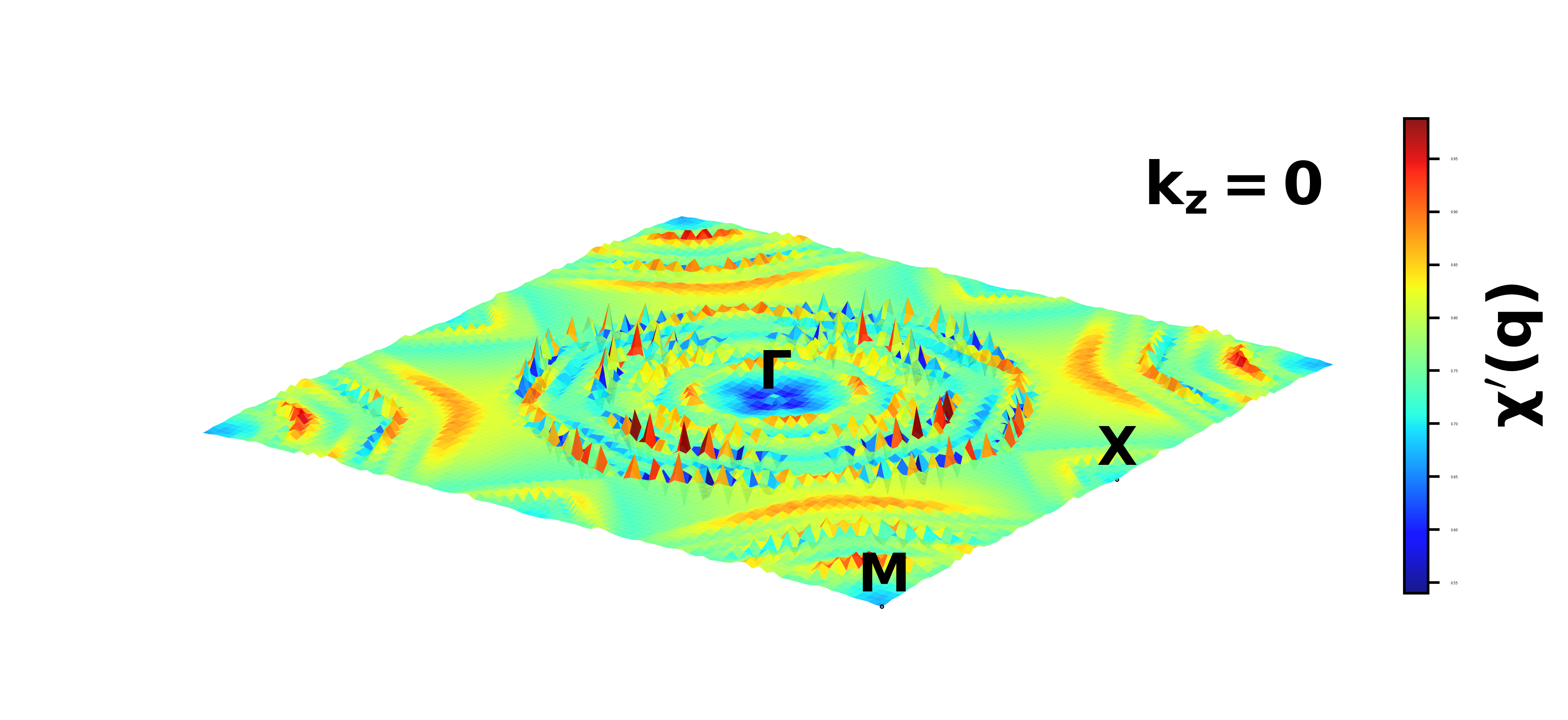}
\includegraphics[width=0.355\linewidth]{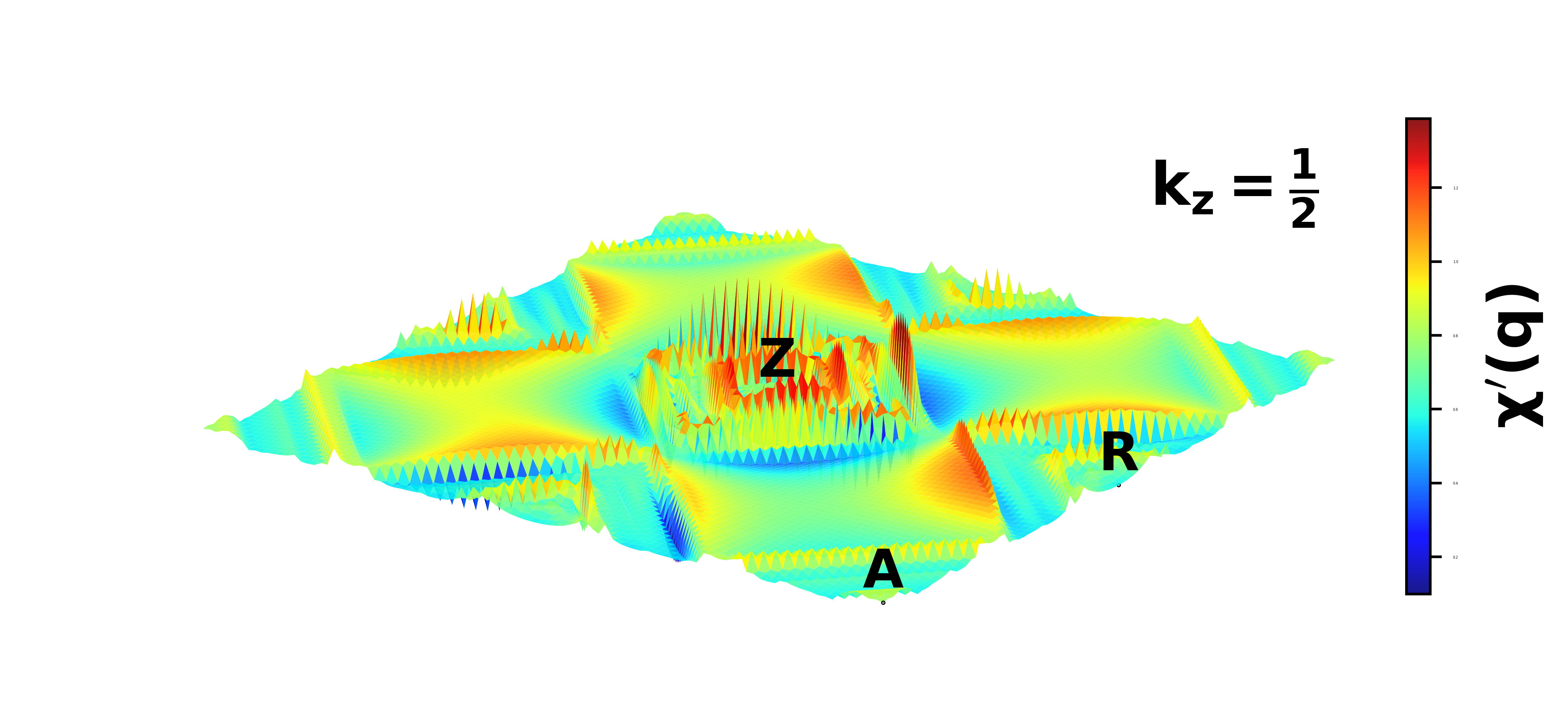}
\begin{center}
        \hspace*{0.01cm}
        \textbf{\small{(d)}}
        \hspace*{4.9cm}
        \textbf{\small{(e)}}
        \hspace*{6.0cm}
        \textbf{\small{(f)}}
    \end{center}
\caption{\label{fig:3} (a) Main: Fermi nesting function of $\mathrm{Ta(MoB)_2}$ at ambient pressure along high symmetry q points, Inset: zoomed version, (b),(c) color plots of Nesting function at $\mathrm{k_z = 0}$ and $\mathrm{k_z = \frac{1}{2}}$ (boundary wall along $\mathrm{k_z}$ axis) planes of first Brillouin zone respectively, (d) Real part of the bare static Lindhard charge susceptibility of $\mathrm{Ta(MoB)_2}$ at ambient pressure along high symmetry q points, Main: for all possible bands combination ($\mathrm{m = n}$ and $\mathrm{m \neq n}$), Left inset: for intra bands only ($\mathrm{m = n}$), Right inset: for inter bands only ($\mathrm{m \neq n}$), (e),(f) color plots of the real part of the bare static Lindhard charge susceptibility at $\mathrm{k_z = 0}$ and $\mathrm{k_z = \frac{1}{2}}$ planes of first Brillouin zone respectively}
\end{figure*}

The Fermi surface nesting has a significant impact on instabilities like charge density waves (CDWs), spin density waves (SDWs), superconductivity, etc. The existence of Fermi surface nesting can be validated by observing peaks in the imaginary part of the bare, static Lindhard charge susceptibility \cite{johannes2008fermi,johannes2006fermi,kaboudvand2022fermi,zhu2015classification}. The Fermi nesting function has been computed using the following formula \cite{ponce2016epw}:

\begin{eqnarray}
\mathrm{\xi_\textbf{q} = \sum_{m,n} \int_{BZ} \frac{d\textbf{k}}{\Omega_{BZ}} \delta(\epsilon_{n\textbf{k}} -\epsilon_F)\delta(\epsilon_{m\textbf{k}+\textbf{q}} - \epsilon_F)}
\label{eq:11}
\end{eqnarray}

implemented in the EPW algorithm \cite{ponce2016epw,giustino2007electron,margine2013anisotropic}. The integration in this case extends throughout the first Brillouin zone. We have illustrated the Fermi nesting function at ambient pressure along high symmetry $\mathrm{q}$ directions in Figure \hyperref[fig:3]{3(a)} and the corresponding colour plots (in same scale) at $\mathrm{k_z = 0}$ and $\mathrm{k_z = \frac{1}{2}}$ planes of the first Brillouin zone in Figures \hyperref[fig:3]{3(b)} and \hyperref[fig:3]{(c)} respectively. It is evident from Figure \hyperref[fig:3]{3(a)}, that the nesting function has comparatively higher value along the $\mathrm{\Gamma-X-M-\Gamma-Z}$ path and several peaks are present along this path. The high peak situated at $\mathrm{\Gamma}$ point, shown in Figure \hyperref[fig:3]{3(a)} and \hyperref[fig:3]{(b)} corresponds to self-nesting (without any momentum translation i.e., $\mathbf{q} = \mathbf{0}$). For a multi-band superconductor, a peak in the $\mathrm{\xi_\textbf{q}}$ at $\mathbf{q} = \mathbf{0}$ point can appear due to two contributions: one is from intra-band ($\mathrm{m = n}$),

\begin{eqnarray}
\mathrm{\xi_\textbf{0} = \sum_{n} \int_{BZ} \frac{d\textbf{k}}{\Omega_{BZ}} \left[\delta(\epsilon_{n\textbf{k}} - \epsilon_F)\right]^2}
\label{eq:12}
\end{eqnarray}

\noindent which can be directly related to high density of state at the Fermi level as confirmed by Figure \hyperref[fig:2]{2(a)} and the other from inter-band ($\mathrm{m \neq n}$),

\begin{eqnarray}
\mathrm{\xi_\textbf{0} = \sum_{m \neq n} \int_{BZ} \frac{d\textbf{k}}{\Omega_{BZ}} \delta(\epsilon_{n\textbf{k}} - \epsilon_F) \delta(\epsilon_{m\textbf{k}} - \epsilon_F)}
\label{eq:13}
\end{eqnarray}

\noindent which originates from the presence of different degenerate or nearly degenerate bands at the Fermi level (see colour plots of Figure 3 in Supplementary Material \cite{supplementary}). The small peaks specifically located along $\mathrm{\Gamma-X-M-\Gamma}$ and $\mathrm{\Gamma-Z}$ path, seen in Figures \hyperref[fig:3]{3(a)} and \hyperref[fig:3]{(b)}, confirm potential existence of weak Fermi surface nesting at non-zero finite momentum. The presence of nesting prompts us to investigate whether a CDW instability exists in the system. Zhu, et al. \cite{zhu2015classification} suggested a categorization of CDWs into three distinct ones based on the mechanisms behind their formation \cite{kaboudvand2022fermi}. In this framework, first type of CDWs correspond to the Peierls picture, where the transition is induced by Fermi surface nesting (FSN). The second category of CDWs are linked to $\mathrm{q}$-dependent electron-phonon interactions, while the third relates to correlated systems exhibiting charge modulations, such as cuprates, where neither FSN nor EPC sufficiently accounts for the transition \cite{zhu2015classification,kaboudvand2022fermi}. Theoretically, the real component of Lindhard electronic susceptibility must diverge to initiate an electronic CDW \cite{johannes2008fermi,johannes2006fermi}. The presence of CDW instability in any real material can be inferred from the emergence of finite peaks in the real part of susceptibility as shown by Johannes et al. \cite{johannes2008fermi,johannes2006fermi}. In the constant matrix element approximation, the real part of bare static Lindhard charge susceptibility can be written as \cite{johannes2008fermi,johannes2006fermi,kaboudvand2022fermi}:

\begin{eqnarray}
\mathrm{\chi^\prime (\textbf{q}) = -\frac{1}{N_\textbf{k}} \sum_{m,n} \sum_{\textbf{k}} \frac{f(\epsilon_{n\textbf{k}}) - f(\epsilon_{m{\textbf{k}} + \textbf{q}})}{\epsilon_{n\textbf{k}} - \epsilon_{m{\textbf{k}} + \textbf{q}}}} 
\label{eq:14}
\end{eqnarray}

\begin{figure*}[t]
\includegraphics[width=0.34\linewidth]{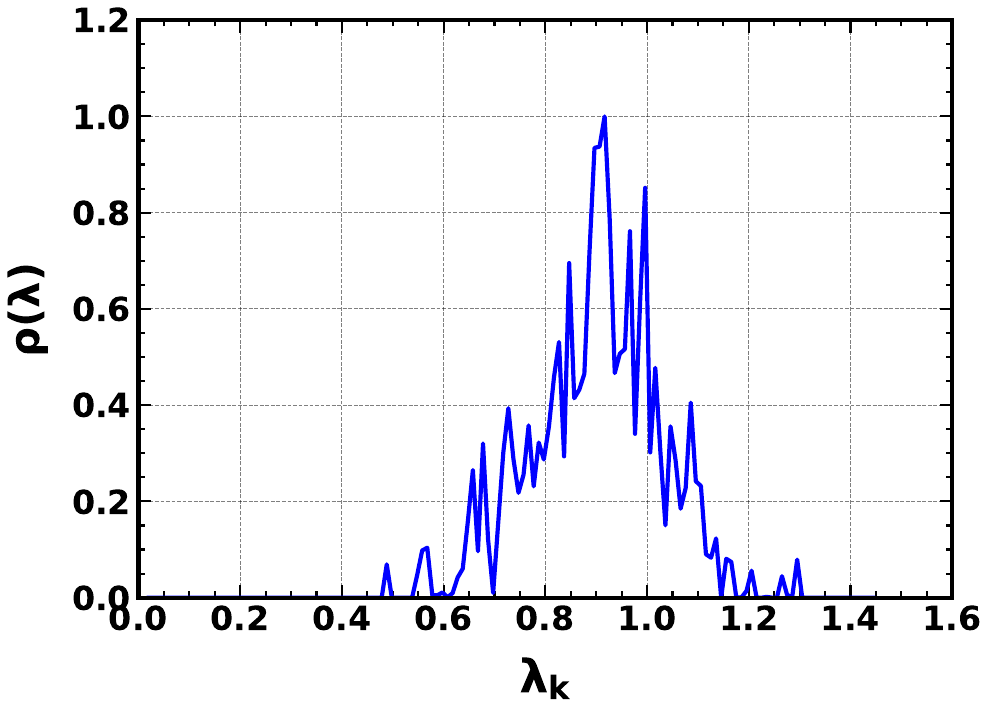}
\includegraphics[width=0.325\linewidth]{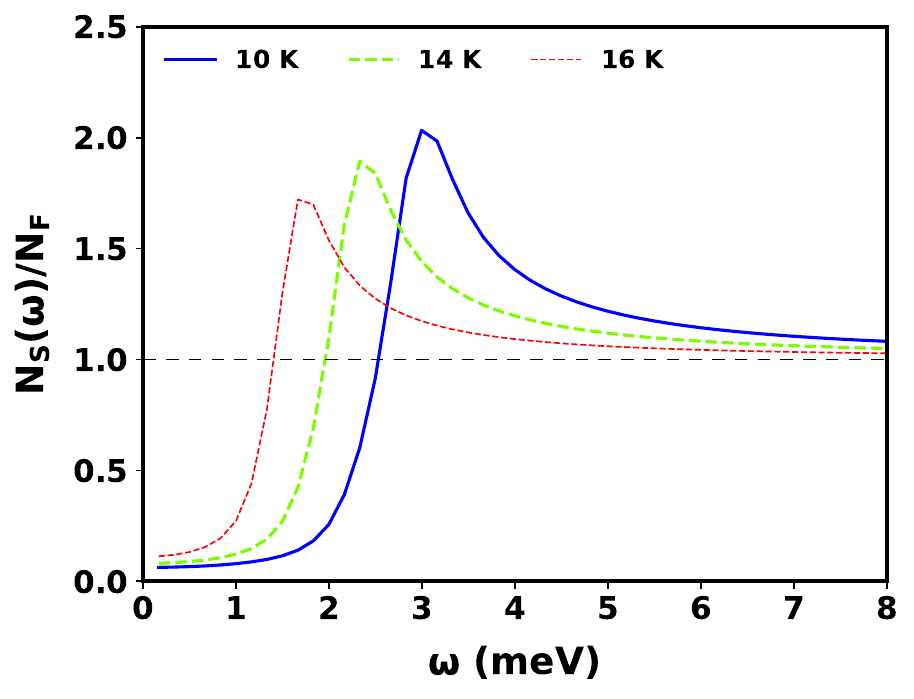}
\includegraphics[width=0.315\linewidth]{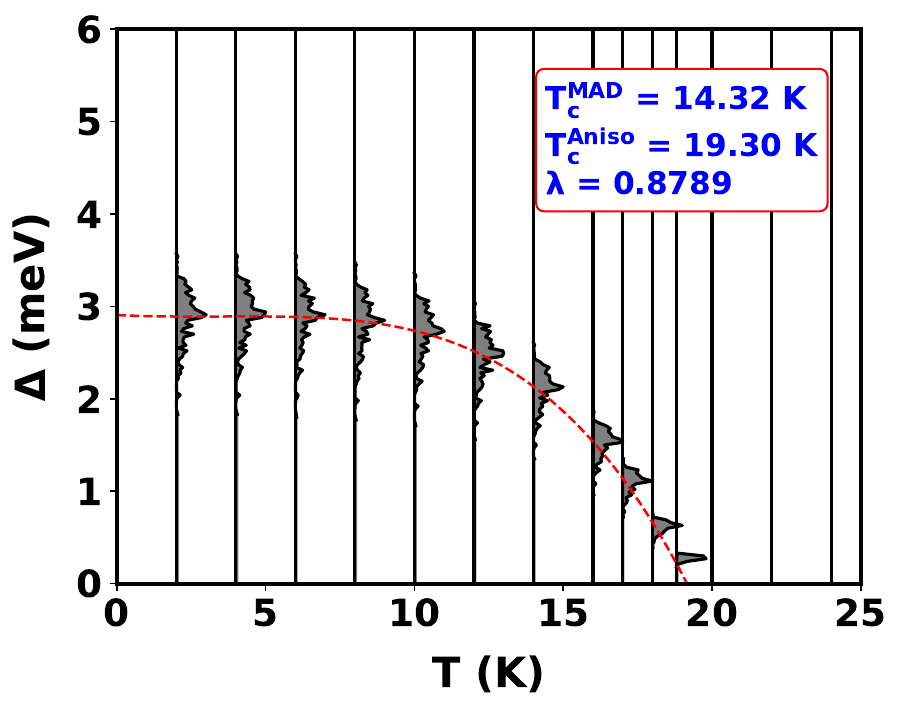}
\begin{center}
        \hspace*{0.5cm}
        \textbf{\small{(a)}}
        \hspace*{5.3cm}
        \textbf{\small{(b)}}
        \hspace*{5.0cm}
        \textbf{\small{(c)}}
    \end{center}
\vspace{-12pt}
\caption{\label{fig:4} (a) Distribution function of electron-phonon coupling strength $\mathrm{\lambda_k}$ for $\mathrm{Ta(MoB)_2}$, (b) quasi-particle density of states in the superconducting state relative to the density of states in the normal state as a function of frequency (For three different temperatures). The dashed black line is the density of states in the normal state, normalized to 1 at the Fermi level, (c) anisotropic superconducting gap on the Fermi surface of $\mathrm{Ta(MoB)_2}$ as a function of temperature}
\end{figure*}

\noindent where, $\mathrm{N_\textbf{k}}$ and $\mathrm{f}$ denote the total number of $\mathrm{\textbf{k}}$ points and the Fermi function, respectively. All the possible intra-band and inter-band cases have been plotted along high symmetry $\mathrm{q}$ directions in the main and inset of Figure \hyperref[fig:3]{3(d)} at temperatures 2 K and 10 K. For $\mathrm{T=2}$ K the corresponding colour plots at $\mathrm{k_z = 0}$ and $\mathrm{k_z = \frac{1}{2}}$ planes of the first Brillouin zone are displayed in Figures \hyperref[fig:3]{3(e)} and \hyperref[fig:3]{(f)} respectively. Despite  small fluctuations around the base line of the real component of susceptibility, there is a lack of substantial peak that could potentially trigger a CDW.  The left inset of Figure \hyperref[fig:3]{3(d)}, representing intra-band contributions, reveals significant peaks along the $\mathrm{\Gamma - Z}$ path, which may serve as an indicator of possible CDW. Our additional calculations, illustrated in Figure 4(b) and 4(c) of Supplementary Material \cite{supplementary}, confirm the absence of significant softening in any phonon modes along this path. The low value of EPC strength along this path, shown in Figure \hyperref[fig:2]{2(b)}, supports the observation of the lack of softening. Consequently, in spite of the presence of weak nesting along $\mathrm{\Gamma - Z}$ path, the poor EPC and the lack of phonon softening confirm an absence of CDW-like distortion in the system. This is also evident from the total susceptibility, reflected in Figure \hyperref[fig:3]{3(d)}, as it does not exhibit any large peak along this path. This also aligns with the previous report by Johannes et al. \cite{johannes2008fermi,johannes2006fermi}, who inferred that Fermi surface nesting could not be the sole driving factor behind the emergence of CDW instability.

\subsection{\label{sec:3.4}Signature of Superconductivity in $\mathbf{Ta(MoB)_2}$ at ambient pressure}

The crystal structure of $\mathrm{Ta(MoB)_2}$ exhibits a configuration (Figure \hyperref[fig:1]{1}), where all Mo atoms are precisely positioned along single x-y plane. Electronic states at the Fermi level show a dominant characteristics of Mo $\mathrm{d_{xy}}$ and $\mathrm{d_{x^2 - y^2}}$ electrons as evident from its high density of states. These electronic states couple strongly with the in-plane vibrations of Mo atoms, namely the LA (longitudinal acoustic) and TA (transverse acoustic) phonon modes, located in the lower frequency region of the phonon dispersion (Figure \hyperref[fig:2]{2(b)}). This coupling is one of the dominant contributions to the superconductivity of $\mathrm{Ta(MoB)_2}$. The phonon modes of the acoustic and optical branch along $\mathrm{\Gamma - X - M - \Gamma}$ path and along $\mathrm{\Gamma - Z}$ path that contribute to the $\mathrm{\lambda_{q\nu} \omega_{q\nu}}$ and hence superconductivity have already been identified in Figure \hyperref[fig:2]{2(b)}. In $\mathrm{MgB_2}$, the distance between two B atoms in the x-y plane is small, which facilitates formation of $\mathrm{\pi}$-bonding states of B $\mathrm{p_z}$-orbitals. These bonding states couple with $\mathrm{E_{2g}}$-phonon modes to engender superconductivity. \cite{choi2002origin,margine2013anisotropic,souma2003origin}. In $\mathrm{Ta(MoB)_2}$, however, the distances between two consecutive and oppositely positioned Mo atoms in the x-y plane are comparatively larger (3.19 Å and 4.52 Å, respectively) making the overlap between two Mo d-orbitals poorer. Right panel of the Figure \hyperref[fig:2]{2(b)} displays our calculated isotropic Eliashberg spectral function and cumulative electron-phonon coupling strength for $\mathrm{Ta(MoB)_2}$. It shows two distinct regimes, one with a prominent peak around 4.95 THz in the lower frequency regime and the other, a smaller peak around 13 THz, in the higher frequency regime. The occurrence of a peak in the lower frequency regime can be attributed to the phonon density of states of Mo atoms in the acoustic mode, which is one of the dominant factors contributing to superconductivity. Unlike $\mathrm{MgB_2}$ \cite{margine2013anisotropic}, the distribution function of EPC strength in Figure \hyperref[fig:4]{4(a)} relates to the coupling between in-plane vibration and electronic states of Mo atoms. The value of the isotropic EPC strength ranges from 0.50 to 1.30 and the cumulative $\mathrm{\lambda}$ is 0.88. By solving Migdal-Eliashberg equation \cite{migdal1958interaction,eliashberg1960interactions} (Equations \hyperref[eq:6]{(6)} and \hyperref[eq:6]{(7)}) along the imaginary energy axis, one can analytically continue these equations along the real energy axis using Pade approximation, implemented in the EPW code \cite{ponce2016epw,giustino2007electron,margine2013anisotropic}. Subsequently, the quasiparticle density of state can be determined by analysing the poles of the normal Green function \cite{margine2013anisotropic}. Figure \hyperref[fig:4]{4(b)} displays the quasiparticle density of states for $\mathrm{Ta(MoB)_2}$, revealing its single-gap characteristics. The superconducting gap in this plot can be determined from the frequency at which the peak appears. As the temperature rises, the superconducting gap decreases. The temperature-dependence of the anisotropic superconducting gap of $\mathrm{Ta(MoB)_2}$ is shown in Figure \hyperref[fig:4]{4(c)}. It exhibits the single gap characteristics of $\mathrm{Ta(MoB)_2}$, which vanishes at the critical temperature $\mathrm{T_c}$ = 19.30 K. Using the Allen-Dynes modified McMillan formula \cite{mcmillan1968transition,allen1975transition}, the critical temperature was determined as $\mathrm{T_c}$ = 14.32 K. The disparity between two critical temperatures is a signature of anisotropy in superconductivity which was also observed in $\mathrm{MgB_2}$ due to the multiband nature of its Fermi surface \cite{liu2001beyond,iavarone2002two,szabo2001evidence,putti2011mgb2,choi2002origin}. Chen et. al. \cite{chen2024high} employed Allen-Dynes modified McMillan formula \cite{mcmillan1968transition,allen1975transition} to estimate the critical temperature of $\mathrm{Ta(MoB)_2}$  and found it to be approximately 12 K.

\begin{table}[h]
\caption{\label{tab:2}%
Variation of critical temperature ($\mathrm{T_c^{Aniso}}$) of $\mathrm{Ta(MoB)_2}$ with effective Coulomb potential. $\mathrm{T_c^{Aniso}}$ decreases with increasing $\mathrm{\mu_c^\ast}$. 
}
\begin{ruledtabular}
\begin{tabular}{cc}
Coulomb Potential ($\mathrm{\mu_c^\ast}$) & Critical Temperature (K) \\
\colrule
0.05 & 23.50 \\
0.10 & 19.30 \\
0.15 & 17.80 \\
\end{tabular}
\end{ruledtabular}
\end{table}

The anisotropic superconducting critical temperatures of $\mathrm{Ta(MoB)_2}$ for the lower, moderate and larger effective Coulomb potential ($\mathrm{\mu_c^\ast}$) are presented in Table \hyperref[tab:2]{II}. The critical temperature of $\mathrm{Ta(MoB)_2}$ decreases as the effective Coulomb potential increases. In the EPW calculation  \cite{ponce2016epw,giustino2007electron,margine2013anisotropic}, a small overestimation of the critical temperature may occur due to the use of isotropic Coulomb parameter and the neglect of non-adiabatic corrections, as reported by Margine et al. \cite{margine2013anisotropic}. 

\subsection{\label{sec:3.5}Evolution of Electronic and Phonon dispersion under hydrostatic pressure}

\begin{figure*}[t]
\includegraphics[width=0.475\linewidth]{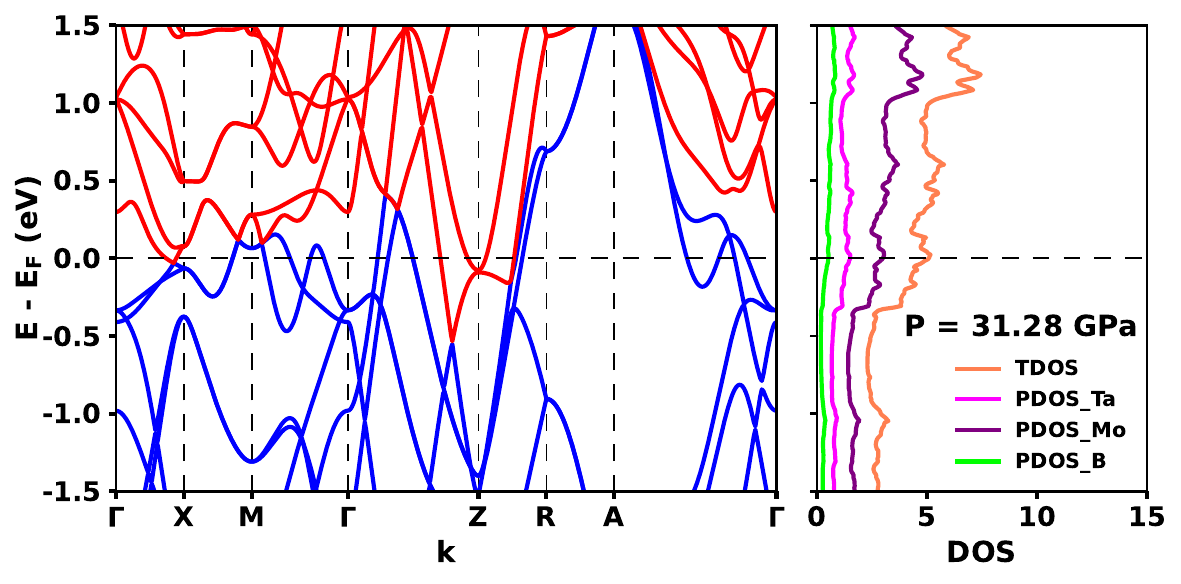}
\includegraphics[width=0.515\linewidth]{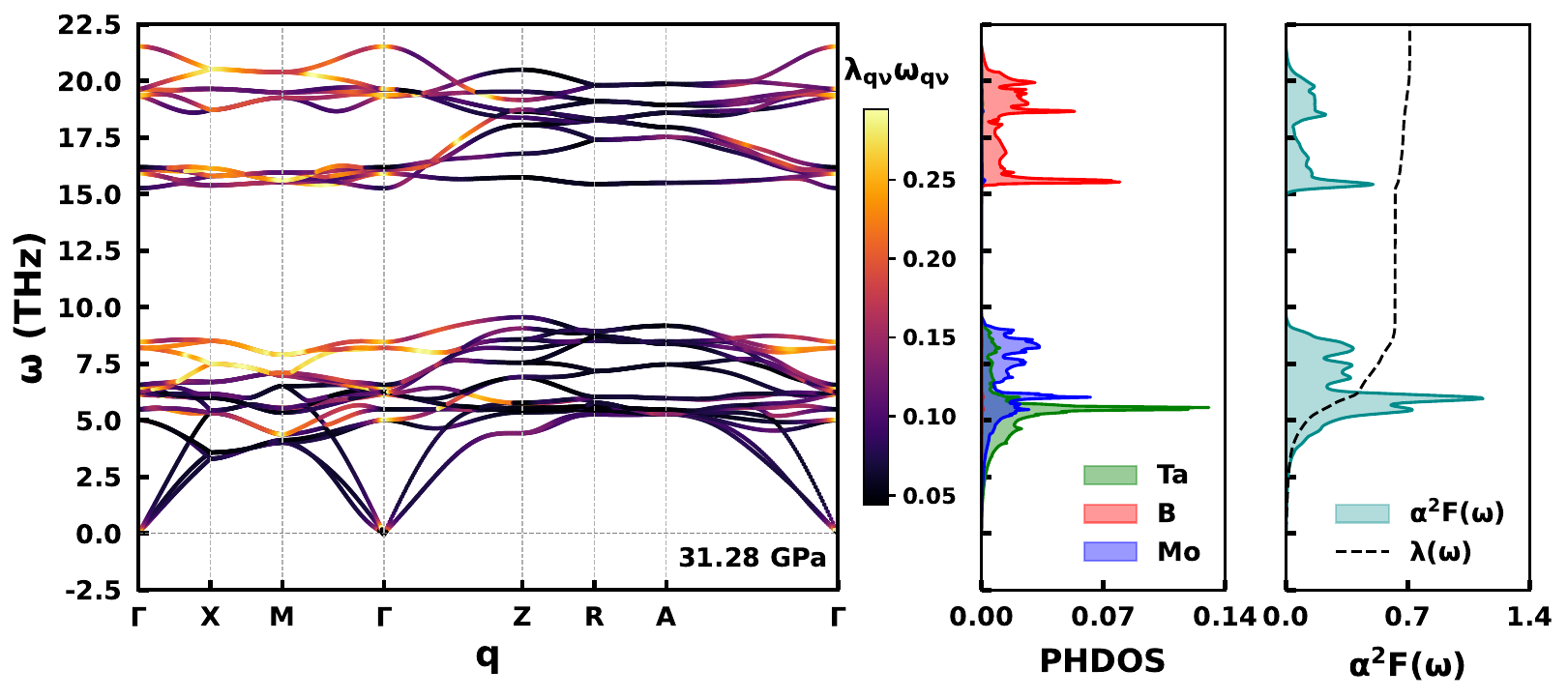}
\vspace*{-0.65cm}
    \begin{center}
        \hspace*{-1.7cm}
        \textbf{\small{(a)}}
        \hspace*{8cm}
        \textbf{\small{(b)}}
    \end{center}
\vspace*{-0.36cm}
\includegraphics[width=0.475\linewidth]{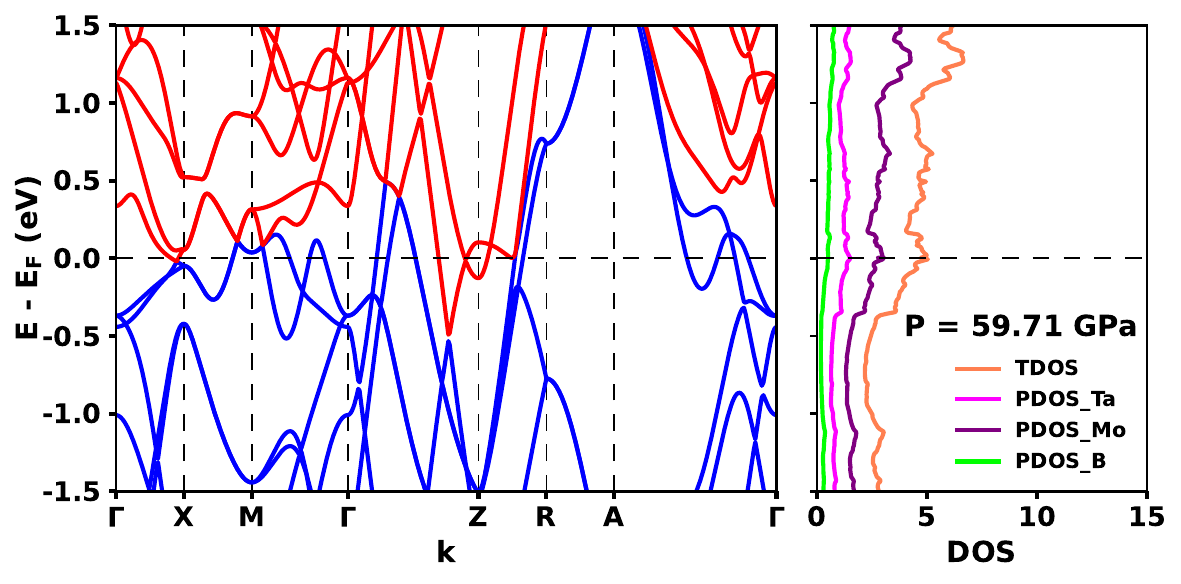}
\includegraphics[width=0.515\linewidth]{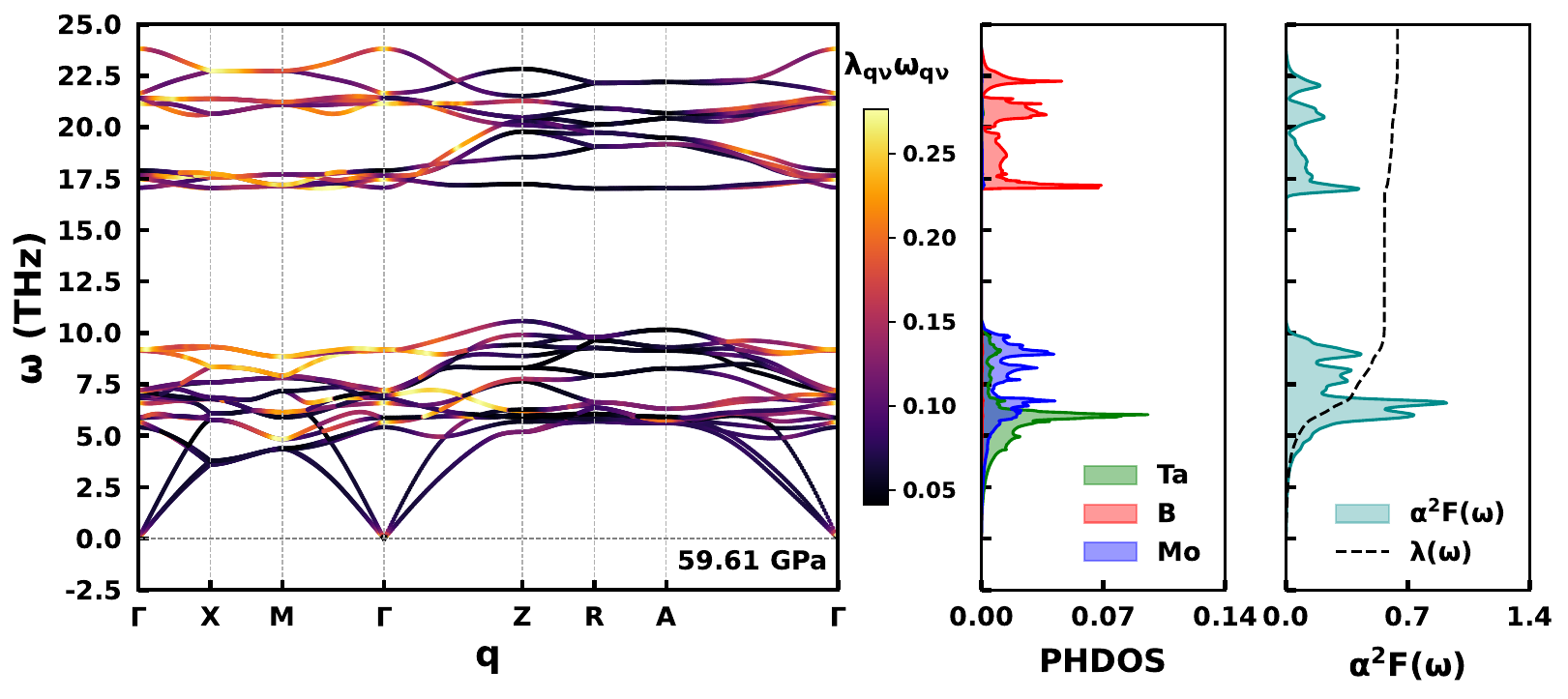}
\vspace*{-0.65cm}
    \begin{center}
        \hspace*{-1.7cm}
        \textbf{\small{(c)}}
        \hspace*{8cm}
        \textbf{\small{(d)}}
    \end{center}
\vspace*{-0.36cm}
\includegraphics[width=0.475\linewidth]{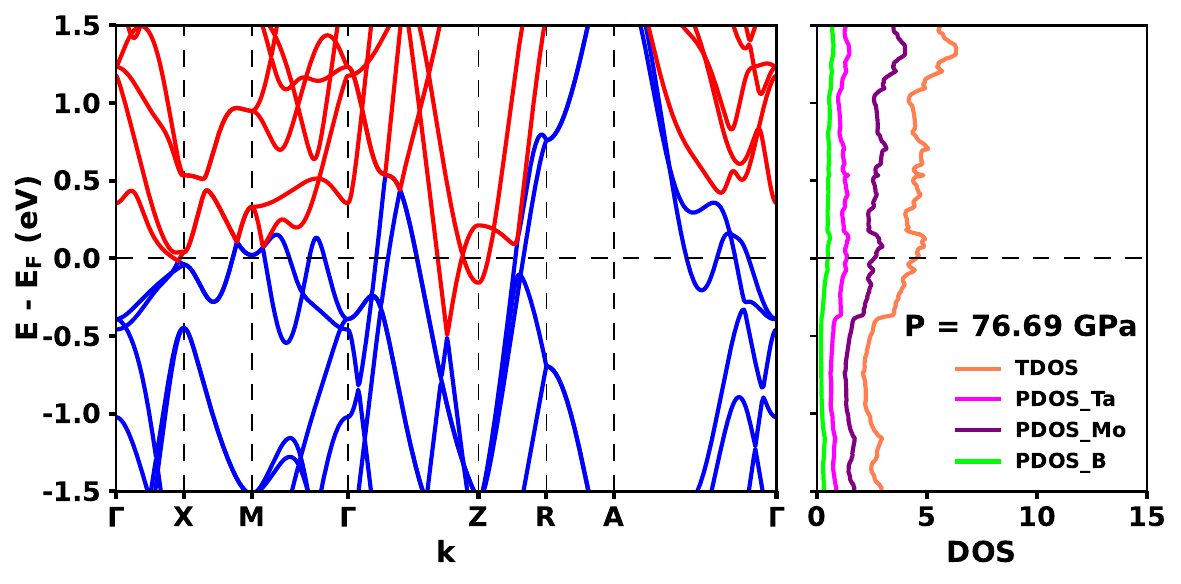}
\includegraphics[width=0.515\linewidth]{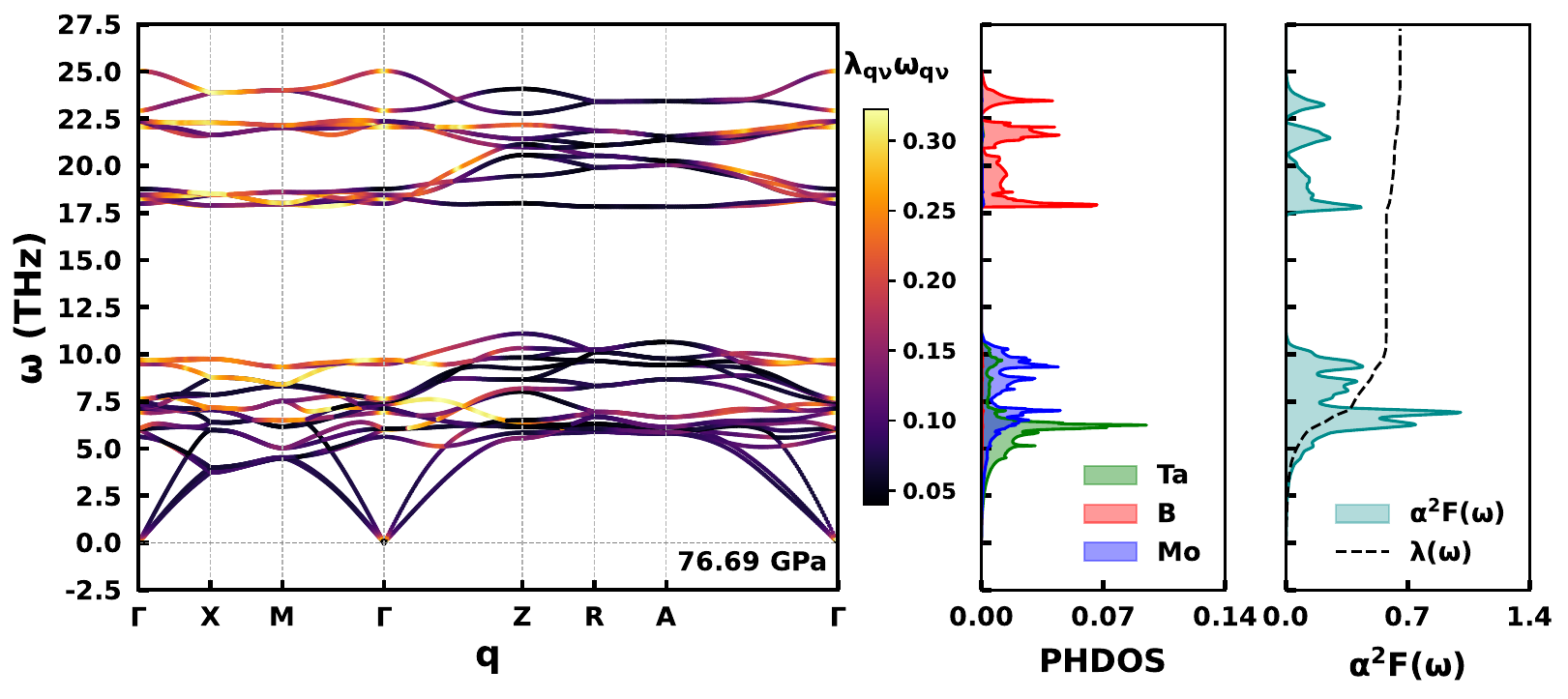}
\vspace*{-0.65cm}
    \begin{center}
        \hspace*{-1.7cm}
        \textbf{\small{(e)}}
        \hspace*{8cm}
        \textbf{\small{(f)}}
    \end{center}
\vspace{-12pt}
\caption{\label{fig:5} Electronic band structure and density of states of $\mathrm{Ta(MoB)_2}$ at different pressure (a) 31.28 GPa, (c) 59.71 GPa and (e) 76.69 GPa. Phonon dispersion, phonon density of states and color plot of $\mathrm{\lambda_{q\nu} \omega_{q\nu}}$ on the phonon bands of $\mathrm{Ta(MoB)_2}$ at different pressure (b) 31.28 GPa, (d) 59.71 GPa and (f) 76.69 GPa.}
\end{figure*}

\begin{figure}[h]

\includegraphics[width=0.30\linewidth]{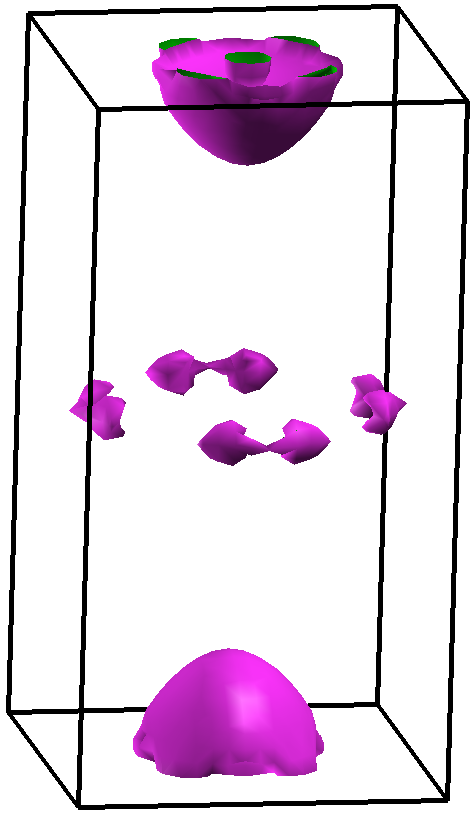}
\includegraphics[width=0.32\linewidth]{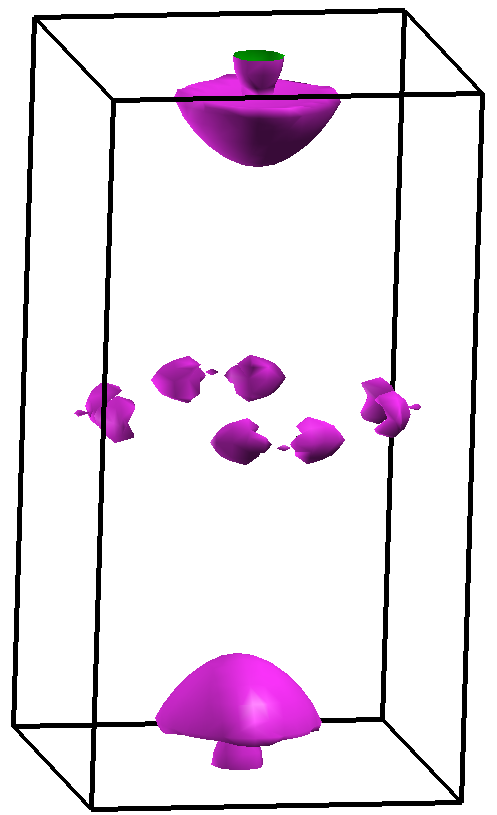}
\vspace*{0.05cm}
    \begin{center}
        \hspace*{-0.3cm}
        \textbf{\small{(a)}}
        \hspace*{2cm}
        \textbf{\small{(b)}}
    \end{center}
\vspace*{0.03cm}
\includegraphics[width=0.30\linewidth]{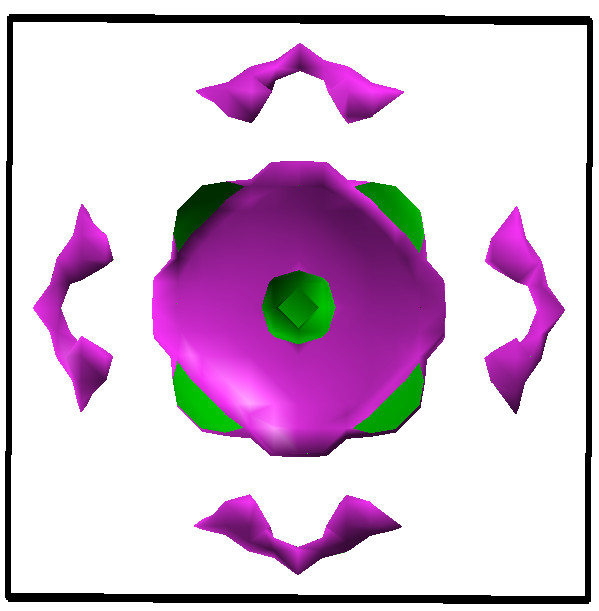}
\includegraphics[width=0.307\linewidth]{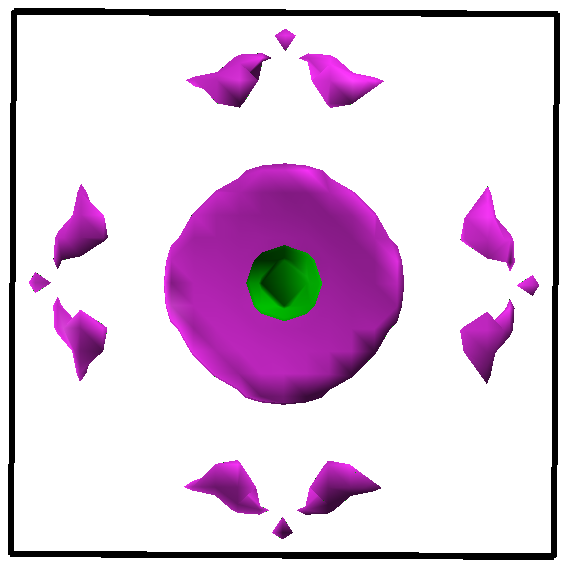}
\vspace*{0.05cm}
    \begin{center}
        \hspace*{-0.3cm}
        \textbf{\small{(c)}}
        \hspace*{2cm}
        \textbf{\small{(d)}}
    \end{center}
\vspace*{0.03cm}
\caption{\label{fig:6} Fermi surface plot of $\mathrm{Ta(MoB)_2}$ corresponding to the third band that intersects the Fermi level (Figure \hyperref[fig:2]{2(f)}) at (a) 59.71 GPa, (b) 76.69 GPa. Top view for (a) 59.71 GPa, (b) 76.69 GPa.}
\end{figure}

Chen et. al. \cite{chen2002effects} presented a theoretical description  which attempts to explain the effect of hydrostatic pressure on EPC. This is directly manifested in the dispersion of electronic and phonon states. Hence it is essential to accurately capture the alteration in the band structure caused by pressure before resolving the issue of superconductivity. Figure \hyperref[fig:5]{5} displays the electronic and phonon dispersion, as well as the color plot of $\mathrm{\lambda_{q\nu} \omega_{q\nu}}$ on the phonon bands along $\mathrm{q}$ points, at three distinct pressures. Under the influence of pressure, Figure \hyperref[fig:5]{5(a)}, \hyperref[fig:5]{(c)} and \hyperref[fig:5]{(e)} capture slight shifts of the electron pocket near the $\mathrm{X}$ point towards the conduction bands, the hole pockets at the $\mathrm{M}$ point towards the valence bands while that along $\mathrm{M - \Gamma}$ and $\mathrm{\Gamma - Z}$ paths towards the conduction band. Since, the phonon modes of acoustic branch and optical branch along the $\mathrm{\Gamma - X - M -\Gamma}$ path and along $\mathrm{\Gamma - Z}$ path are the contributors to superconductivity, (shown in Figure \hyperref[fig:2]{2(b)}), modifications of the bands that couple  to those modes primarily affect superconductivity. Due to the shifts of electron and hole pockets, the total electron concentration with respect to hole decreases with pressure up to 59.71 GPa and increases again at 76.69 GPa. Under ambient pressure, there are two electron pockets located around the Z point. The electron pocket corresponding to the fourth band that crosses the Fermi level (Figure \hyperref[fig:2]{2(g)}) rises above the Fermi level at a pressure 76.69 GPa. This is easily comprehended when viewed in the band-wise colour plot of Figure 3 in the supplementary material \cite{supplementary}. The right panels of Figure \hyperref[fig:5]{5(a)}, \hyperref[fig:5]{(c)}, and \hyperref[fig:5]{(e)} illustrate a gradual decrease in the total density of states and projected density of states on Mo atoms at the Fermi level under pressure. A decrease in the projected density of states on Mo atoms at the Fermi level, which plays a dominant role in superconductivity, is indicative of a decrease in superconductivity \cite{mcmillan1968transition}. Figure \hyperref[fig:5]{5(b)}, \hyperref[fig:5]{(d)}, and \hyperref[fig:5]{(f)} provide evidence of the dynamical stability of $\mathrm{Ta(MoB)_2}$ under high pressure. An increase in pressure results in the stiffening of phonon modes. Phonon modes from the acoustic and optical branch, that contribute to superconductivity along the $\mathrm{\Gamma - X - M - \Gamma}$ path and along the $\mathrm{\Gamma - Z}$ path tend to move up in frequency. Phonon stiffening is a reliable signature of a decrease in superconductivity \cite{mcmillan1968transition,wang2009superconductivity}, whereas softening enhances it \cite{mcmillan1968transition,allen1975superconductivity}. The phonon density of states and the Eliashberg spectral function plotted in the right panels of Figure \hyperref[fig:5]{5(b)}, \hyperref[fig:5]{(d)}, and \hyperref[fig:5]{(f)} also confirms the phonon modes of the acoustic and optical branch, which are contributing to superconductivity, stiffen gradually with pressure. The color plots of $\mathrm{\lambda_{q\nu} \omega_{q\nu}}$ in the phonon bands of Figure \hyperref[fig:5]{5(b)}, \hyperref[fig:5]{(d)}, and \hyperref[fig:5]{(f)} illustrates that it decreases until the pressure reaches 59.71 GPa, and thereafter it starts increasing. This also gets support from the inset of Figure \hyperref[fig:7]{7(d)} where the cumulative EPC strength is plotted against applied pressure. There are two substantial modifications that occur when the pressure is increased from 59.71 GPa to 76.69 GPa. Up to 59.71 GPa, the number of bands that cross the Fermi level is four and it decreases to three at a 76.69 GPa pressure. The Fermi surface plots of $\mathrm{Ta(MoB)_2}$ at pressures 59.71 and 76.69 GPa corresponding to the third band (Figure \hyperref[fig:2]{2(f)}) that intersects the Fermi level is illustrated in Figure \hyperref[fig:6]{6(a)} and \hyperref[fig:6]{(b)} with their top view in Figure \hyperref[fig:6]{6(c)} and \hyperref[fig:6]{(d)} respectively. As the pressure increases from 0 to 59.71 GPa, the open, hemispherical, bowl-shaped electron pocket, located around the Z point, progressively transitions to a closed one. When the pressure reaches 76.69 GPa, this hemispherical electron pocket becomes entirely closed. Adding a short tubular sheet at the center of the hemisphere transforms it into a mushroom-shaped electron pocket, having a smoother surface compared to the hemispherical one. Again, at 76.69 GPa, the electron-like dumbbell-shaped Fermi sheet around the $\mathrm{X}$ point (along the $\mathrm{\Gamma - X}$ path) gets divided into three smaller segments. In Figure \hyperref[fig:6]{6}, two opposite faces of the Fermi surface are shown using pink and green colors. Such a change in the topology of the Fermi surface leads to a Lifshitz transition \cite{wang2024theoretical,feng2022superconductivity,lifshitz1960anomalies}. These transitions can occur due to the change of external parameters like hydrostatic pressure, doping, magnetic field etc, resulting in the alterations of the electronic characteristics of the material, including changes in the density of states at the Fermi level, electrical conductivity, nesting condition etc.\cite{wang2024theoretical,feng2022superconductivity,lifshitz1960anomalies}.

\subsection{\label{sec:3.6}Impact of Pressure on Superconducting Critical Temperature in $\mathbf{Ta(MoB)_2}$}

\begin{figure*}[t]
\includegraphics[width=0.36\linewidth]{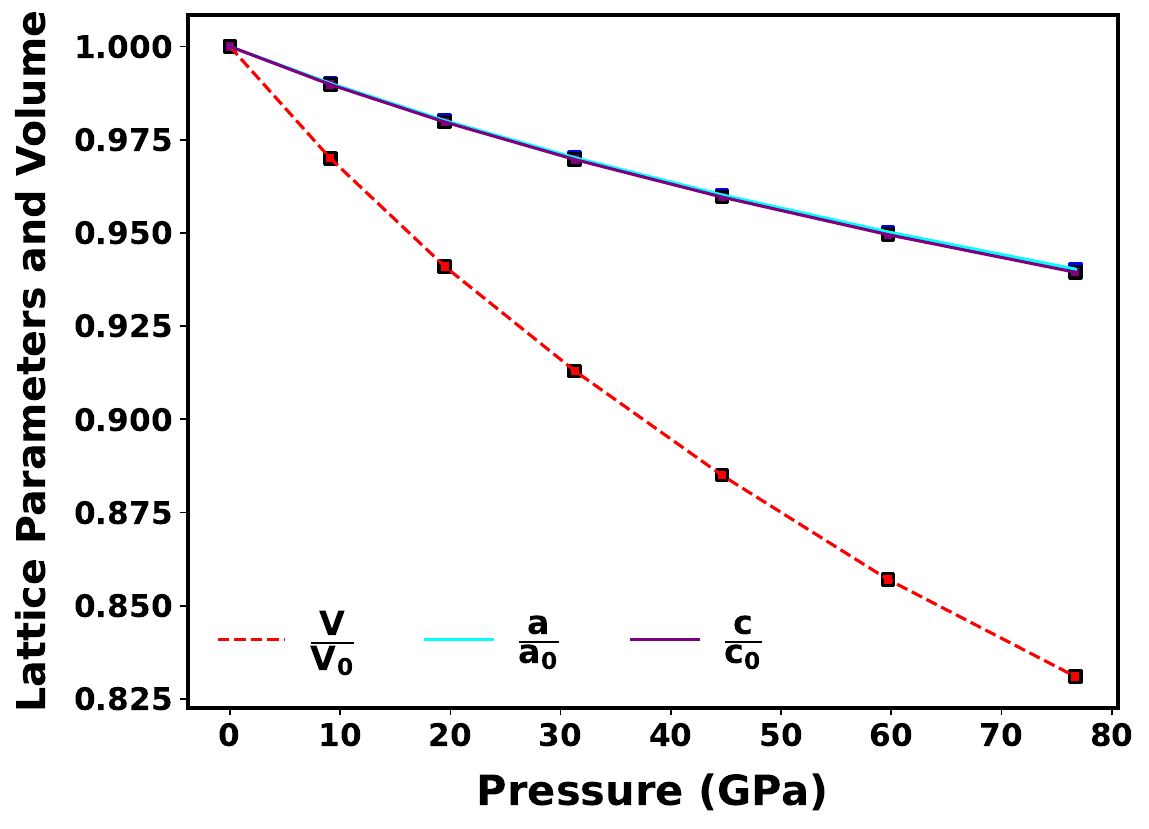}
\includegraphics[width=0.348\linewidth]{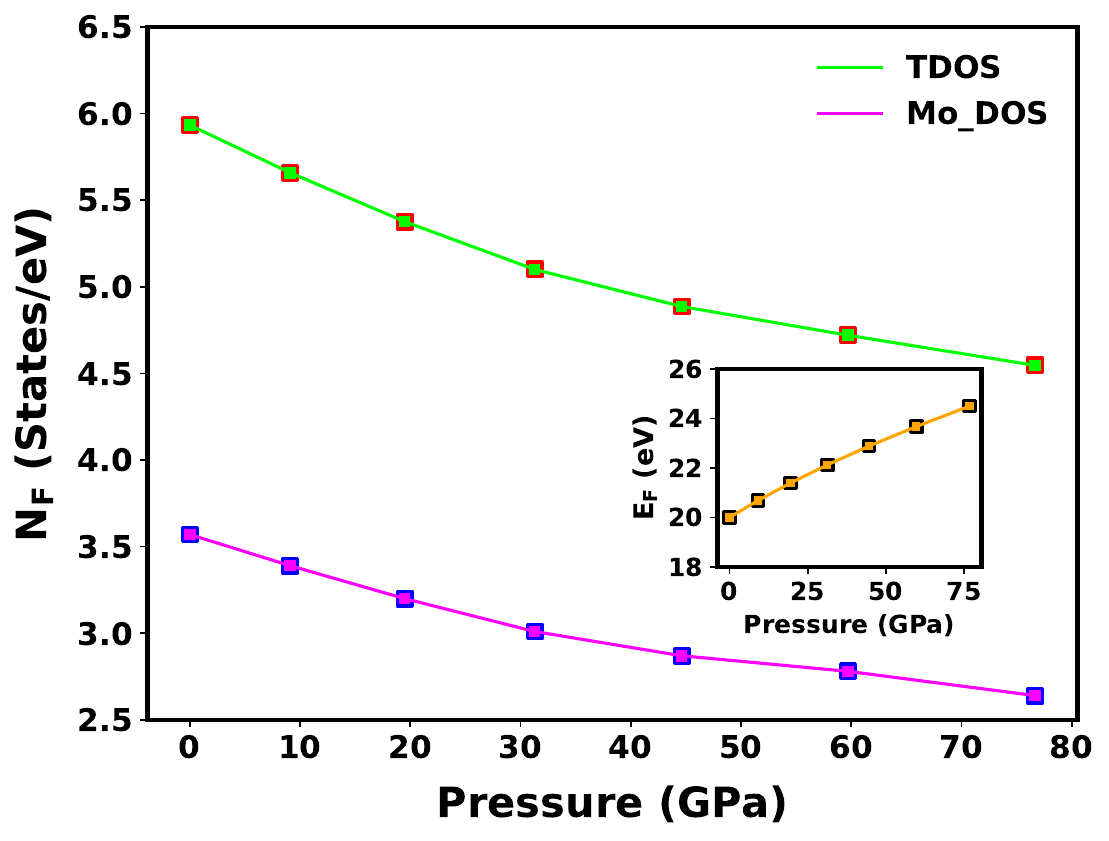}
\vspace{-6pt}
\begin{center}
        \hspace*{1.2cm}
        \textbf{\small{(a)}}
        \hspace*{5.6cm}
        \textbf{\small{(b)}}
    \end{center}
\vspace{-6pt}
\includegraphics[width=0.354\linewidth]{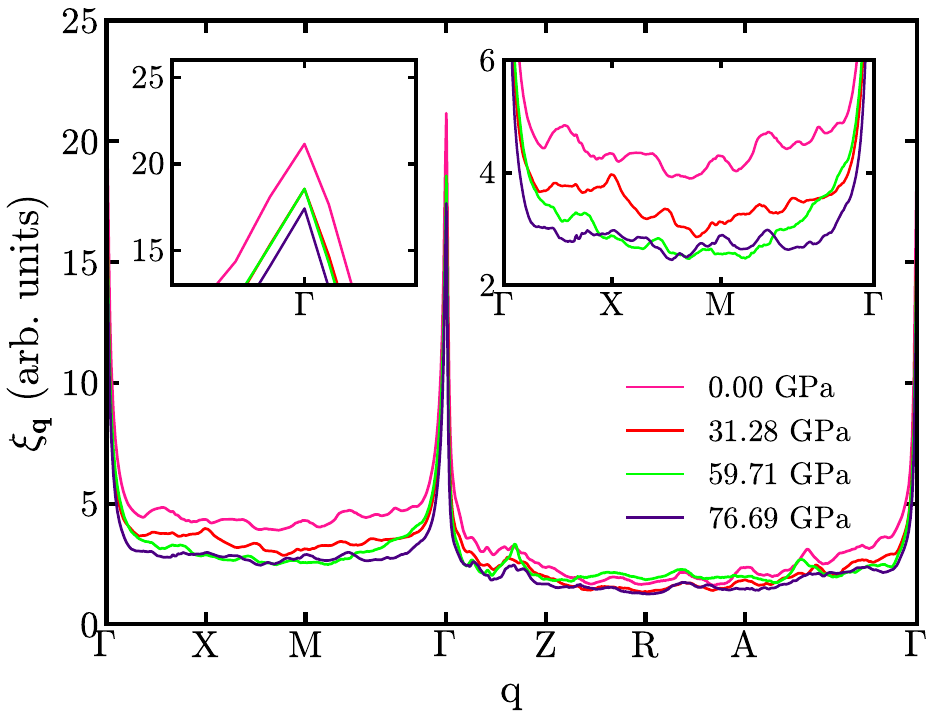}
\includegraphics[width=0.354\linewidth]{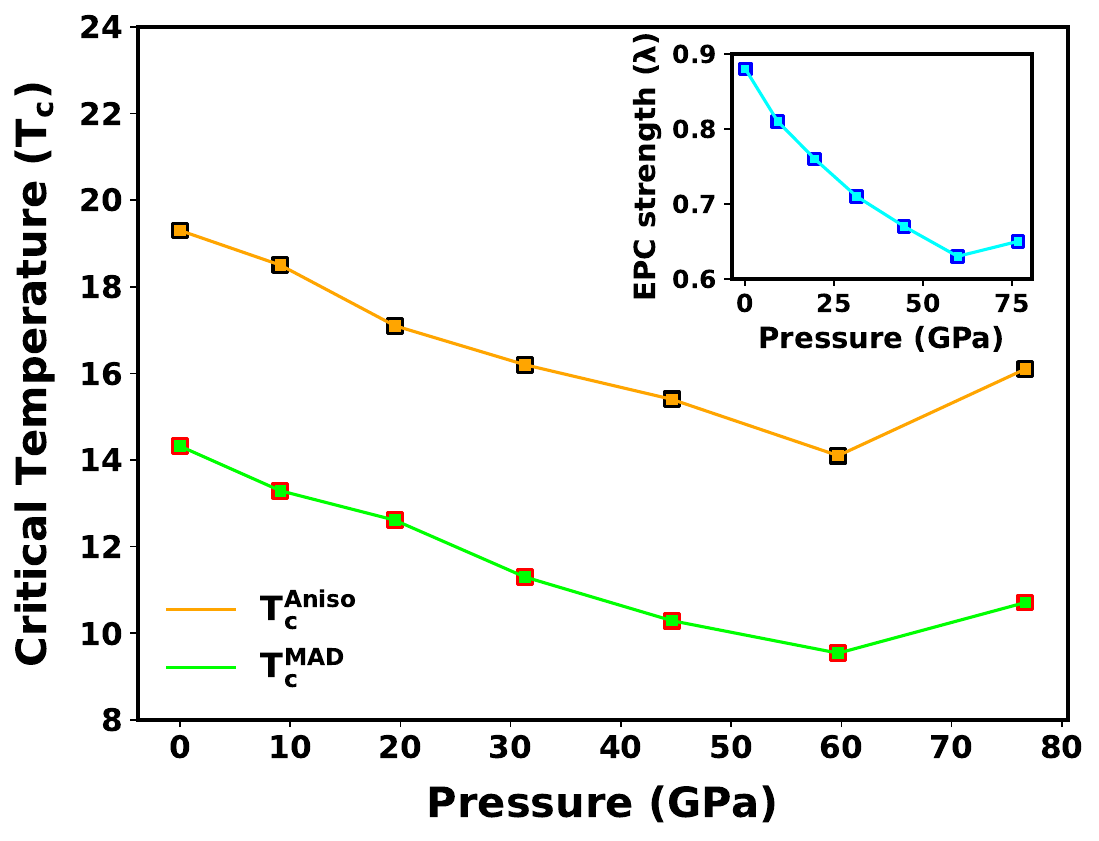}
\vspace{-6pt}
\begin{center}
        \hspace*{0.8cm}
        \textbf{\small{(c)}}
        \hspace*{5.6cm}
        \textbf{\small{(d)}}
    \end{center}
\vspace{-8pt}
\caption{\label{fig:7} (a) Change in lattice parameters and volume of unit cell with pressure, (b) Main: Variation of density of states at the Fermi level with pressure, Inset: Plot of Fermi energy vs pressure, (c) Main: Fermi nesting function for different pressures along high symmetry q points, Inset: zoomed version of the main figure, (d) Main: Variation of superconducting critical temperature with pressure, Inset: Plot of EPC strength versus pressure}
\end{figure*}

The lattice parameters and volume of the unit cell of $\mathrm{Ta(MoB)_2}$ are reducing under compressive pressure, illustrated in Figure \hyperref[fig:7]{7(a)}. The absence of any discontinuity in the volume indicates the absence of structural phase transition up to 76.69 GPa. However, further clarification should be achieved through future experiments. We have examined the stability, electronic and superconducting properties of $\mathrm{Ta(MoB)_2}$ upto 76.69 GPa. Although unconventional superconductivity involves pairing mechanisms that extend beyond electron-phonon interactions, such as spin fluctuations, electron-electron interactions (e.g., in heavy-fermion compounds), conventional superconductivity explicitly involves electron-phonon coupling. The McMillan Hopfield equation can be used to calculate the EPC strength as follows \cite{mcmillan1968transition}:

\begin{eqnarray}
\mathrm{\lambda = \frac{N_F \textlangle I^2 \textrangle}{M \textlangle \omega^2 \textrangle}}
\label{eq:15}
\end{eqnarray}

where $\mathrm{N_F}$ is the density of states at the Fermi level, $\mathrm{\textlangle I^2 \textrangle}$ is the average of the square of electron-phonon matrix element, $\mathrm{\textlangle \omega^2 \textrangle}$ is the averaged square of the phonon frequency and $\mathrm{M}$ is the mass of the ion involved. The total and the projected density of states on Mo atoms at the Fermi level decreases almost linearly at a rate of -0.0179 states/eV/GPa and -0.0119 states/eV/GPa, respectively, as the pressure increases (main part of Figure \hyperref[fig:7]{7(b)}). Figure \hyperref[fig:5]{5(b)}, \hyperref[fig:5]{(d)} and \hyperref[fig:5]{(f)} depict how the frequency of those phonon modes in the acoustic and optical branch along the $\mathrm{\Gamma - X - M - \Gamma}$ path and along $\mathrm{\Gamma - Z}$ path which contribute to EPC strength and superconductivity, increases as pressure increases. Equation \hyperref[eq:15]{(15)} suggests that these factors contribute to a decrease in EPC strength under pressure. The inset of Figure \hyperref[fig:7]{7(b)} shows an increase in Fermi energy as pressure is applied. The effect of this increase is also evident in the electronic band structure under high pressure, already discussed in the preceding section. The main and inset of Figure \hyperref[fig:7]{7(d)} shows the response of superconducting critical temperature ($\mathrm{T_c}$) and EPC strength ($\mathrm{\lambda}$) with applied hydrostatic pressure, respectively. It shows that the EPC strength, and hence the superconducting critical temperature, drop almost linearly with pressure up to 59.71 GPa, as supported by the McMillan Hopfield equation \cite{mcmillan1968transition}. However, when the pressure is raised beyond 59.71 GPa, up to 76.69 GPa, both the EPC strength and superconductivity increase once more. This increase serves as an indication of an electronic phase transition. In addition to the governing factors mentioned in Equation \hyperref[eq:15]{(15)}, Fermi nesting is another significant component that can influence the EPC strength and superconductivity indirectly by altering the density of states at the Fermi level, EPC matrix elements and other related parameters. In Figure \hyperref[fig:7]{7(c)}, the values of Fermi nesting function along $\mathrm{q}$ points for various pressures is displayed. The peak corresponding to self-nesting (located at the $\mathrm{\Gamma}$ point) and the small peaks corresponding to weak nestings located along $\mathrm{\Gamma-X-M-\Gamma}$ path, fall with pressure up to 59.71 GPa. The reduction of the peak corresponding to self-nesting is possibly due to intra-band contribution that is reflected through the decline in density of states at the Fermi level with pressure, shown in Figure \hyperref[fig:7]{7(b)}. The peaks associated with weak nesting diminish with pressure up to 59.71 GPa due to band shifts as captured in Figure \hyperref[fig:5]{5(a)},\hyperref[fig:5]{(c)},\hyperref[fig:5]{(e)} resulting in a reduction of the number of $\mathrm{k}$ points available for nesting. In addition to what Equation \hyperref[eq:15]{(15)} has already revealed, a decrease in the peak of the nesting function up to 59.71 GPa also suggests a decline in superconductivity. Increasing the pressure from 59.71 GPa to 76.69 GPa further makes the peak corresponding to weak nesting sharper along $\mathrm{\Gamma-X-M-\Gamma}$ path. This leads to an increase in EPC strength and superconductivity, shown in Figure \hyperref[fig:7]{7(d)}. Fermi nesting has the potential to induce several instabilities, such as charge density wave (CDW), superconductivity (SC), spin density waves in a system \cite{johannes2008fermi,johannes2006fermi,kaboudvand2022fermi}. The abrupt enhancement of the weak nesting peak along $\mathrm{\Gamma-X-M-\Gamma}$ path indicates the increase in number of $\mathrm{k}$ points available for nesting. This is actually related to the alteration in the topology of the Fermi surface, namely the Lifshitz transition, occurring at a 76.69 GPa pressure. Thus, the superconducting T$_c$ exhibits a V-shaped pattern under applied pressure. Yue et. al. have also noticed similar contrasting response of the Fermi nesting function and the V-shaped superconductivity under pressure due to structural phase transition \cite{yue2018electron}. 

\section{\label{sec:4} Summary and Conclusions}

To summarise our results, a Migdal Eliashberg theory is used to investigate the anisotropic phonon-mediated superconductivity in $\mathrm{Ta(MoB)_2}$. Some vibrational modes of acoustic and optical branch along $\mathrm{\Gamma - X - M - \Gamma}$ and $\mathrm{\Gamma - Z}$ path contribute to superconductivity of $\mathrm{Ta(MoB)_2}$. A significant contribution is coming from the coupling between electronic states of Mo $\mathrm{d_{xy}}$,$\mathrm{d_{x^2 - y^2}}$ orbitals and the in-plane vibrational modes of Mo atoms. The anisotropy in superconductivity here owes its origin to the multi-band nature of the Fermi surface. Our findings suggest that CDW instability is unlikely here due to weak Fermi surface nesting and a small EPC. The absence of a notable peak in the real part of the total Lindhard susceptibility and the lack of phonon softening assures this. Furthermore, We have applied hydrostatic pressure up to 76.69 GPa due to its metastability and low bulk modulus of $\mathrm{Ta(MoB)_2}$. A linear decrease trend in the EPC strength and superconductivity up to 59.71 GPa is observed. This is attributed to a consistent reduction in the density of states at the Fermi level, the stiffening of superconductivity contributing phonon modes, and the gradual decrease in the self-nesting, weak-nesting peaks. As the pressure increases from 59.71 GPa to 76.69 GPa, the weak-nesting peak along $\mathrm{\Gamma - X - M - \Gamma}$ path are found to be sharped due to the modification of the Fermi surface topology. This makes a sudden increase in the EPC strength and superconductivity at 76.69 GPa. Thus, our investigation reveals that superconductivity of $\mathrm{Ta(MoB)_2}$ under pressure has a V-shaped response followed by an electronic phase change from the Lifshitz transition. We believe that future experiments at high pressure may shed more light on this. 

\begin{acknowledgments}
\vspace{-2pt}
The National Supercomputing Mission (NSM) is gratefully acknowledged for granting us access to the PARAM Shakti computing resources at IIT Kharagpur. This infrastructure, which is being implemented by C-DAC, is being supported by the Ministry of Electronics and Information Technology (MeitY) and the Department of Science and Technology (DST), governments of India. High performance computing facility of HRI is also acknowledged. S.P. is thankful to Dr. Swapnil Deshpande and Dr. Shubham Patel for useful discussion. S.P. also acknowledges IIT Kharagpur for providing research fellowship. We would also like to thank the anonymous reviewers for their insightful suggestions.
\end{acknowledgments}

\nocite{*}
\bibliography{apssamp}

@article{Yao2021,
title = {Superconducting materials: Challenges and opportunities for large-scale applications},
journal = {iScience},
volume = {24},
number = {6},
pages = {102541},
year = {2021},
issn = {2589-0042},
doi = {https://doi.org/10.1016/j.isci.2021.102541},
url = {https://www.sciencedirect.com/science/article/pii/S2589004221005095},
author = {Chao Yao and Yanwei Ma}
}

@article{bardeen1957theory,
  title={Theory of superconductivity},
  author={Bardeen, John and Cooper, Leon N and Schrieffer, John Robert},
  journal={Physical review},
  volume={108},
  number={5},
  pages={1175},
  year={1957},
  publisher={APS},
  url={https://journals.aps.org/pr/abstract/10.1103/PhysRev.108.1175}
}

@article{mcmillan1968transition,
  title={Transition temperature of strong-coupled superconductors},
  author={McMillan, WL},
  journal={Physical Review},
  volume={167},
  number={2},
  pages={331},
  year={1968},
  publisher={APS},
  url={https://journals.aps.org/pr/abstract/10.1103/PhysRev.167.331}
}

@article{allen1975transition,
  title={Transition temperature of strong-coupled superconductors reanalyzed},
  author={Allen, Ph B and Dynes, RC},
  journal={Physical Review B},
  volume={12},
  number={3},
  pages={905},
  year={1975},
  publisher={APS},
  url={https://journals.aps.org/prb/abstract/10.1103/PhysRevB.12.905}
}

@article{migdal1958interaction,
  title={Interaction between electrons and lattice vibrations in a normal metal},
  author={Migdal, AB},
  journal={Sov. Phys. JETP},
  volume={7},
  number={6},
  pages={996--1001},
  year={1958},
  url={http://jetp.ras.ru/cgi-bin/e/index/e/7/6/p996?a=list}
}

@article{eliashberg1960interactions,
  title={Interactions between electrons and lattice vibrations in a superconductor},
  author={Eliashberg, GM},
  journal={Sov. Phys. JETP},
  volume={11},
  number={3},
  pages={696--702},
  year={1960},
  url={http://jetp.ras.ru/cgi-bin/e/index/e/11/3/p696?a=list}
}

@article{hohenberg1964inhomogeneous,
  title={Inhomogeneous electron gas},
  author={Hohenberg, Pierre and Kohn, Walter},
  journal={Physical review},
  volume={136},
  number={3B},
  pages={B864},
  year={1964},
  publisher={APS},
  url={https://journals.aps.org/pr/abstract/10.1103/PhysRev.136.B864}
}

@article{oliveira1988density,
  title={Density-functional theory for superconductors},
  author={Oliveira, Luiz Nunes de and Gross, EKU and Kohn, W},
  journal={Physical review letters},
  volume={60},
  number={23},
  pages={2430},
  year={1988},
  publisher={APS},
  url={https://journals.aps.org/prl/abstract/10.1103/PhysRevLett.60.2430}
}

@article{nagamatsu2001superconductivity,
  title={Superconductivity at 39 K in magnesium diboride},
  author={Nagamatsu, Jun and Nakagawa, Norimasa and Muranaka, Takahiro and Zenitani, Yuji and Akimitsu, Jun},
  journal={nature},
  volume={410},
  number={6824},
  pages={63--64},
  year={2001},
  publisher={Nature Publishing Group UK London},
  url={https://www.nature.com/articles/35065039}
}

@article{kong2001electron,
  title={Electron-phonon interaction in the normal and superconducting states of $\mathrm{MgB_2}$},
  author={Kong, Y and Dolgov, OV and Jepsen, O and Andersen, OK},
  journal={Physical Review B: Condensed Matter and Materials Physics},
  volume={64},
  number={2},
  year={2001},
  publisher={American Physical Society},
  url={https://journals.aps.org/prb/abstract/10.1103/PhysRevB.64.020501}
}

@article{kortus2001superconductivity,
  title={Superconductivity of metallic boron in $\mathrm{MgB_2}$},
  author={Kortus, Jens and Mazin, II and Belashchenko, Kirill D and Antropov, Vladimir P and Boyer, LL},
  journal={Physical Review Letters},
  volume={86},
  number={20},
  pages={4656},
  year={2001},
  publisher={APS},
  url={https://journals.aps.org/prl/abstract/10.1103/PhysRevLett.86.4656}
}

@article{liu2001beyond,
  title={Beyond Eliashberg superconductivity in $\mathrm{MgB_2}$: anharmonicity, two-phonon scattering, and multiple gaps},
  author={Liu, Amy Y and Mazin, II and Kortus, Jens},
  journal={Physical Review Letters},
  volume={87},
  number={8},
  pages={087005},
  year={2001},
  publisher={APS},
  url={https://journals.aps.org/prl/abstract/10.1103/PhysRevLett.87.087005}
}

@article{iavarone2002two,
  title={Two-Band Superconductivity in $\mathrm{MgB_2}$},
  author={Iavarone, Maria and Karapetrov, G and Koshelev, AE and Kwok, WK and Crabtree, GW and Hinks, DG and Kang, WN and Choi, Eun-Mi and Kim, Hyun Jung and Kim, Hyeong-Jin and others},
  journal={Physical review letters},
  volume={89},
  number={18},
  pages={187002},
  year={2002},
  publisher={APS},
  url={https://journals.aps.org/prl/abstract/10.1103/PhysRevLett.89.187002}
}

@article{szabo2001evidence,
  title={Evidence for two superconducting energy gaps in $\mathrm{MgB_2}$ by point-contact spectroscopy},
  author={Szab{\'o}, P and Samuely, P and Ka{\v{c}}mar{\v{c}}{\'\i}k, J and Klein, Thierry and Marcus, J and Fruchart, Daniel and Miraglia, Salvatore and Marcenat, C and Jansen, AGM},
  journal={Physical review letters},
  volume={87},
  number={13},
  pages={137005},
  year={2001},
  publisher={APS},
  url={https://journals.aps.org/prl/abstract/10.1103/PhysRevLett.87.137005}
}

@article{putti2011mgb2,
  title={$\mathrm{MgB_2}$, a two-gap superconductor for practical applications},
  author={Putti, Marina and Grasso, Giovanni},
  journal={MRS bulletin},
  volume={36},
  number={8},
  pages={608--613},
  year={2011},
  publisher={Cambridge University Press},
  url={https://www.cambridge.org/core/journals/mrs-bulletin/article/mgb2-a-twogap-superconductor-for-practical-applications/6914D14489A795DF08BD5A2B33B858FD}
}

@article{choi2002origin,
  title={The origin of the anomalous superconducting properties of $\mathrm{MgB_2}$},
  author={Choi, Hyoung Joon and Roundy, David and Sun, Hong and Cohen, Marvin L and Louie, Steven G},
  journal={Nature},
  volume={418},
  number={6899},
  pages={758--760},
  year={2002},
  publisher={Nature Publishing Group UK London},
  url={https://www.nature.com/articles/nature00898}
}

@article{margine2013anisotropic,
  title={Anisotropic migdal-eliashberg theory using wannier functions},
  author={Margine, Elena Roxana and Giustino, Feliciano},
  journal={Physical Review B—Condensed Matter and Materials Physics},
  volume={87},
  number={2},
  pages={024505},
  year={2013},
  publisher={APS},
  url={https://journals.aps.org/prb/abstract/10.1103/PhysRevB.87.024505}
}

@article{souma2003origin,
  title={The origin of multiple superconducting gaps in $\mathrm{MgB_2}$},
  author={Souma, S and Machida, Y and Sato, T and Takahashi, T and Matsui, H and Wang, S-C and Ding, H and Kaminski, A and Campuzano, JC and Sasaki, S and others},
  journal={Nature},
  volume={423},
  number={6935},
  pages={65--67},
  year={2003},
  publisher={Nature Publishing Group UK London},
  url={https://www.nature.com/articles/nature01619}
}

@article{akimitsu2005superconductivity,
  title={Superconductivity in $\mathrm{MgB_2}$ and its related materials},
  author={Akimitsu, Jun and Akutagawa, Satoshi and Kawashima, Kenji and Muranaka, Takahiro},
  journal={Progress of Theoretical Physics Supplement},
  volume={159},
  pages={326--337},
  year={2005},
  publisher={Oxford Academic},
  url={https://academic.oup.com/ptps/article/doi/10.1143/PTPS.159.326/1871795}
}

@article{katsura2010possibility,
  title={On the possibility of $\mathrm{MgB_2}$-like superconductivity in potassium hexaboride},
  author={Katsura, Yukari and Yamamoto, Ayako and Ogino, Hiraku and Horii, Shigeru and Shimoyama, Jun-ichi and Kishio, Kohji and Takagi, Hidenori},
  journal={Physica C: Superconductivity and its applications},
  volume={470},
  pages={S633--S634},
  year={2010},
  publisher={Elsevier},
  url={https://www.sciencedirect.com/science/article/pii/S0921453410000055}
}

@article{yu2022superconductive,
  title={Superconductive materials with $\mathrm{MgB_2}$-like structures from data-driven screening},
  author={Yu, Ze and Bo, Tao and Liu, Bo and Fu, Zhendong and Wang, Huan and Xu, Sheng and Xia, Tianlong and Li, Shiliang and Meng, Sheng and Liu, Miao},
  journal={Physical Review B},
  volume={105},
  number={21},
  pages={214517},
  year={2022},
  publisher={APS},
  url={https://journals.aps.org/prb/abstract/10.1103/PhysRevB.105.214517}
}

@article{miao2016first,
  title={First-principles prediction of $\mathrm{MgB_2}$-like $\mathrm{NaBC}$: A more promising high-temperature superconducting material than $\mathrm{LiBC}$},
  author={Miao, Rende and Huang, Guiqin and Yang, Jun},
  journal={Solid State Communications},
  volume={233},
  pages={30--34},
  year={2016},
  publisher={Elsevier},
  url={https://www.sciencedirect.com/science/article/pii/S0038109816000612}
}

@article{an2021superconductivity,
  title={Superconductivity and topological properties of $\mathrm{MgB_2}$-type diborides from first principles},
  author={An, Yipeng and Li, Jie and Wang, Kun and Wang, Guangtao and Gong, Shijing and Ma, Chunlan and Wang, Tianxing and Jiao, Zhaoyong and Dong, Xiao and Xu, Guoliang and others},
  journal={Physical Review B},
  volume={104},
  number={13},
  pages={134510},
  year={2021},
  publisher={APS},
  url={https://journals.aps.org/prb/abstract/10.1103/PhysRevB.104.134510}
}

@article{wang2023covalent,
  title={Covalent bond inducing strong electron-phonon coupling superconductivity in $\mathrm{MgB_2}$-type transition metal diboride $\mathrm{WB_2}$},
  author={Wang, Jiajun and Wang, Muyao and Liu, Xiaohan and Jiang, Man and Liu, Liangliang},
  journal={Physical Review Materials},
  volume={7},
  number={7},
  pages={074804},
  year={2023},
  publisher={APS},
  url={https://journals.aps.org/prmaterials/abstract/10.1103/PhysRevMaterials.7.074804}
}

@article{pallecchi2009investigation,
  title={Investigation of $\mathrm{Li}$-doped $\mathrm{MgB_2}$},
  author={Pallecchi, I and Brotto, P and Ferdeghini, C and Putti, Marina and Palenzona, A and Manfrinetti, P and Lehmann, A Geddo and Orecchini, Andrea and Petrillo, Caterina and Sacchetti, Francesco and others},
  journal={Superconductor Science and Technology},
  volume={22},
  number={9},
  pages={095014},
  year={2009},
  publisher={IOP Publishing},
  url={https://iopscience.iop.org/article/10.1088/0953-2048/22/9/095014/meta?casa_token=18kFJH8X_vAAAAAA:cS-z4yWlMgT4jNh5G3GOzMpzGM97TAU7N0GQQIH832MJl9c5e2CqtOLxCIWSVRtra3RGaUZxrE3wSr7UgE2GQjMhVc4YwA}
}

@article{feng2002enhanced,
  title={Enhanced flux pinning in $\mathrm{Zr}$-doped $\mathrm{MgB_2}$ bulk superconductors prepared at ambient pressure},
  author={Feng, Y and Zhao, Yong and Pradhan, AK and Cheng, CH and Yau, JKF and Zhou, L and Koshizuka, N and Murakami, Masato},
  journal={Journal of applied physics},
  volume={92},
  number={5},
  pages={2614--2619},
  year={2002},
  publisher={American Institute of Physics},
  url={https://pubs.aip.org/aip/jap/article/92/5/2614/472195/Enhanced-flux-pinning-in-Zr-doped-MgB2-bulk}
}

@article{pei2023pressure,
  title={Pressure-induced superconductivity at 32 K in $\mathrm{MoB_2}$},
  author={Pei, Cuiying and Zhang, Jianfeng and Wang, Qi and Zhao, Yi and Gao, Lingling and Gong, Chunsheng and Tian, Shangjie and Luo, Ruitao and Li, Mingtao and Yang, Wenge and others},
  journal={National Science Review},
  volume={10},
  number={5},
  pages={nwad034},
  year={2023},
  publisher={Oxford University Press},
  url={https://academic.oup.com/nsr/article/10/5/nwad034/7036811}
}

@article{lim2022creating,
  title={Creating superconductivity in $\mathrm{WB_2}$ through pressure-induced metastable planar defects},
  author={Lim, J and Hire, AC and Quan, Y and Kim, JS and Xie, SR and Sinha, S and Kumar, RS and Popov, D and Park, C and Hemley, RJ and others},
  journal={Nature communications},
  volume={13},
  number={1},
  pages={7901},
  year={2022},
  publisher={Nature Publishing Group UK London},
  url={https://www.nature.com/articles/s41467-022-35191-8}
}

@article{chen2024high,
  title={High-throughput screening for boride superconductors},
  author={Chen, Shiya and Wu, Zepeng and Zhang, Zhen and Wu, Shunqing and Ho, Kai-Ming and Antropov, Vladimir and Sun, Yang},
  journal={Inorganic Chemistry},
  volume={63},
  number={19},
  pages={8654--8663},
  year={2024},
  publisher={ACS Publications},
  url={https://pubs.acs.org/doi/full/10.1021/acs.inorgchem.4c00159}
}

@article{wudil2022hydrostatic,
  title={Hydrostatic pressure-tuning of thermoelectric properties of $\mathrm{CsSnI_3}$ perovskite by first-principles calculations},
  author={Wudil, YS and Peng, Q and Alsayoud, AQ and Gondal, MA},
  journal={Computational Materials Science},
  volume={201},
  pages={110917},
  year={2022},
  publisher={Elsevier},
  url={https://www.sciencedirect.com/science/article/pii/S0927025621006200}
}

@article{zhang2021hydrostatic,
  title={Hydrostatic pressure tuning of thermal conductivity for $\mathrm{PbTe}$ and $\mathrm{PbSe}$ considering pressure-induced phase transitions},
  author={Zhang, Min and Tang, Guihua and Li, Yifei},
  journal={ACS omega},
  volume={6},
  number={5},
  pages={3980--3990},
  year={2021},
  publisher={ACS Publications},
  url={https://pubs.acs.org/doi/full/10.1021/acsomega.0c05907}
}

@article{hemme2023tuning,
  title={Tuning the Multiferroic Properties of $\mathrm{BiFeO_3}$ under Uniaxial Strain},
  author={Hemme, P and Philippe, JC and Medeiros, A and Alekhin, A and Houver, S and Gallais, Y and Sacuto, A and Forget, A and Colson, D and Mantri, S and others},
  journal={Physical Review Letters},
  volume={131},
  number={11},
  pages={116801},
  year={2023},
  publisher={APS},
  url={https://journals.aps.org/prl/abstract/10.1103/PhysRevLett.131.116801}
}

@article{wang2023controlled,
  title={Controlled frustration release on the kagome lattice by uniaxial-strain tuning},
  author={Wang, Jierong and Spitaler, M and Su, Y-S and Zoch, KM and Krellner, C and Puphal, P and Brown, Stuart E and Pustogow, A},
  journal={Physical Review Letters},
  volume={131},
  number={25},
  pages={256501},
  year={2023},
  publisher={APS},
  url={https://journals.aps.org/prl/abstract/10.1103/PhysRevLett.131.256501}
}

@article{frisenda2017biaxial,
  title={Biaxial strain tuning of the optical properties of single-layer transition metal dichalcogenides},
  author={Frisenda, Riccardo and Dr{\"u}ppel, Matthias and Schmidt, Robert and Michaelis de Vasconcellos, Steffen and Perez de Lara, David and Bratschitsch, Rudolf and Rohlfing, Michael and Castellanos-Gomez, Andres},
  journal={npj 2D Materials and Applications},
  volume={1},
  number={1},
  pages={10},
  year={2017},
  publisher={Nature Publishing Group UK London},
  url={https://www.nature.com/articles/s41699-017-0013-7}
}

@article{gofryk2012electronic,
  title={Electronic tuning and uniform superconductivity in $\mathrm{CeCoIn_5}$},
  author={Gofryk, K and Ronning, Filip and Zhu, J-X and Ou, MN and Tobash, Paul H and Stoyko, SS and Lu, X and Mar, A and Park, <? format?> T and Bauer, ED and others},
  journal={Physical Review Letters},
  volume={109},
  number={18},
  pages={186402},
  year={2012},
  publisher={APS},
  url={https://journals.aps.org/prl/abstract/10.1103/PhysRevLett.109.186402}
}

@article{bi2022giant,
  title={Giant enhancement of superconducting critical temperature in substitutional alloy ($\mathrm{La}$, $\mathrm{Ce}$) $\mathrm{H_9}$},
  author={Bi, Jingkai and Nakamoto, Yuki and Zhang, Peiyu and Shimizu, Katsuya and Zou, Bo and Liu, Hanyu and Zhou, Mi and Liu, Guangtao and Wang, Hongbo and Ma, Yanming},
  journal={Nature Communications},
  volume={13},
  number={1},
  pages={5952},
  year={2022},
  publisher={Nature Publishing Group UK London},
  url={https://www.nature.com/articles/s41467-022-33743-6}
}

@article{razavi2002effect,
  title={Effect of pressure on the superconductivity of $\mathrm{MgB_2}$},
  author={Razavi, FS and Bose, SK and Ploczek, H},
  journal={Physica C: Superconductivity},
  volume={366},
  number={2},
  pages={73--79},
  year={2002},
  publisher={Elsevier},
  url={https://www.sciencedirect.com/science/article/pii/S0921453401011170}
}

@article{prassides2001compressibility,
  title={Compressibility of the $\mathrm{MgB_2}$ superconductor},
  author={Prassides, K and Iwasa, Y and Ito, T and Chi, Dam H and Uehara, K and Nishibori, E and Takata, M and Sakata, M and Ohishi, Y and Shimomura, O and others},
  journal={Physical Review B},
  volume={64},
  number={1},
  pages={012509},
  year={2001},
  publisher={APS},
  url={https://journals.aps.org/prb/abstract/10.1103/PhysRevB.64.012509}
}

@article{wang2009superconductivity,
  title={Superconductivity of $\mathrm{MgB_2}$ under ultrahigh pressure: A first-principles study},
  author={Wang, Yanchao and Lv, Jian and Ma, Yanming and Cui, Tian and Zou, Guangtian},
  journal={Physical Review B—Condensed Matter and Materials Physics},
  volume={80},
  number={9},
  pages={092505},
  year={2009},
  publisher={APS},
  url={https://journals.aps.org/prb/abstract/10.1103/PhysRevB.80.092505}
}

@article{shao2004pressure,
  title={The pressure dependence of the structure and superconducting transition temperature of $\mathrm{MgB_2}$},
  author={Shao, Yang and Zhang, X},
  journal={Journal of Physics: Condensed Matter},
  volume={16},
  number={7},
  pages={1103},
  year={2004},
  publisher={IOP Publishing},
  url={https://iopscience.iop.org/article/10.1088/0953-8984/16/7/010/meta?casa_token=8dedCaRkLY0AAAAA:zBN6Li0enmw4p_Gv_5yrNeGrGmrxsPRbOf-DUmUDDJWtjWPjCNRqli4kVfPUl3VG8V6EBpbHqd4a5h0wyW-DLxx5ywLMWw}
}

@article{ma2009absence,
  title={Absence of superconductivity in the high-pressure polymorph of $\mathrm{MgB_2}$},
  author={Ma, Yanming and Wang, Yanchao and Oganov, Artem R},
  journal={Physical Review B—Condensed Matter and Materials Physics},
  volume={79},
  number={5},
  pages={054101},
  year={2009},
  publisher={APS},
  url={https://journals.aps.org/prb/abstract/10.1103/PhysRevB.79.054101}
}

@article{hu2013pressure,
  title={Pressure-induced stabilization and insulator-superconductor transition of $\mathrm{BH}$},
  author={Hu, Chao-Hao and Oganov, Artem R and Zhu, Qiang and Qian, Guang-Rui and Frapper, Gilles and Lyakhov, Andriy O and Zhou, Huai-Ying},
  journal={Physical review letters},
  volume={110},
  number={16},
  pages={165504},
  year={2013},
  publisher={APS},
  url={https://journals.aps.org/prl/abstract/10.1103/PhysRevLett.110.165504}
}

@article{tafti2015universal,
  title={Universal V-shaped temperature-pressure phase diagram in the iron-based superconductors $\mathrm{KFe_2As_2}$, $\mathrm{RbFe_2As_2}$, and $\mathrm{CsFe_2As_2}$},
  author={Tafti, FF and Ouellet, A and Juneau-Fecteau, A and Faucher, S and Lapointe-Major, M and Doiron-Leyraud, N and Wang, AF and Luo, X-G and Chen, XH and Taillefer, Louis},
  journal={Physical Review B},
  volume={91},
  number={5},
  pages={054511},
  year={2015},
  publisher={APS},
  url={https://journals.aps.org/prb/abstract/10.1103/PhysRevB.91.054511}
}

@article{yue2018electron,
  title={Electron--phonon interaction and superconductivity in the high-pressure c/16 phase of lithium from first principles},
  author={Yue, Sheng-Ying and Cheng, Long and Liao, Bolin and Hu, Ming},
  journal={Physical Chemistry Chemical Physics},
  volume={20},
  number={42},
  pages={27125--27130},
  year={2018},
  publisher={Royal Society of Chemistry},
  url={https://pubs.rsc.org/en/content/articlehtml/2018/cp/c8cp05455j}
}

@article{zhao2021inverted,
  title={Inverted V-shaped evolution of superconducting temperature in $\mathrm{SrBC}$ under pressure},
  author={Zhao, Ru-Yi and Yan, Xun-Wang and Gao, Miao},
  journal={Chinese Physics B},
  volume={30},
  number={7},
  pages={076301},
  year={2021},
  publisher={IOP Publishing},
  url={https://iopscience.iop.org/article/10.1088/1674-1056/abfbcc/meta?casa_token=fPIYB1IztH4AAAAA:MLXJZchJTdsc3U1xUmtlMKdO1kV2gId1bnH7-pZTH0P9p5gtWdC1jqXyn8QsndlxdzL6D0NBK7rDkpT_XN9Ble86eA3kUw}
}

@article{pan2015pressure,
  title={Pressure-driven dome-shaped superconductivity and electronic structural evolution in tungsten ditelluride},
  author={Pan, Xing-Chen and Chen, Xuliang and Liu, Huimei and Feng, Yanqing and Wei, Zhongxia and Zhou, Yonghui and Chi, Zhenhua and Pi, Li and Yen, Fei and Song, Fengqi and others},
  journal={Nature communications},
  volume={6},
  number={1},
  pages={7805},
  year={2015},
  publisher={Nature Publishing Group UK London},
  url={https://www.nature.com/articles/ncomms8805}
}

@article{zhu2013superconductivity,
  title={Superconductivity in topological insulator $\mathrm{Sb_2Te_3}$ induced by pressure},
  author={Zhu, J and Zhang, JL and Kong, PP and Zhang, SJ and Yu, XH and Zhu, JL and Liu, QQ and Li, X and Yu, RC and Ahuja, Rajeev and others},
  journal={Scientific Reports},
  volume={3},
  number={1},
  pages={2016},
  year={2013},
  publisher={Nature Publishing Group UK London},
  url={https://www.nature.com/articles/srep02016}
}

@article{matsubayashi2014superconductivity,
  title={Superconductivity in the topological insulator $\mathrm{Bi_2Te_3}$ under hydrostatic pressure},
  author={Matsubayashi, K and Terai, T and Zhou, JS and Uwatoko, Y},
  journal={Physical Review B},
  volume={90},
  number={12},
  pages={125126},
  year={2014},
  publisher={APS},
  url={https://journals.aps.org/prb/abstract/10.1103/PhysRevB.90.125126}
}

@article{mohanta2015multiband,
  title={Multiband theory of superconductivity at the $\mathrm{LaAlO_3}$/$\mathrm{SrTiO_3}$ interface},
  author={Mohanta, N and Taraphder, A},
  journal={Physical Review B},
  volume={92},
  number={17},
  pages={174531},
  year={2015},
  publisher={APS},
  url={https://journals.aps.org/prb/abstract/10.1103/PhysRevB.92.174531}
}

@article{meng2024layer,
  title={Layer-Dependent Superconductivity in Iron-Based Superconductors $\mathrm{CsCa_2Fe_4As_4F_2}$ and $\mathrm{CaKFe_4As_4}$},
  author={Meng, Ke and Zhang, Xu and Song, Boqin and Li, Bai Zhuo and Kong, Xiangming and Huang, Sicheng and Yang, Xiaofan and Jin, Xiaobo and Wu, Yiyuan and Nie, Jiaying and others},
  journal={Nano Letters},
  year={2024},
  publisher={ACS Publications},
  url={https://pubs.acs.org/doi/full/10.1021/acs.nanolett.4c01725}
}

@article{patel2024electron,
  title={Electron-phonon coupling, critical temperatures, and gaps in $\mathrm{NbSe_2}$/$\mathrm{MoS_2}$ Ising superconductors},
  author={Patel, Shubham and Jena, Soumyasree and Taraphder, A},
  journal={Physical Review B},
  volume={110},
  number={1},
  pages={014507},
  year={2024},
  publisher={APS},
  url={https://journals.aps.org/prb/abstract/10.1103/PhysRevB.110.014507}
}

@article{zhao2022synthesis,
  title={Synthesis, characterization, and first-principles analysis of the $\mathrm{MAB}$-like ternary transition-metal boride $\mathrm{Fe(MoB)_2}$},
  author={Zhao, Xingbin and Zhou, Chao and Bao, Kuo and Zhu, Pinwen and Ma, Shuailing and Tao, Qiang and Cui, Tian},
  journal={Inorganic Chemistry},
  volume={61},
  number={29},
  pages={11046--11056},
  year={2022},
  publisher={ACS Publications},
  url={https://pubs.acs.org/doi/full/10.1021/acs.inorgchem.2c00635}
}

@article{wang2021nd2fe14b,
  title={$\mathrm{Nd_2Fe_{14}B}$ hard magnetic powders: Chemical synthesis and mechanism of coercivity},
  author={Wang, Xiaobai and Zhu, Kai and Li, Wei and Xu, Junjie and Ali, Zeeshan and Hou, Yanglong},
  journal={Journal of Magnetism and Magnetic Materials},
  volume={518},
  pages={167384},
  year={2021},
  publisher={Elsevier},
  url={https://www.sciencedirect.com/science/article/pii/S0304885320323519}
}

@article{jatmika2020superconducting,
  title={Superconducting properties of the ternary boride $\mathrm{YRh_4B_4}$},
  author={Jatmika, Jumaeda and Maruyama, Hiroshi and Rahman, Md Shahidur and Sakai, Akito and Nakatsuji, Satoru and Iyo, Akira and Ebihara, Takao},
  journal={Superconductor Science and Technology},
  volume={33},
  number={12},
  pages={125006},
  year={2020},
  publisher={IOP Publishing},
  url={https://iopscience.iop.org/article/10.1088/1361-6668/abbb18/meta}
}

@article{kanoun2012structure,
  title={Structure, Elastic Stiffness, and Hardness of $\mathrm{Os_{1-x}Ru_{x}B_2}$ Solid Solution Transition-Metal Diborides},
  author={Kanoun, Mohammed Benali and Hermet, Patrick and Goumri-Said, Souraya},
  journal={The Journal of Physical Chemistry C},
  volume={116},
  number={21},
  pages={11746--11751},
  year={2012},
  publisher={ACS Publications}, 
  url={https://pubs.acs.org/doi/full/10.1021/jp3015657}
}

@article{sobolev1968phase,
  title={Phase equilibria in tantalum—Titanium—Boron and tantalum—Molybdenum—Boron systems},
  author={Sobolev, AS and Kuz'ma, Yu B and Soboleva, TE and Fedorov, TF},
  journal={Soviet Powder Metallurgy and Metal Ceramics},
  volume={7},
  number={1},
  pages={48--51},
  year={1968},
  publisher={Springer},
  url={https://link.springer.com/article/10.1007/BF00776109}
}

@article{kresse1993ab,
  title={Ab initio molecular dynamics for liquid metals},
  author={Kresse, Georg and Hafner, J{\"u}rgen},
  journal={Physical review B},
  volume={47},
  number={1},
  pages={558},
  year={1993},
  publisher={APS},
  url={https://journals.aps.org/prb/abstract/10.1103/PhysRevB.47.558}
}

@article{kresse1996efficiency,
  title={Efficiency of ab-initio total energy calculations for metals and semiconductors using a plane-wave basis set},
  author={Kresse, Georg and Furthm{\"u}ller, J{\"u}rgen},
  journal={Computational materials science},
  volume={6},
  number={1},
  pages={15--50},
  year={1996},
  publisher={Elsevier},
  url={https://www.sciencedirect.com/science/article/pii/0927025696000080}
}

@article{perdew1996generalized,
  title={Generalized gradient approximation made simple},
  author={Perdew, John P and Burke, Kieron and Ernzerhof, Matthias},
  journal={Physical review letters},
  volume={77},
  number={18},
  pages={3865},
  year={1996},
  publisher={APS},
  url={https://journals.aps.org/prl/abstract/10.1103/PhysRevLett.77.3865}
}

@article{blochl1994projector,
  title={Projector augmented-wave method},
  author={Bl{\"o}chl, Peter E},
  journal={Physical review B},
  volume={50},
  number={24},
  pages={17953},
  year={1994},
  publisher={APS},
  url={https://journals.aps.org/prb/abstract/10.1103/PhysRevB.50.17953}
}

@article{kresse1999ultrasoft,
  title={From ultrasoft pseudopotentials to the projector augmented-wave method},
  author={Kresse, Georg and Joubert, Daniel},
  journal={Physical review b},
  volume={59},
  number={3},
  pages={1758},
  year={1999},
  publisher={APS},
  url={https://journals.aps.org/prb/abstract/10.1103/PhysRevB.59.1758}
}

@article{giannozzi2009quantum,
  title={QUANTUM ESPRESSO: a modular and open-source software project for quantum simulations of materials},
  author={Giannozzi, Paolo and Baroni, Stefano and Bonini, Nicola and Calandra, Matteo and Car, Roberto and Cavazzoni, Carlo and Ceresoli, Davide and Chiarotti, Guido L and Cococcioni, Matteo and Dabo, Ismaila and others},
  journal={Journal of physics: Condensed matter},
  volume={21},
  number={39},
  pages={395502},
  year={2009},
  publisher={IOP Publishing},
  url={https://iopscience.iop.org/article/10.1088/0953-8984/21/39/395502/meta?casa_token=Q1o-SgXOfIcAAAAA:AcKJZWmcxLyoqhoYTaLk4S2CTaWOy8JFIW15ZHlRs-1g5kVszaGoluAO6jyfxDKDJryLvvCNnY_YisQo2JnFDWw-3L24Yg}
}

@article{giannozzi2017advanced,
  title={Advanced capabilities for materials modelling with Quantum ESPRESSO},
  author={Giannozzi, Paolo and Andreussi, Oliviero and Brumme, Thomas and Bunau, Oana and Nardelli, M Buongiorno and Calandra, Matteo and Car, Roberto and Cavazzoni, Carlo and Ceresoli, Davide and Cococcioni, Matteo and others},
  journal={Journal of physics: Condensed matter},
  volume={29},
  number={46},
  pages={465901},
  year={2017},
  publisher={IOP Publishing},
  url={https://iopscience.iop.org/article/10.1088/1361-648X/aa8f79/meta?casa_token=VhYftN1l7f0AAAAA:afQJ6dF7XyCU7OB2opIvMCIfkwBih8W7FSzRLPYAWddxt_Bai0a91dKgOIwIMDClgXOiQydmnahrzbRQhdP1_jrkixOeOw}
}

@article{giannozzi2020quantum,
  title={Quantum ESPRESSO toward the exascale},
  author={Giannozzi, Paolo and Baseggio, Oscar and Bonf{\`a}, Pietro and Brunato, Davide and Car, Roberto and Carnimeo, Ivan and Cavazzoni, Carlo and De Gironcoli, Stefano and Delugas, Pietro and Ferrari Ruffino, Fabrizio and others},
  journal={The Journal of chemical physics},
  volume={152},
  number={15},
  year={2020},
  publisher={AIP Publishing},
  url={https://pubs.aip.org/aip/jcp/article/152/15/154105/1058748}
}

@article{garrity2014pseudopotentials,
  title={Pseudopotentials for high-throughput DFT calculations},
  author={Garrity, Kevin F and Bennett, Joseph W and Rabe, Karin M and Vanderbilt, David},
  journal={Computational Materials Science},
  volume={81},
  pages={446--452},
  year={2014},
  publisher={Elsevier},
  url={https://www.sciencedirect.com/science/article/pii/S0927025613005077}
}

@article{ponce2016epw,
  title={EPW: Electron--phonon coupling, transport and superconducting properties using maximally localized Wannier functions},
  author={Ponc{\'e}, Samuel and Margine, Elena R and Verdi, Carla and Giustino, Feliciano},
  journal={Computer Physics Communications},
  volume={209},
  pages={116--133},
  year={2016},
  publisher={Elsevier},
  url={https://www.sciencedirect.com/science/article/pii/S0010465516302260}
}

@article{allen1983theory,
  title={Theory of superconducting $\mathrm{T_c}$},
  author={Allen, Philip B and Mitrovi{\'c}, Bo{\v{z}}idar},
  journal={Solid state physics},
  volume={37},
  pages={1--92},
  year={1983},
  publisher={Elsevier},
  url={https://www.sciencedirect.com/science/article/abs/pii/S0081194708606657}                       
}

@article{giustino2007electron,
  title={Electron-phonon interaction using Wannier functions},
  author={Giustino, Feliciano and Cohen, Marvin L and Louie, Steven G},
  journal={Physical Review B—Condensed Matter and Materials Physics},
  volume={76},
  number={16},
  pages={165108},
  year={2007},
  publisher={APS},
  url={https://journals.aps.org/prb/abstract/10.1103/PhysRevB.76.165108}
}

@article{momma2011vesta,
  title={VESTA 3 for three-dimensional visualization of crystal, volumetric and morphology data},
  author={Momma, Koichi and Izumi, Fujio},
  journal={Journal of applied crystallography},
  volume={44},
  number={6},
  pages={1272--1276},
  year={2011},
  publisher={International Union of Crystallography},
  url={https://journals.iucr.org/paper?db5098}
}

@article{kokalj1999xcrysden,
  title={XCrySDen—a new program for displaying crystalline structures and electron densities},
  author={Kokalj, Anton},
  journal={Journal of Molecular Graphics and Modelling},
  volume={17},
  number={3-4},
  pages={176--179},
  year={1999},
  publisher={Elsevier},
  url={https://www.sciencedirect.com/science/article/pii/S1093326399000285}
}

@article{jain2013commentary,
  title={Commentary: The Materials Project: A materials genome approach to accelerating materials innovation},
  author={Jain, Anubhav and Ong, Shyue Ping and Hautier, Geoffroy and Chen, Wei and Richards, William Davidson and Dacek, Stephen and Cholia, Shreyas and Gunter, Dan and Skinner, David and Ceder, Gerbrand and others},
  journal={APL materials},
  volume={1},
  number={1},
  year={2013},
  publisher={AIP Publishing},
  url={https://pubs.aip.org/aip/apm/article/1/1/011002/119685}
}

@article{peterson2021materials,
  title={Materials discovery through machine learning formation energy},
  author={Peterson, Gordon GC and Brgoch, Jakoah},
  journal={Journal of Physics: Energy},
  volume={3},
  number={2},
  pages={022002},
  year={2021},
  publisher={IOP Publishing}, 
  url={https://iopscience.iop.org/article/10.1088/2515-7655/abe425/meta}
}

@article{twyman2022environmental,
  title={Environmental stability of crystals: a greedy screening},
  author={Twyman, Nicholas M and Walsh, Aron and Buonassisi, Tonio},
  journal={Chemistry of Materials},
  volume={34},
  number={6},
  pages={2545--2552},
  year={2022},
  publisher={ACS Publications},
  url={https://pubs.acs.org/doi/full/10.1021/acs.chemmater.1c02644}
}

@article{curtarolo2005accuracy,
  title={Accuracy of ab initio methods in predicting the crystal structures of metals: A review of 80 binary alloys},
  author={Curtarolo, Stefano and Morgan, Dane and Ceder, Gerbrand},
  journal={Calphad},
  volume={29},
  number={3},
  pages={163--211},
  year={2005},
  publisher={Elsevier},
  url={https://www.sciencedirect.com/science/article/pii/S0364591605000064}
}

@article{zheng2023superconductivity,
  title={Superconductivity in the $\mathrm{Li-BC}$ system at 100 GPa},
  author={Zheng, Feng and Sun, Yang and Wang, Renhai and Fang, Yimei and Zhang, Feng and Wu, Shunqing and Wang, Cai-Zhuang and Antropov, Vladimir and Ho, Kai-Ming},
  journal={Physical Review B},
  volume={107},
  number={1},
  pages={014508},
  year={2023},
  publisher={APS},
  url={https://journals.aps.org/prb/abstract/10.1103/PhysRevB.107.014508}
}

@article{arakcheeva2003commensurate,
  title={The commensurate composite $\mathrm{\sigma}$-structure of $\mathrm{\beta}$-tantalum},
  author={Arakcheeva, Alla and Chapuis, Gervais and Birkedal, Henrik and Pattison, Phil and Grinevitch, Vladimir},
  journal={Structural Science},
  volume={59},
  number={3},
  pages={324--336},
  year={2003},
  publisher={International Union of Crystallography},
  url={https://journals.iucr.org/paper?S0108768103009005}
}

@article{ross1963high,
  title={High temperature X-ray metallography: I. A new debye-scherrer camera for use at very high temperatures II. A new parafocusing camera III. Applications to the study of chromium, hafnium, molybdenum, rhodium, ruthenium and tungsten},
  author={Ross, RG and Hume-Rothery, W},
  journal={Journal of the Less Common Metals},
  volume={5},
  number={3},
  pages={258--270},
  year={1963},
  publisher={Elsevier},
  url={https://www.sciencedirect.com/science/article/pii/0022508863900316}
}

@article{decker1959crystal,
  title={The crystal structure of a simple rhombohedral form of boron},
  author={Decker, BF and Kasper, JS},
  journal={Acta Crystallographica},
  volume={12},
  number={7},
  pages={503--506},
  year={1959},
  publisher={International Union of Crystallography},
  url={https://journals.iucr.org/paper?S0365110X59001529}
}

@article{quan2021mob,
  title={MoB 2 under pressure: Superconducting Mo enhanced by boron},
  author={Quan, Yundi and Lee, Kwan-Woo and Pickett, Warren E},
  journal={Physical Review B},
  volume={104},
  number={22},
  pages={224504},
  year={2021},
  publisher={APS},
  url={https://journals.aps.org/prb/abstract/10.1103/PhysRevB.104.224504}
}

@UNPUBLISHED{supplementary,
  title={See supplementary material at http://link.aps.org/supplemental/10.1103/b627-nvxl for computational details and remaining plots including {E}liashberg spectral function, quasiparticle desnity of states in the superconducting states with respect to normal states and anisotropic superconducting gap on the {F}ermi surface at each pressure, which includes Ref \cite{cococcioni2005linear}}
 
}

@article{birch1947finite,
  title={Finite elastic strain of cubic crystals},
  author={Birch, Francis},
  journal={Physical review},
  volume={71},
  number={11},
  pages={809},
  year={1947},
  publisher={APS},
  url={https://journals.aps.org/pr/abstract/10.1103/PhysRev.71.809}
}

@article{baroni2001phonons,
  title={Phonons and related crystal properties from density-functional perturbation theory},
  author={Baroni, Stefano and De Gironcoli, Stefano and Dal Corso, Andrea and Giannozzi, Paolo},
  journal={Reviews of modern Physics},
  volume={73},
  number={2},
  pages={515},
  year={2001},
  publisher={APS},
  url={https://journals.aps.org/rmp/abstract/10.1103/RevModPhys.73.515}
}

@article{johannes2008fermi,
  title={Fermi surface nesting and the origin of charge density waves in metals},
  author={Johannes, MD and Mazin, II},
  journal={Physical Review B—Condensed Matter and Materials Physics},
  volume={77},
  number={16},
  pages={165135},
  year={2008},
  publisher={APS},
  url={https://journals.aps.org/prb/abstract/10.1103/PhysRevB.77.165135}
}

@article{johannes2006fermi,
  title={Fermi-surface nesting and the origin of the charge-density wave in $\mathrm{NbSe_2}$},
  author={Johannes, MD and Mazin, II and Howells, CA},
  journal={Physical Review B—Condensed Matter and Materials Physics},
  volume={73},
  number={20},
  pages={205102},
  year={2006},
  publisher={APS},
  url={https://journals.aps.org/prb/abstract/10.1103/PhysRevB.73.205102}
}

@article{kaboudvand2022fermi,
  title={Fermi surface nesting and the Lindhard response function in the kagome superconductor $\mathrm{CsV_3Sb_5}$},
  author={Kaboudvand, Farnaz and Teicher, Samuel ML and Wilson, Stephen D and Seshadri, Ram and Johannes, Michelle D},
  journal={Applied Physics Letters},
  volume={120},
  number={11},
  year={2022},
  publisher={AIP Publishing},
  url={https://pubs.aip.org/aip/apl/article/120/11/111901/2833165}
}

@article{zhu2015classification,
  title={Classification of charge density waves based on their nature},
  author={Zhu, Xuetao and Cao, Yanwei and Zhang, Jiandi and Plummer, EW and Guo, Jiandong},
  journal={Proceedings of the National Academy of Sciences},
  volume={112},
  number={8},
  pages={2367--2371},
  year={2015},
  publisher={National Acad Sciences},
  url={https://www.pnas.org/doi/abs/10.1073/pnas.1424791112}
}

@article{chen2002effects,
  title={Effects of pressure on the superconducting properties of magnesium diboride},
  author={Chen, XJ and Zhang, H and Habermeier, H-U},
  journal={Physical Review B},
  volume={65},
  number={14},
  pages={144514},
  year={2002},
  publisher={APS},
  url={https://journals.aps.org/prb/abstract/10.1103/PhysRevB.65.144514}
}

@article{allen1975superconductivity,
  title={Superconductivity and phonon softening: II. Lead alloys},
  author={Allen, PB and Dynes, RC},
  journal={Physical Review B},
  volume={11},
  number={5},
  pages={1895},
  year={1975},
  publisher={APS},
  url={https://journals.aps.org/prb/abstract/10.1103/PhysRevB.11.1895}
}

@article{wang2024theoretical,
  title={A theoretical study of Lifshitz transition for $\mathrm{2H-TaS_2}$},
  author={Wang, Wenxuan and Jiang, Zhenyi and Zhang, Xiaodong and Zheng, Jiming and Du, Hongwei and Zhang, Zhiyong},
  journal={Physical Chemistry Chemical Physics},
  volume={26},
  number={22},
  pages={15868--15876},
  year={2024},
  publisher={Royal Society of Chemistry},
  url={https://pubs.rsc.org/en/content/articlehtml/2024/cp/d4cp00977k}
}

@article{feng2022superconductivity,
  title={Superconductivity induced by Lifshitz transition in pristine $\mathrm{SnS_2}$ under high pressure},
  author={Feng, Jiajia and Li, Cong and Deng, Wen and Lin, Bencheng and Liu, Wenhui and Susilo, Resta A and Dong, Hongliang and Chen, Zhiqiang and Zhou, Nan and Yi, Xiaolei and others},
  journal={The Journal of Physical Chemistry Letters},
  volume={13},
  number={40},
  pages={9404--9410},
  year={2022},
  publisher={ACS Publications},
  url={https://pubs.acs.org/doi/full/10.1021/acs.jpclett.2c02580}
}

@article{lifshitz1960anomalies,
  title={Anomalies of electron characteristics of a metal in the high pressure region},
  author={Lifshitz, IM and others},
  journal={Sov. Phys. JETP},
  volume={11},
  number={5},
  pages={1130--1135},
  year={1960},
  url={http://jetp.ras.ru/cgi-bin/e/index/e/11/5/p1130?a=list}
}

@article{cococcioni2005linear,
  title={Linear response approach to the calculation of the effective interaction parameters in the $\mathrm{LDA+ U}$ method},
  author={Cococcioni, Matteo and De Gironcoli, Stefano},
  journal={Physical Review B—Condensed Matter and Materials Physics},
  volume={71},
  number={3},
  pages={035105},
  year={2005},
  publisher={APS},
  url={https://journals.aps.org/prb/abstract/10.1103/PhysRevB.71.035105}
}

\end{document}


\vspace*{0.8cm}
\begin{center}
    \large\textbf{Supplementary Material}
\end{center}

\preprint{APS/123-QED}

\title{Pressure induced evolution of anisotropic superconductivity and Fermi surface nesting in a ternary boride}

\author{Subhajit Pramanick\textsuperscript{1}}
\email{subhajitbhu@kgpian.iitkgp.ac.in}
\author{Sudip Chakraborty\textsuperscript{2}}
\email{sudipchakraborty@hri.res.in}
\author{A. Taraphder\textsuperscript{1}}
\email{arghya@phy.iitkgp.ac.in}

\affiliation{\textsuperscript{1}Department of Physics, Indian Institute of Technology Kharagpur, Kharagpur 721302, India}
\affiliation{\textsuperscript{2}Materials Theory for Energy Scavenging Lab, Harish-Chandra Research Institute, A CI of Homi Bhabha National Institute, Chhatnag Road, Jhunsi, Prayagraj 211019, India}

\date{\today}

\maketitle


\section{\label{sec:1}Thermodynamic stability calculation\protect\\ }

The formation energy for all the competing binary and ternary borides has been computed in Quantum Espresso (QE) package \cite{giannozzi2009quantum,giannozzi2017advanced,giannozzi2020quantum} using PBE functional \cite{perdew1996generalized}. For this purpose, we have considered Ta (mp-569794) \cite{arakcheeva2003commensurate}, Mo (mp-129) \cite{ross1963high}, B (mp-160) \cite{decker1959crystal} as the reference phases of Ta, Mo and B respectively.

\begin{figure*}[h]
\begin{center}
    \includegraphics[width=0.78\linewidth]{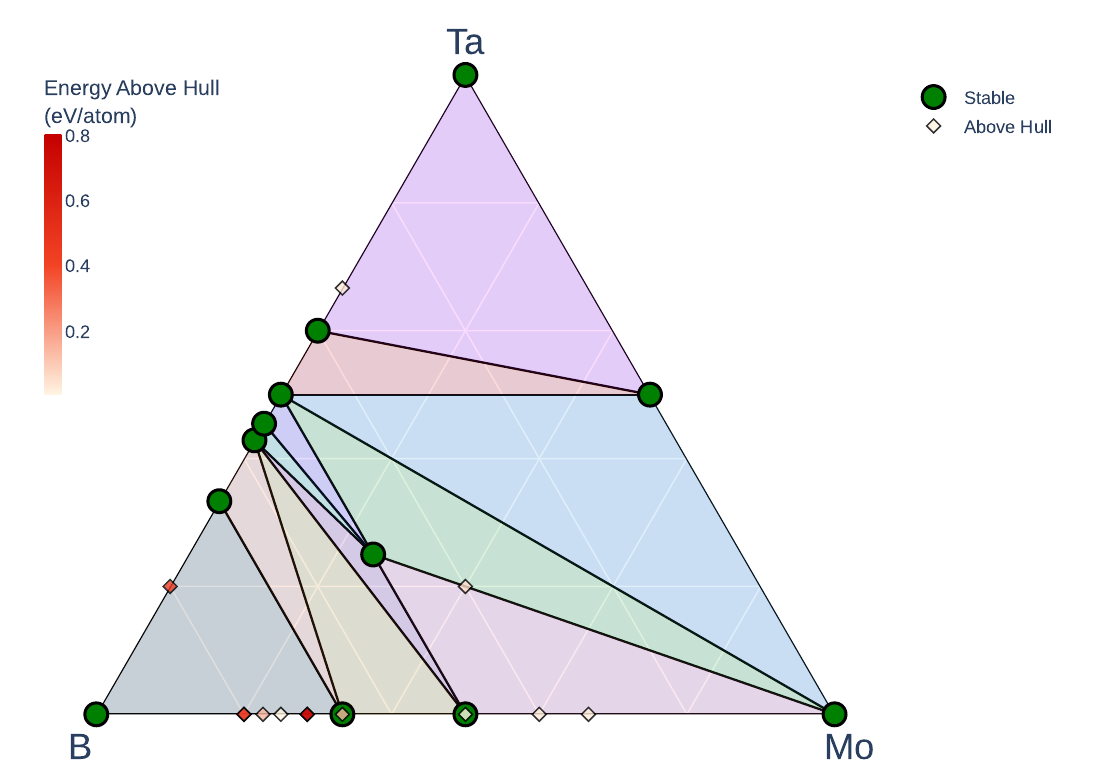}
\end{center}
\vspace{-0.5cm}
\caption{\label{fig:1} Phase diagram of binary and ternary borides at ambient pressure}
\end{figure*}

Taking the values of formation energies, we have calculated hull energy ($\mathrm{E_{hull}}$) using pymatgen library of python code. At ambient pressure, the calculated values of formation energy and hull energy of all binary and ternary borides

\begin{table}[h!]
\caption{\label{tab:1}Formation energy and hull energy of various binary and ternary borides at ambient pressure calculated using PBE functional in Quantum Espresso}
\vspace{0.5cm}
\renewcommand{\arraystretch}{1.5}  
\centering
\begin{tabular}{|c|c|c|}
\hline
\textrm{Material (Materials project ID)} & \textrm{Formation Energy (eV/atom)} & \textrm{Hull Energy (eV/atom)} \\ \hline
Ta (mp-569794) & 0.0 & 0.0 \\ \hline
Mo (mp-129) & 0.0 & 0.0 \\ \hline
B (mp-160) & 0.0 & 0.0 \\ \hline
TaMo (mp-1217895) & -0.095 & 0.0 \\ \hline
$\mathrm{Ta_3B_2}$ (mp-13415) & -0.677 & 0.0 \\ \hline
$\mathrm{Ta_3B_4}$ (mp-10142) & -0.775 & 0.0 \\ \hline
$\mathrm{Ta_5B_6}$ (mp-28629) & -0.793 & 0.0 \\ \hline
TaB (mp-1097) & -0.821 & 0.0 \\ \hline
$\mathrm{TaB_2}$ (mp-1108) & -0.651 & 0.0 \\ \hline
$\mathrm{Ta_2B}$ (mp-1032) & -0.534 & +0.030 \\ \hline
$\mathrm{TaB_4}$ (mp-1189303) & +0.071 & +0.462 \\ \hline
$\mathrm{B_2Mo}$ (mp-2331) & -0.444 & 0.0 \\ \hline
$\mathrm{B_2Mo}$ (mp-960) & -0.291 & +0.153 \\ \hline
$\mathrm{B_2Mo_3}$ (mp-1079053) & -0.381 & +0.027 \\ \hline
$\mathrm{B_3Mo}$ (mp-1080111) & -0.326 & +0.007 \\ \hline
BMo (mp-1890) & -0.510 & 0.0 \\ \hline
BMo (mp-999198) & -0.499 & +0.011 \\ \hline
$\mathrm{BMo_2}$ (mp-2501) & -0.318 & +0.022 \\ \hline
$\mathrm{B_{24}Mo_7}$ (mp-1228730) & -0.150 & +0.151 \\ \hline
$\mathrm{B_4Mo}$ (mp-1228687) & -0.008 & +0.258 \\ \hline
$\mathrm{B_4Mo}$ (mp-1106346) & +0.264 & +0.530 \\ \hline
$\mathrm{B_5Mo_2}$ (mp-7229) & +0.059 & +0.440 \\ \hline
$\mathrm{B_5Mo_2}$ (mp-1104594) & +0.422 & +0.803 \\ \hline
$\mathrm{TaB_2Mo}$ (mp-1217965) & -0.672 & 0.0 \\ \hline
$\mathrm{Ta(MoB)_2}$ (mp-1079659) & -0.480 & +0.058 \\ \hline
\end{tabular}
\end{table}

have been attached in Table \hyperref[tab:1]{I}. We have plotted the phase diagram of binary and ternary borides at ambient pressure in Figure \hyperref[fig:1]{1}. From the all possible competing phases, hull energy indicates how distant any material is from being thermodynamically stable. Materials having 0 eV/atom hull energy shows thermodynamic stability among all competing phases at 0 K \cite{chen2024high,twyman2022environmental}. Materials having hull energy less than +0.1 eV/atom is known as metastable \cite{twyman2022environmental}. They exhibit kinetic stability, indicating that substantial energy barriers inhibit their disintegration into more stable phases. On the other hand, hull energy larger than +0.1 eV/atom signifies poorer stability and often instability \cite{chen2024high,twyman2022environmental}.
\\

At ambient pressure, the calculated hull energy of $\mathrm{Ta(MoB)_2}$ is +0.058 eV/atom. Such a small positive value suggests that $\mathrm{Ta(MoB)_2}$ is metastable. The computed values of hull energy of $\mathrm{Ta(MoB)_2}$ at each pressure have been shown in Table I of the main text. The formation enrgies of $\mathrm{Ta(MoB)_2}$ under pressure have also been attached in Table I of the main text calculated in two ways. For calculating $\mathrm{E_{form_1}}$, we have always kept the reference phases of Ta, Mo and B at ambient pressure and for calculating $\mathrm{E_{form_2}}$, we have kept the reference phases at the same pressure at which formation energy is being calculated. The values of hull energy of $\mathrm{Ta(MoB)_2}$ indicate its metastability under a pressure upto 76.69 GPa.
\\

Conventional and primitive cell of $\mathrm{Ta(MoB)_2}$ both are tetragonal (space group: p4/mbm). The details of lattice parameters and atomic positions of the primitive cell of $\mathrm{Ta(MoB)_2}$ has been attached in POSCAR format below: 
\begin{verbatim}
Primitive Cell
  1.000000
    6.10597543706210    0.00000000000000    0.00000000000000
    0.00000000000000    6.10597543706210    0.00000000000000
    0.00000000000000    0.00000000000000    3.17475666993849
  Ta  B   Mo
   2   4   4
DIRECT
    0.5000000000000000    0.5000000000000000    0.0000000000000000    Ta1
    0.0000000000000000    0.0000000000000000    0.0000000000000000    Ta2
    0.6100332061614568    0.1100332061614569    0.0000000000000000     B1
    0.3899667938385432    0.8899667938385432    0.0000000000000000     B2
    0.8899667938385432    0.6100332061614568    0.0000000000000000     B3
    0.1100332061614569    0.3899667938385432    0.0000000000000000     B4
    0.1745584190508767    0.6745584190508767    0.5000000000000000    Mo1
    0.8254415809491233    0.3254415809491233    0.5000000000000000    Mo2
    0.3254415809491233    0.1745584190508767    0.5000000000000000    Mo3
    0.6745584190508767    0.8254415809491233    0.5000000000000000    Mo4

\end{verbatim}

\section{\label{sec:2}Electronic and phononic dispersion calculation\protect\\ }

\begin{figure*}[h]
\includegraphics[width=0.28\linewidth]{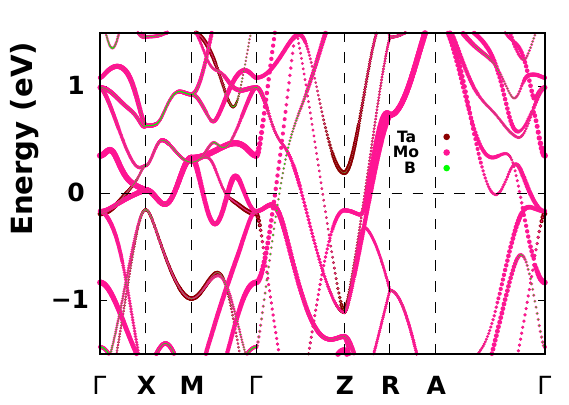}
\includegraphics[width=0.31\linewidth]{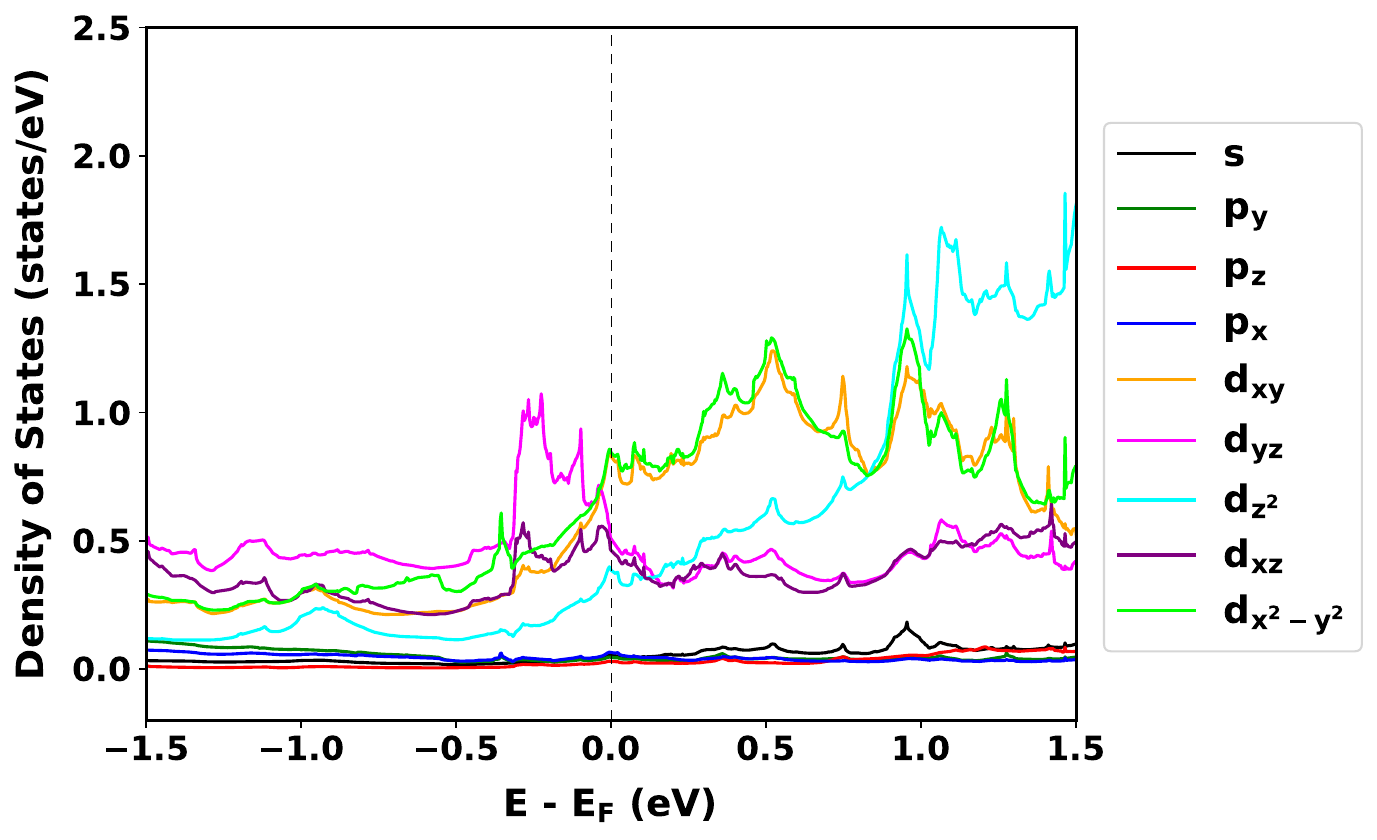}
\includegraphics[width=0.38\linewidth]{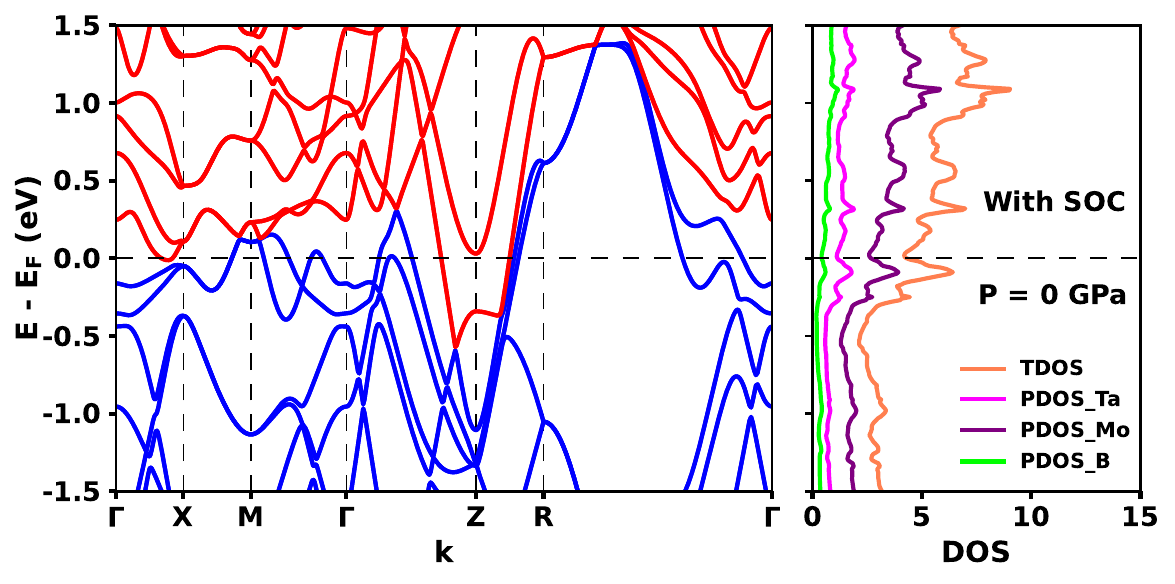}
\vspace*{-0.20cm}
    \begin{center}
        \hspace*{-0.5cm}
        \textbf{\small{(a)}}
        \hspace*{3.8cm}
        \textbf{\small{(b)}}
        \hspace*{5.6cm}
        \textbf{\small{(c)}}
    \end{center}
\caption{\label{fig:2} (a) Projected electronic band structure of $\mathrm{Ta(MoB)_2}$ at ambient pressure calculated using VASP Package \cite{kresse1993ab,kresse1996efficiency}, (b) Orbital projected density of states on Mo atoms. Both of these case, SOC has not been considered, (c) Electronic band structure and density of states of $\mathrm{Ta(MoB)_2}$ at ambient pressure considering SOC calculated using QE package \cite{giannozzi2009quantum,giannozzi2017advanced,giannozzi2020quantum}.}
\end{figure*}

The Vienna Ab initio Simulation Package (VASP) \cite{kresse1993ab,kresse1996efficiency} has also been used to calculate the electronic band structure and density of states of $\mathrm{Ta(MoB)_2}$ at the ambient pressure, as shown in Figure \hyperref[fig:2]{2(a)}. The Quantum Espresso (QE) package \cite{giannozzi2009quantum,giannozzi2017advanced,giannozzi2020quantum} was employed to calculate the electronic bands depicted in the main text (Figure 2(a) for ambient pressure and Figure 5(a),(c),(e) for high pressure). We have disregarded spin-orbit coupling (SOC) in both calculations due to the computational complexity associated with the computation of phonon dispersion considering SOC. In Figure \hyperref[fig:2]{2(a)}, the different colours are used to denote the contribution of each type of atom to the band structure. Clearly it has been captured that Mo atoms dominantly contribute to bands around the Fermi level. The Fermi level in the bands calculated by VASP \cite{kresse1993ab,kresse1996efficiency} is slightly different, but the pattern of both bands (calculated by two different packages) is nearly identical. The number of electron and hole pockets in those bands, as determined by two distinct packages, are not equal as a result of this difference.

\begin{figure*}[h]
\includegraphics[width=0.325\linewidth]{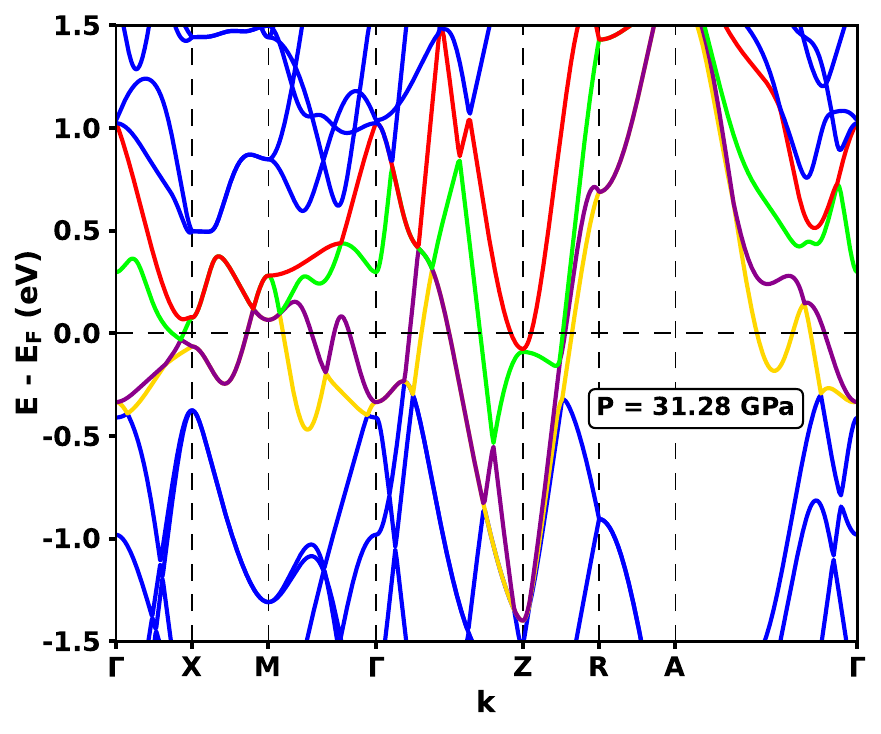}
\includegraphics[width=0.325\linewidth]{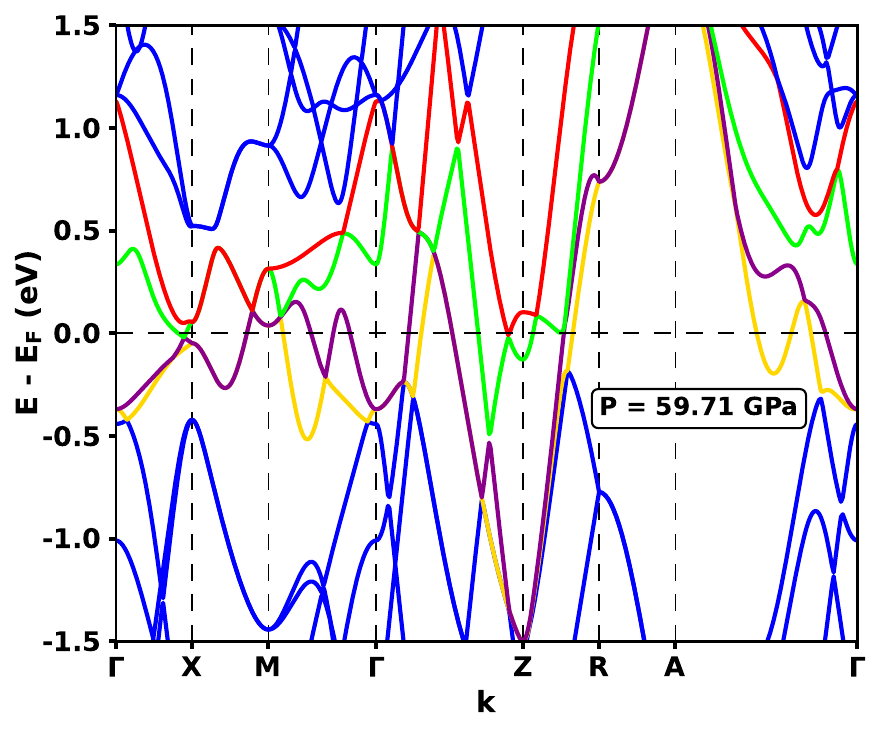}
\includegraphics[width=0.325\linewidth]{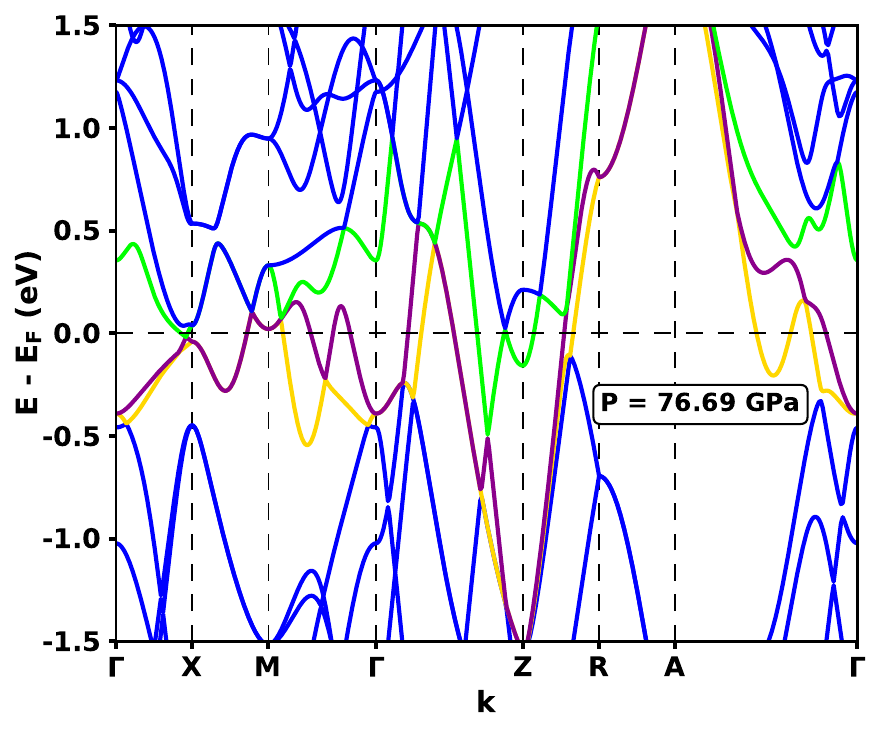}
\vspace*{-0.20cm}
    \begin{center}
        \hspace*{0.4cm}
        \textbf{\small{(a)}}
        \hspace*{5.1cm}
        \textbf{\small{(b)}}
        \hspace*{5.1cm}
        \textbf{\small{(c)}}
    \end{center}
\caption{\label{fig:3} Colour plots of the four bands within band structure that intersect the Fermi level of $\mathrm{Ta(MoB)_2}$ for three different pressure (a) 31.28 GPa, (b) 59.71 GPa, (c) 76.69 GPa}
\end{figure*}

 However, our whole calculations including high pressure study have been completed using QE package \cite{giannozzi2009quantum,giannozzi2017advanced,giannozzi2020quantum} which was used in the main text. The bands structure of $\mathrm{Ta(MoB)_2}$, which was determined using VASP \cite{kresse1993ab,kresse1996efficiency}, is attached here for fat-band analysis and verification purposes only. The methodology for calculating electronic and phononic bands structure using those two packages have been already discussed in the main text. We have calculated orbital projected density of states on Mo atoms in the Figure \hyperref[fig:2]{2(b)}. It emphasizes how largely the density of states at the Fermi level results from the d-orbitals of Mo atoms. We have also calculated electronic band structure and density of states of $\mathrm{Ta(MoB)_2}$ at ambient pressure considering SOC by QE package \cite{giannozzi2009quantum,giannozzi2017advanced,giannozzi2020quantum} in the Figure \hyperref[fig:2]{2(c)}. Owing to the maintenance of both inversion and time reversal symmetry, SOC can not lift the degeneracy of bands for up and down spin states. Atomic SOC has removed certain orbital degeneracies of Ta and Mo atoms. Notably, there are no substantial changes within the Fermi window except from this. Therefore we believe SOC can not alter the superconducting properties in a large extent. Figure \hyperref[fig:3]{3 (a-c)} displays the bands that intersect the Fermi level, represented by various colors, to illustrate the variation in the total number of bands crossing the Fermi level with pressure. The colors red, lime, deepmagenta, and gold are employed to represent the four bands that intersect the Fermi level, whereas the color blue is used to indicate the other remaining bands.

Figure \hyperref[fig:3]{3 (a-c)} effectively shows how one of the four bands crossing the Fermi level rises entirely above the Fermi level at a pressure of 76.69 GPa. In our computation, the number of bands crossing the Fermi level stays four upto 59.71 GPa pressure; it becomes three when we progressively raise the pressure from 59.71 GPa to 76.69 GPa. This is one of the salient features of the Lifshitz transition, which we believe to be occurring at 76.69 GPa.

\begin{figure*}[h]
\includegraphics[width=0.333\linewidth]{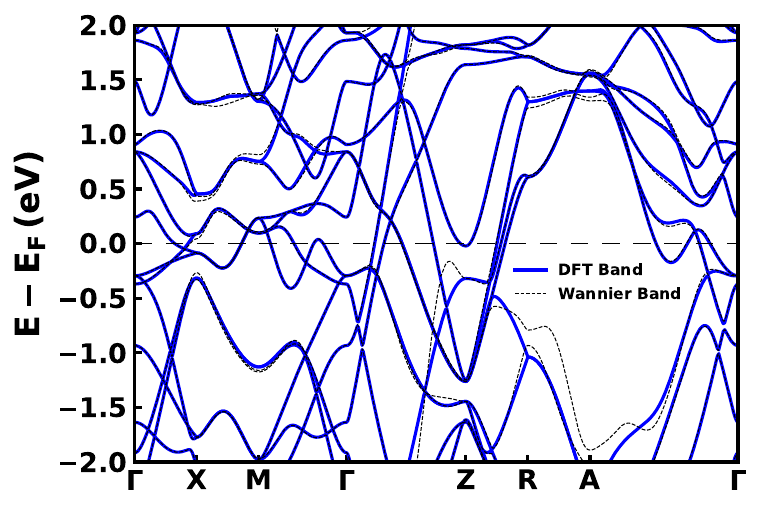}
\includegraphics[width=0.3025\linewidth]{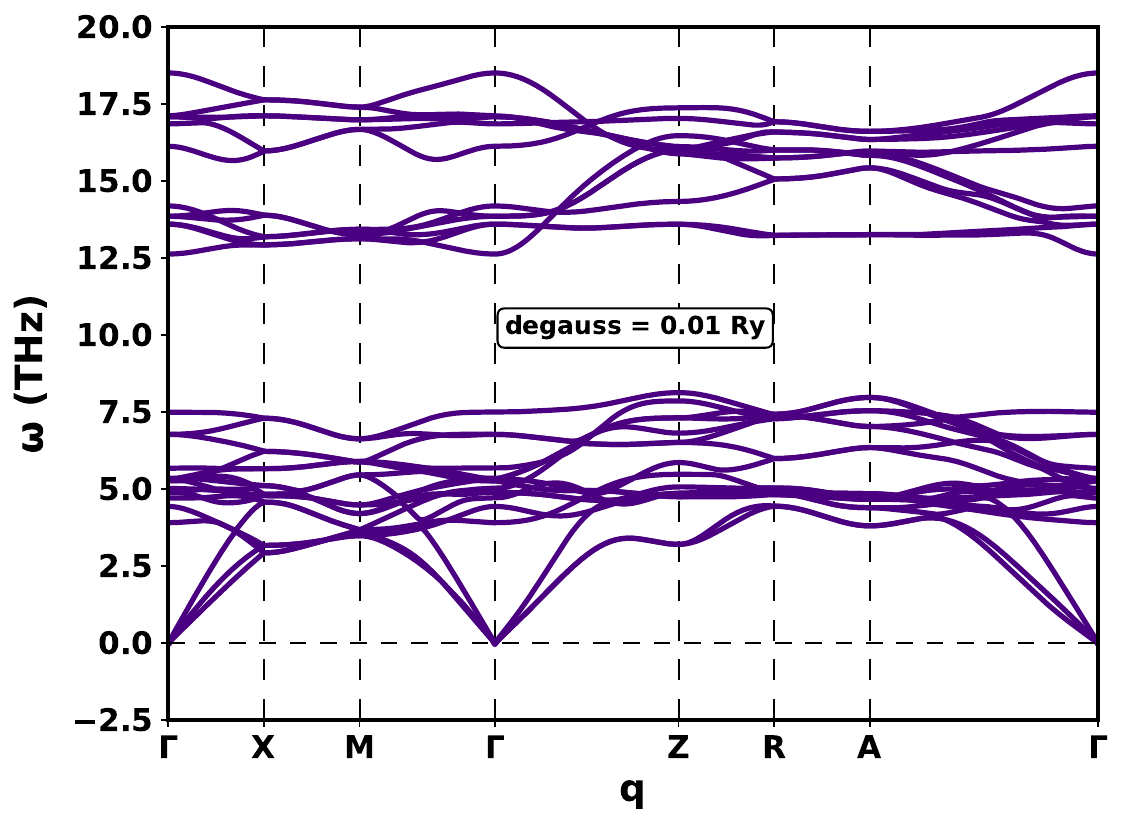}
\includegraphics[width=0.3025\linewidth]{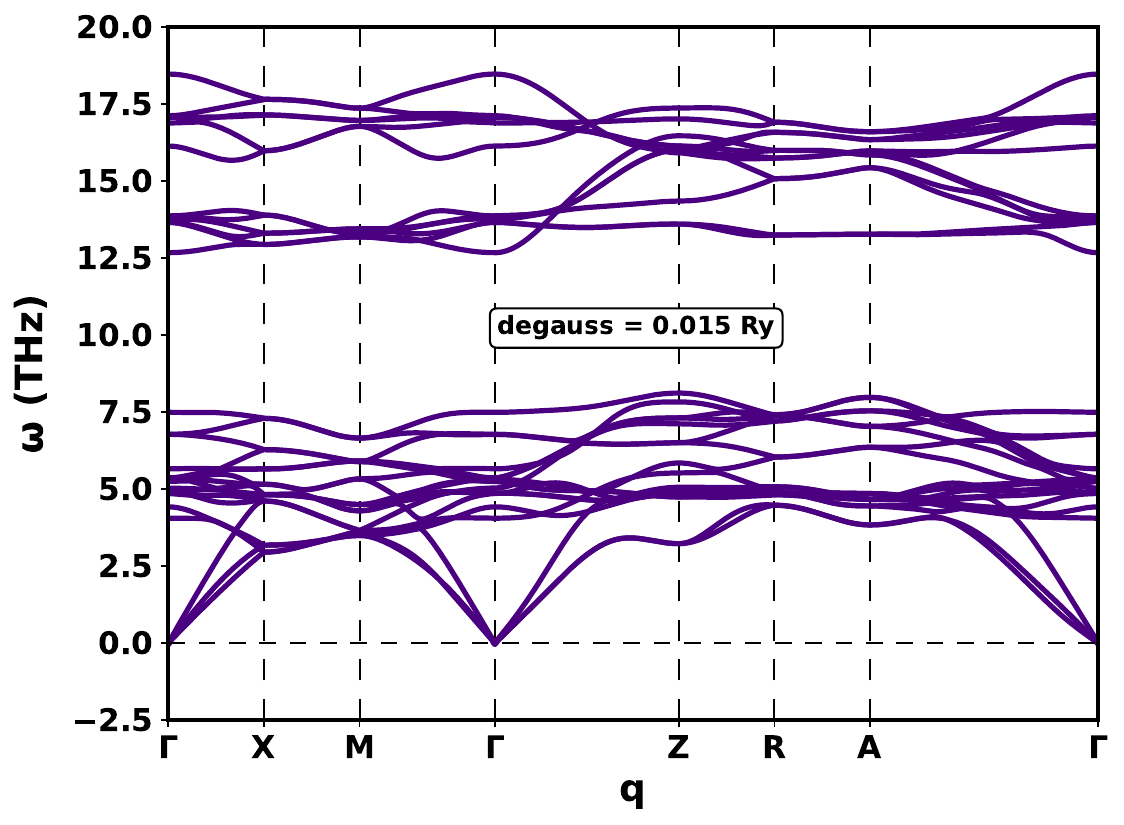}
\vspace*{-0.20cm}
    \begin{center}
        \hspace*{0.4cm}
        \textbf{\small{(a)}}
        \hspace*{5.1cm}
        \textbf{\small{(b)}}
        \hspace*{5.1cm}
        \textbf{\small{(c)}}
    \end{center}
\caption{\label{fig:4} (a) Comparison between DFT bands and Wannier bands of $\mathrm{Ta(MoB)_2}$ at ambient pressure; Phonon dispersion of $\mathrm{Ta(MoB)_2}$ at ambient pressure for a smearing width (b) 0.01 Ry, (c) 0.015 Ry}
\end{figure*}

\begin{figure*}[h]
\includegraphics[width=0.475\linewidth]{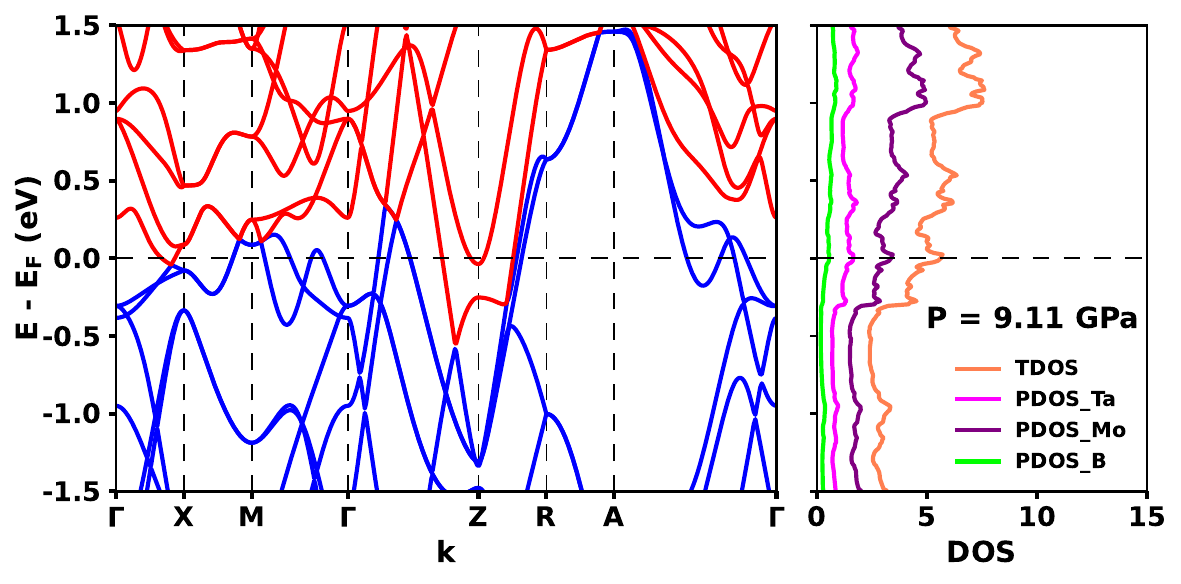}
\includegraphics[width=0.515\linewidth]{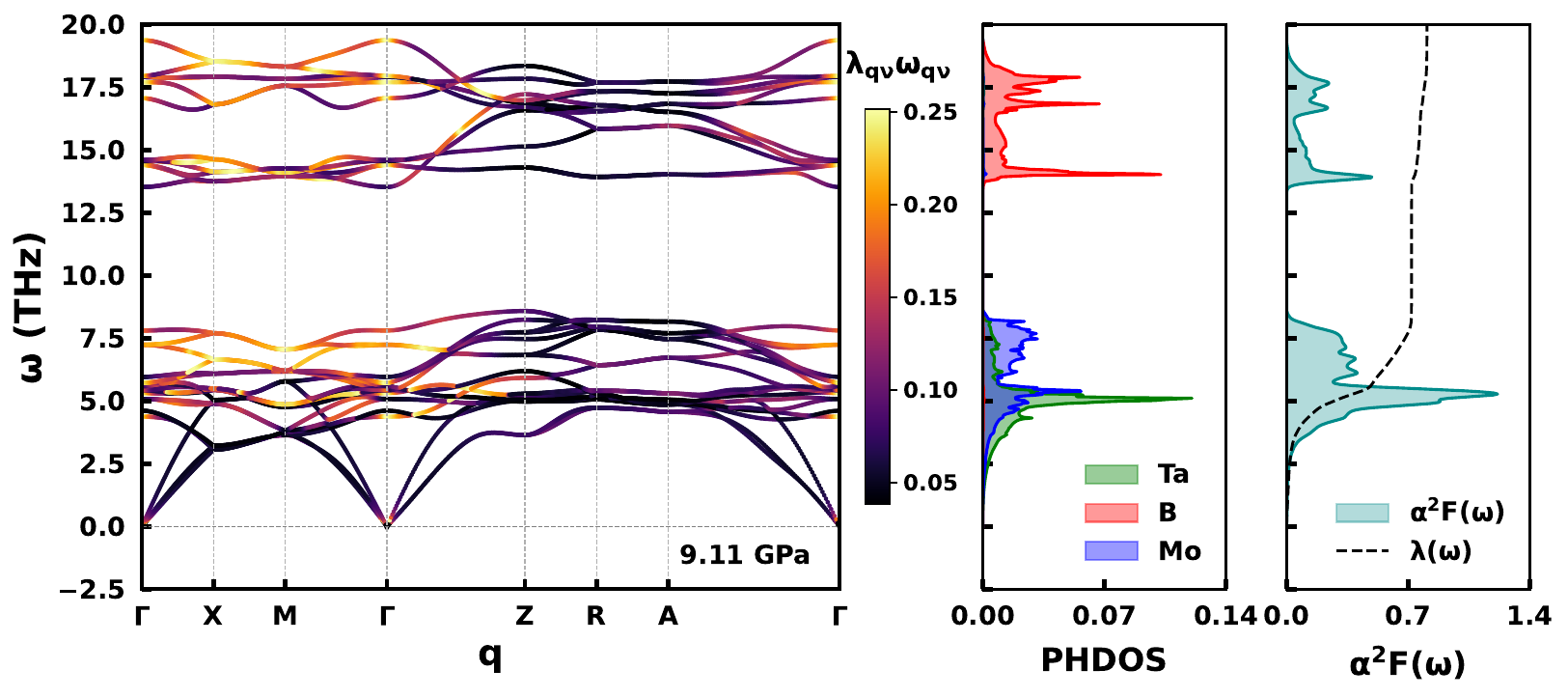}
\vspace*{-0.65cm}
    \begin{center}
        \hspace*{-1.7cm}
        \textbf{\small{(a)}}
        \hspace*{8cm}
        \textbf{\small{(b)}}
    \end{center}
\vspace*{-0.25cm}
\includegraphics[width=0.475\linewidth]{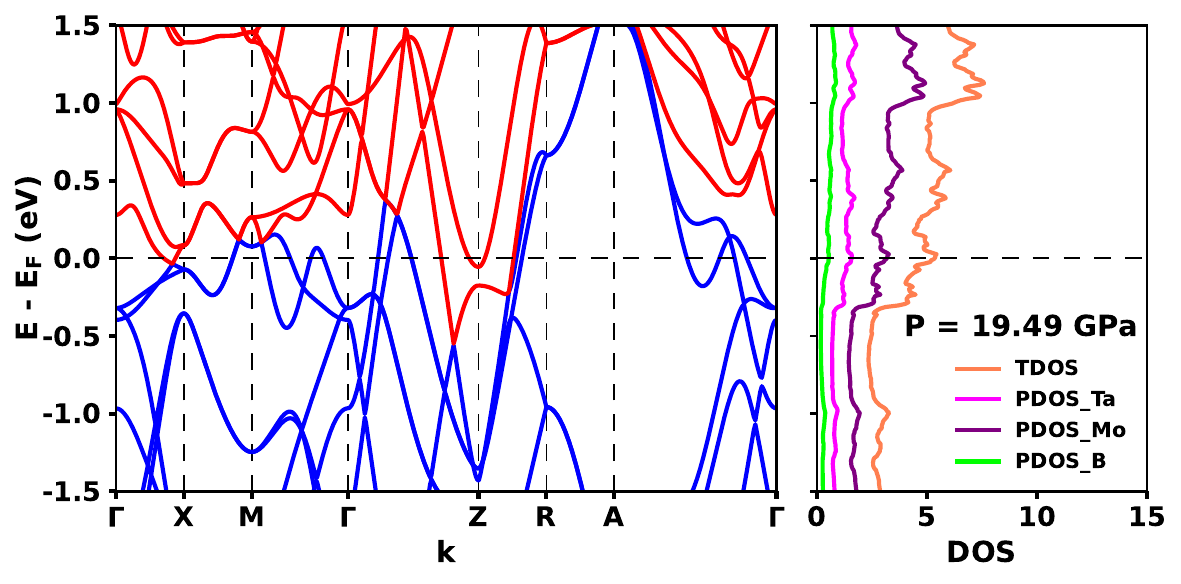}
\includegraphics[width=0.515\linewidth]{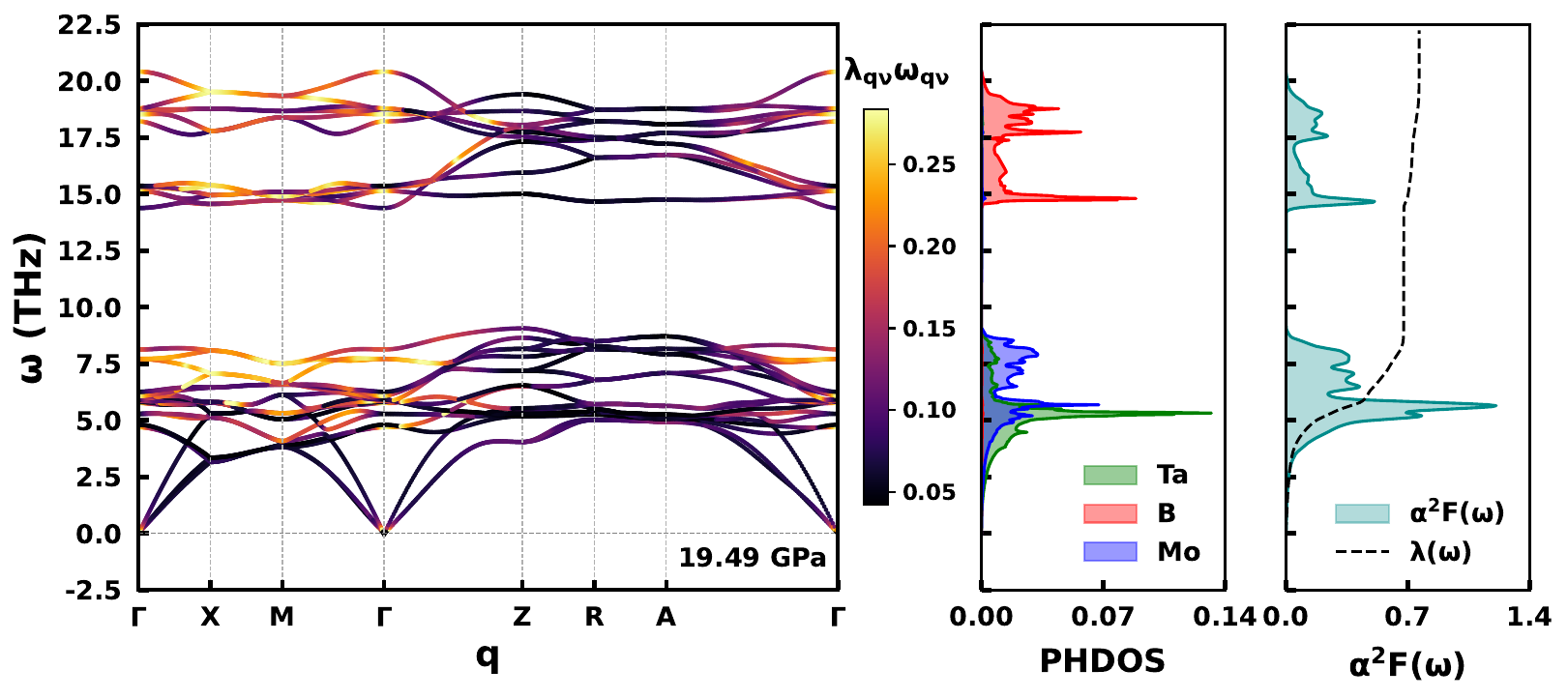}
\vspace*{-0.65cm}
    \begin{center}
        \hspace*{-1.7cm}
        \textbf{\small{(c)}}
        \hspace*{8cm}
        \textbf{\small{(d)}}
    \end{center}
\vspace*{-0.25cm}    
\includegraphics[width=0.475\linewidth]{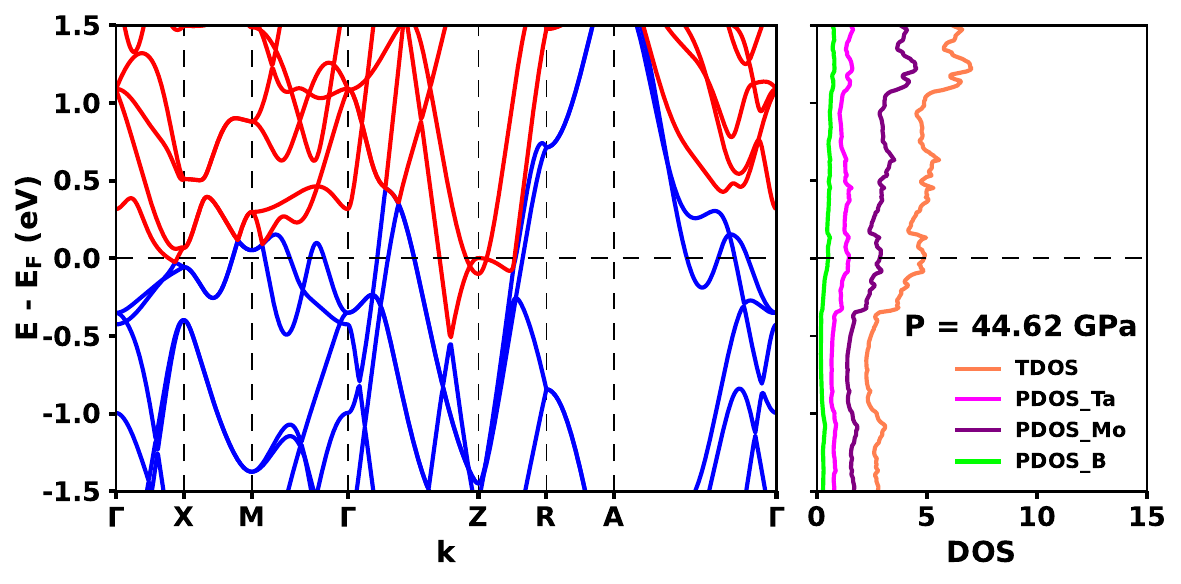}
\includegraphics[width=0.515\linewidth]{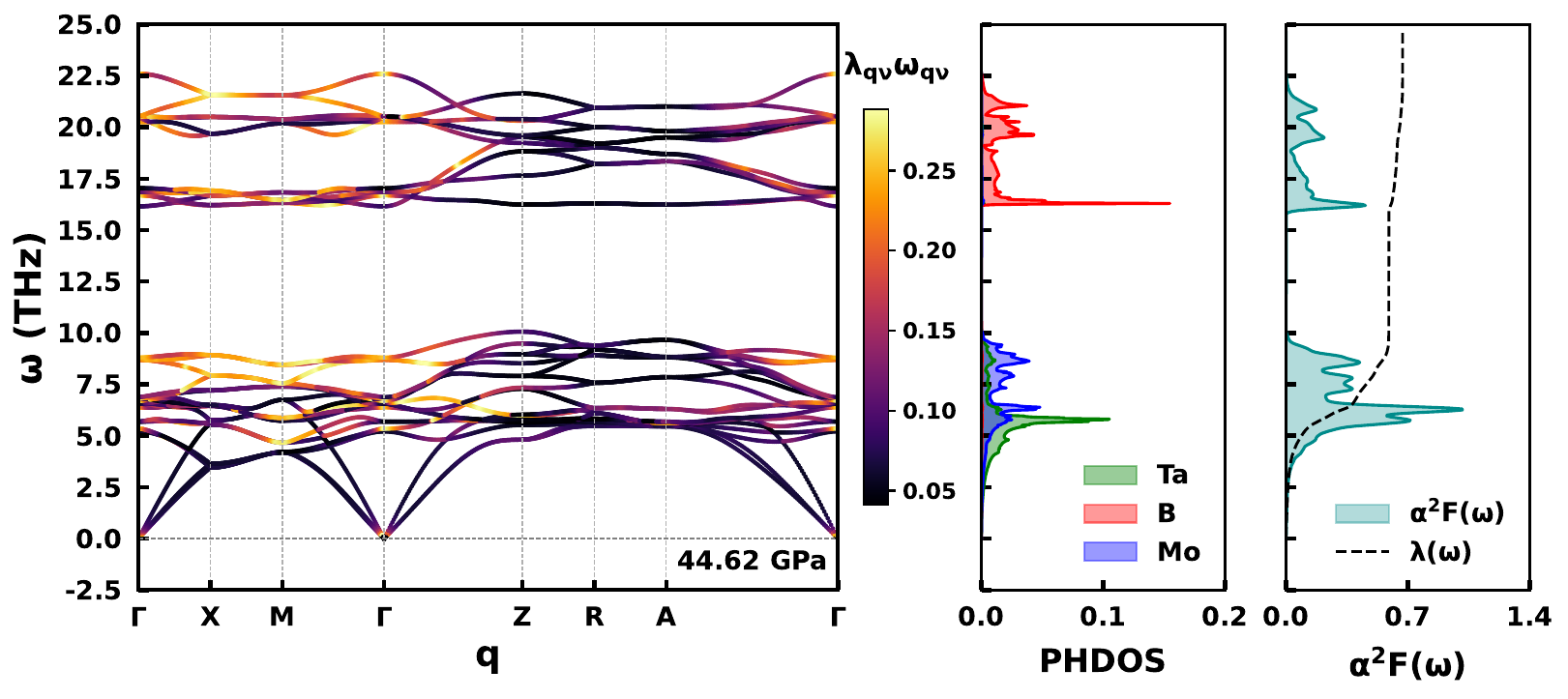}
\vspace*{-0.65cm}
    \begin{center}
        \hspace*{-1.7cm}
        \textbf{\small{(e)}}
        \hspace*{8cm}
        \textbf{\small{(f)}}
    \end{center}
\vspace*{-0.25cm}    
\caption{\label{fig:5} Electronic band structure, total density of states and projected density of states on each atoms of $\mathrm{Ta(MoB)_2}$ at different pressure (a) 9.11 GPa, (c) 19.49 GPa and (e) 44.62 GPa. Phonon dispersion, phonon density of states and color plot of $\mathrm{\lambda_{q\nu} \omega_{q\nu}}$ on phonon bands of $\mathrm{Ta(MoB)_2}$ at different pressure (b) 9.11 GPa, (d) 19.49 GPa and (f) 44.62 GPa. Here, Quantum Espresso \cite{giannozzi2009quantum,giannozzi2017advanced,giannozzi2020quantum} has been used like in the main text}
\end{figure*}

Figure \hyperref[fig:4]{4(a)} illustrates the contrast between the DFT bands computed using the Quantum Espresso software \cite{giannozzi2009quantum,giannozzi2017advanced,giannozzi2020quantum} and the wannier bands computed using the EPW software \cite{ponce2016epw,giustino2007electron,margine2013anisotropic} under ambient pressure. Wannierization allows for the interpolation of electronic band structures and electron-phonon matrix elements on a much finer grid of k-points and q-points than what can be practically achieved by direct DFT computations. A wannier spread of less than or equal to 5 $\mathrm{{\AA}^2}$ and a match of Wannier bands with DFT bands indicate good wannierization. All of our calculated wannier spreads are less than or equal to 5 $\mathrm{{\AA}^2}$, and we have consistently maintained this across all pressure conditions. In Figure \hyperref[fig:4]{4(b),(c)}, we have plotted phonon dispersion of $\mathrm{Ta(MoB)_2}$ at ambient pressure for different smearing width (degauss of 0.01 Ry and 0.015 Ry respectively). The absence of significant phonon softening in phonon bands specifically at the phonon wave vector corresponding to weak nesting is clearly visible from these plots. Figure \hyperref[fig:5]{5} displays the electronic and phononic dispersions of $\mathrm{Ta(MoB)_2}$ at three remaining pressures (\hyperref[fig:5]{5(a)} 9.11 GPa, \hyperref[fig:5]{5(b)} 19.49 GPa, and \hyperref[fig:5]{5(c)} 44.62 GPa) that were not included in the main text's Figure 5. The band structures at

these three different pressures shown in Figure \hyperref[fig:5]{5} also depict the gradual shift of the bands along $\mathrm{\Gamma - X}$ and $\mathrm{M - Z}$ path, a decline in the projected density of states on Mo atoms at the Fermi level and a gradual stiffening of the phonon modes of acoustic and optical branch specifically along $\mathrm{\Gamma - X - M - \Gamma}$ path which contribute to the superconductivity. In the main text, we have already explained the changes in the dispersion and their impact on superconductivity under pressure in detail. The Eliashberg spectral function is depicted in the right panel of Figure \hyperref[fig:5]{5(b)},\hyperref[fig:5]{(d)},\hyperref[fig:5]{(f)} as a function of frequency at each of the applied pressure points. The Eliashberg spectral function is composed of two clusters. The cluster associated with the lower frequency, or the acoustic branch, corresponds to the modes primarily influenced by Mo atoms and consequently the coupling between in-plane stretching with d-orbitals states. The maximum value of that cluster lowers as pressure increases upto 59.71 GPa and again rises when pressure has been further increased from 59.71 GPa to 76.69 GPa. Frequency scale of these figures also indicate phonon stiffening with increase in pressure.

\section{\label{sec:3}Spin-polarized calculation with DFT+U method\protect\\ }

At first, we have performed spin polarized calculation using VASP \cite{kresse1993ab,kresse1996efficiency} for various collinear magnetic orderings (FM, A-AFM, C-AFM and G-AFM) without including Hubbard U parameter. For each pressure, ground state energies of these different magnetic orderings have been shown in the Table \hyperref[tab:2]{II}. From here, it is noticable that energy values of different orderings are nearly identical for any particular pressure. In the output file (OUTCAR file), there are no magnetic moments induced in any atoms (Ta and Mo atoms).

\begin{table}[h!]
\caption{\label{tab:2}The values of ground state energies at different pressures for various magnetic phases without Hubbard $\mathrm{U_{eff}}$ ($\mathrm{= U - J}$)}
\renewcommand{\arraystretch}{2.1}  
\centering
\vspace{10pt}
\begin{tabular}{|c|c|c|c|c|c|}
\hline
\multirow{2}{*}{Pressure (GPa)} & \multicolumn{5}{c|}{Energy (eV)} \\
\cline{2-6}
 & NM & FM & A-AFM & C-AFM & G-AFM \\
\hline
0.00   & -99.075 & -99.075 & -99.075 & -99.075 & -99.075 \\
\hline
9.11   & -98.999 & -98.999 & -98.999 & -98.999 & -98.999 \\
\hline
19.49  & -98.711 & -98.711 & -98.711 & -98.711 & -98.711 \\
\hline
31.28  & -98.190 & -98.190 & -98.190 & -98.190 & -98.190 \\
\hline
44.62  & -97.412 & -97.412 & -97.412 & -97.412 & -97.412 \\
\hline
59.61  & -96.356 & -96.356 & -96.356 & -96.356 & -96.356 \\
\hline
76.69  & -94.996 & -94.996 & -94.996 & -94.996 & -94.996 \\
\hline
\end{tabular}
\end{table}

\begin{figure*}[h!]
\begin{center}
    \includegraphics[width=0.315\linewidth]{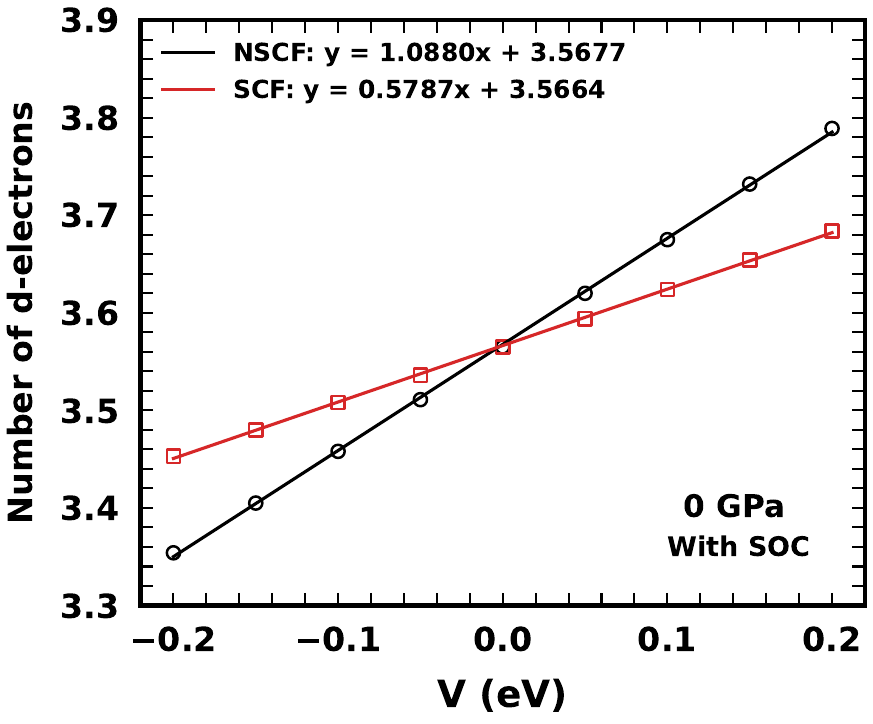}
\hspace{10pt}
\includegraphics[width= 0.315\linewidth]{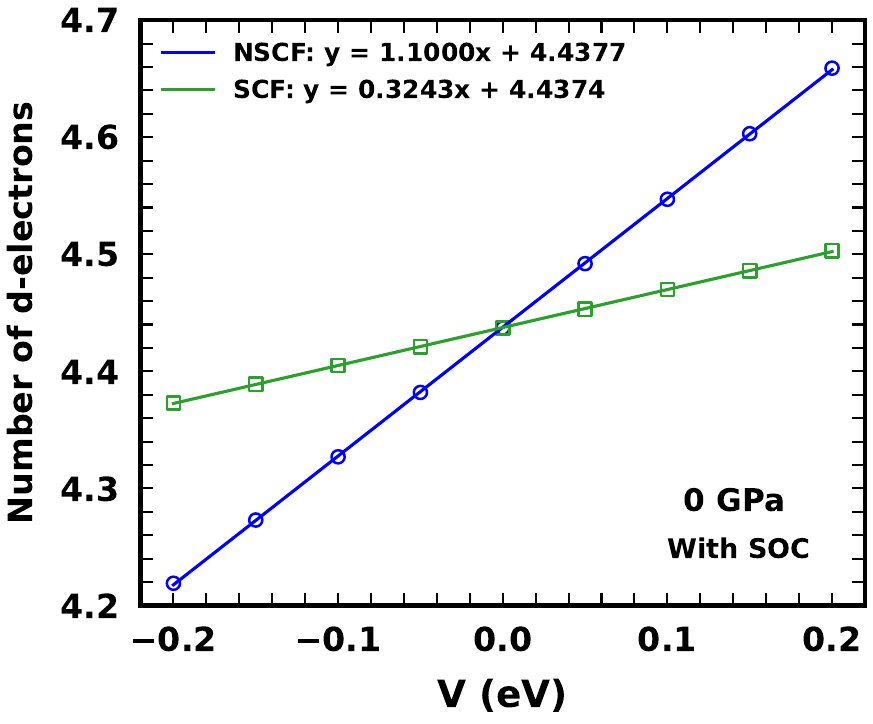}
\end{center}
\vspace*{-0.65cm}
\begin{center}
    \hspace*{0.8cm}
    \textbf{\small{(a)}}
    \hspace*{5.6cm}
    \textbf{\small{(b)}}
\end{center}

\begin{center}
    \includegraphics[width=0.315\linewidth]{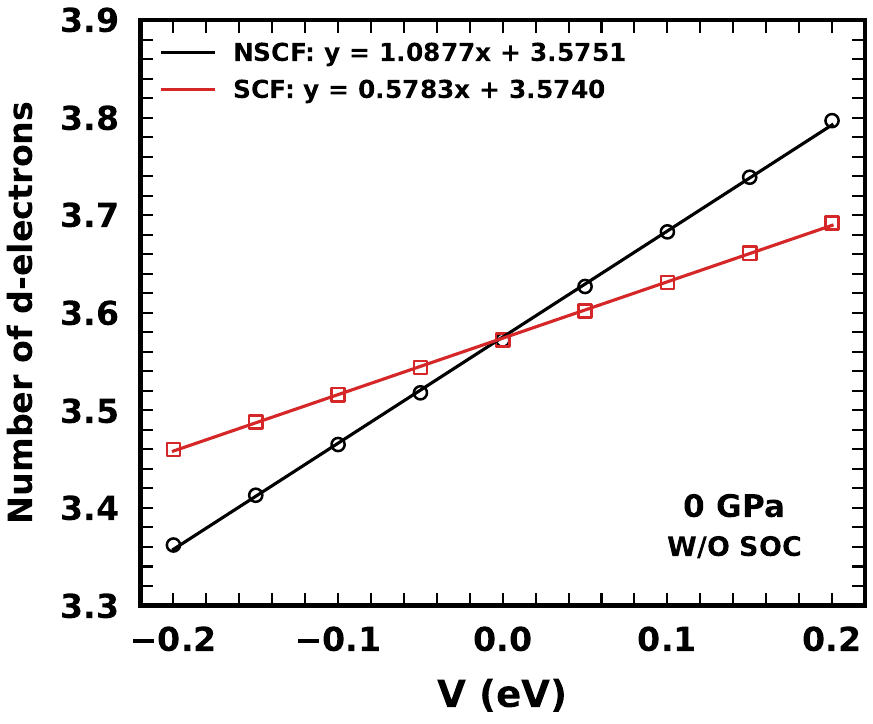}
\hspace{10pt}
\includegraphics[width= 0.315\linewidth]{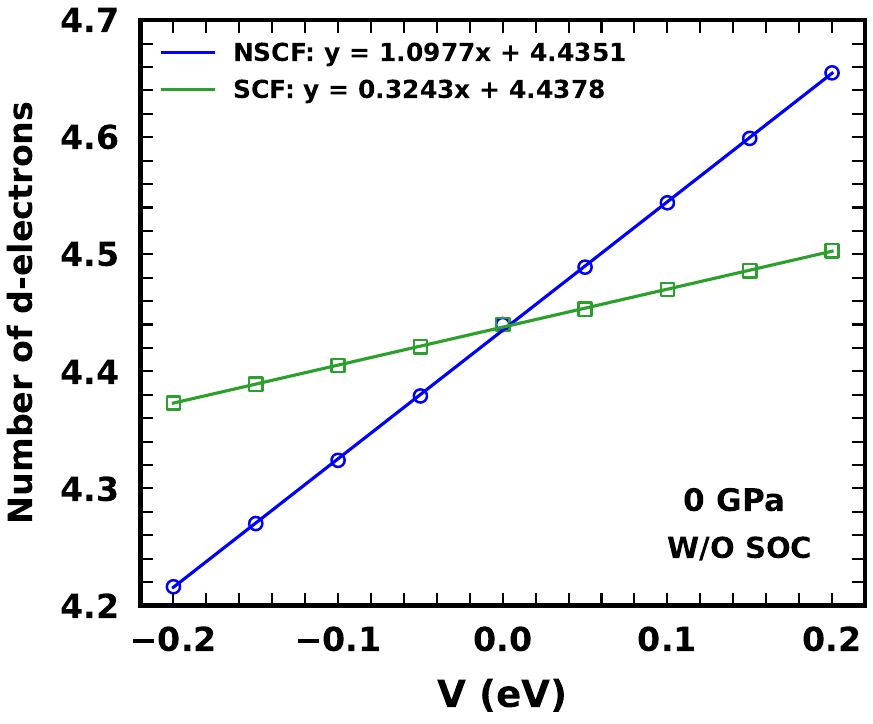}
\end{center}
\vspace*{-0.65cm}
\begin{center}
    \hspace*{0.8cm}
    \textbf{\small{(c)}}
    \hspace*{5.6cm}
    \textbf{\small{(d)}}
\end{center}
\vspace*{-0.55cm}
\caption{\label{fig:6} Linear fit of the number of d-electrons as a function of the additional potential V to find the response function in linear response ansatz method proposed by Cococcioni et al. \cite{cococcioni2005linear} at ambient pressure (a) on the site of Ta atom (with SOC), (b) on the site of Mo atom (with SOC), (c) on the site of Ta atom (without SOC),(d) on the site of Mo atom (without SOC)}
\end{figure*}

\begin{figure*}[t]
\begin{center}
    \includegraphics[width=0.27\linewidth]{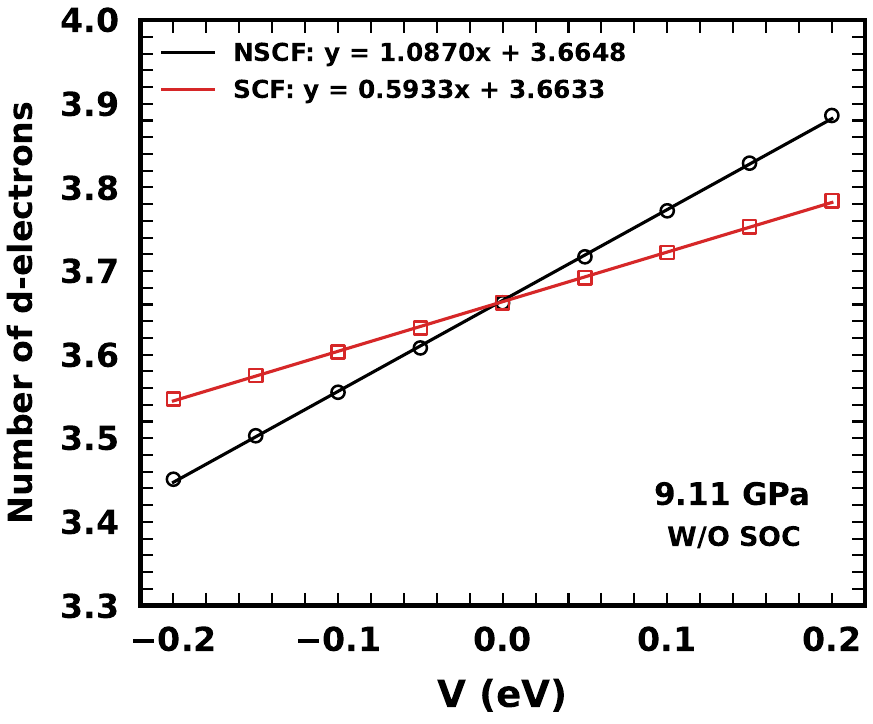}
    \includegraphics[width= 0.27\linewidth]{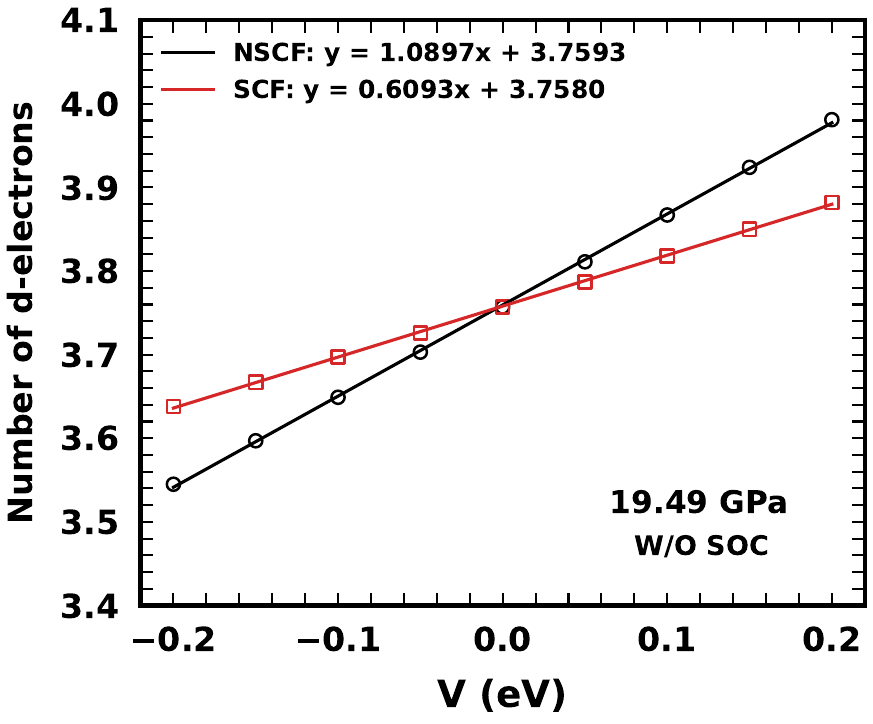}
    \includegraphics[width= 0.27\linewidth]{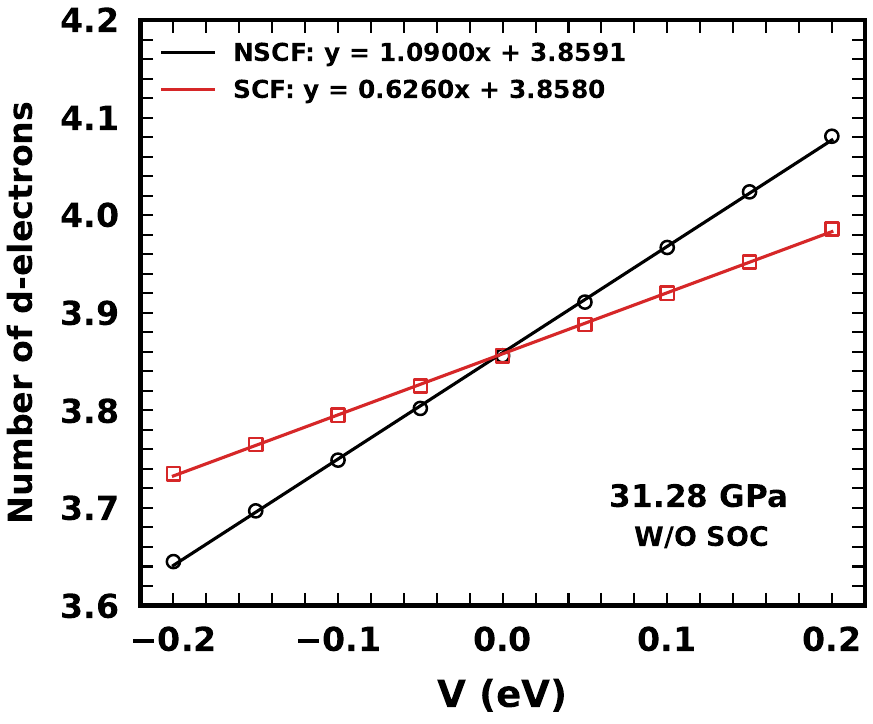}
\end{center}
\vspace*{-0.65cm}
\begin{center}
    \hspace*{0.7cm}
    \textbf{\small{(a)}}
    \hspace*{4.2cm}
    \textbf{\small{(b)}}
    \hspace*{4.3cm}
    \textbf{\small{(c)}}
\end{center}
\vspace*{-0.45cm}
\begin{center}
    \includegraphics[width=0.27\linewidth]{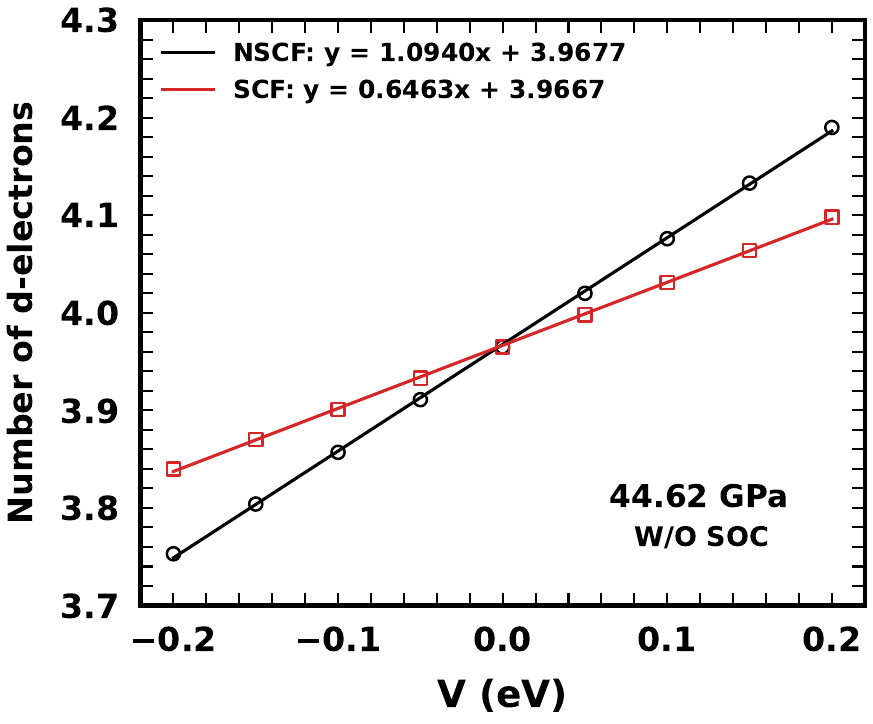}
    \includegraphics[width= 0.27\linewidth]{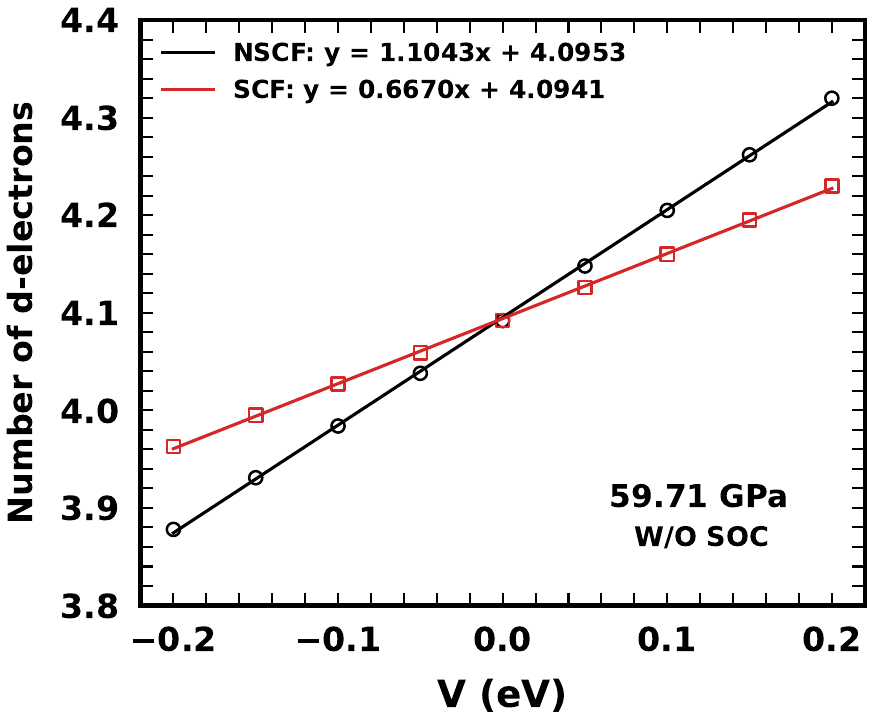}
    \includegraphics[width= 0.27\linewidth]{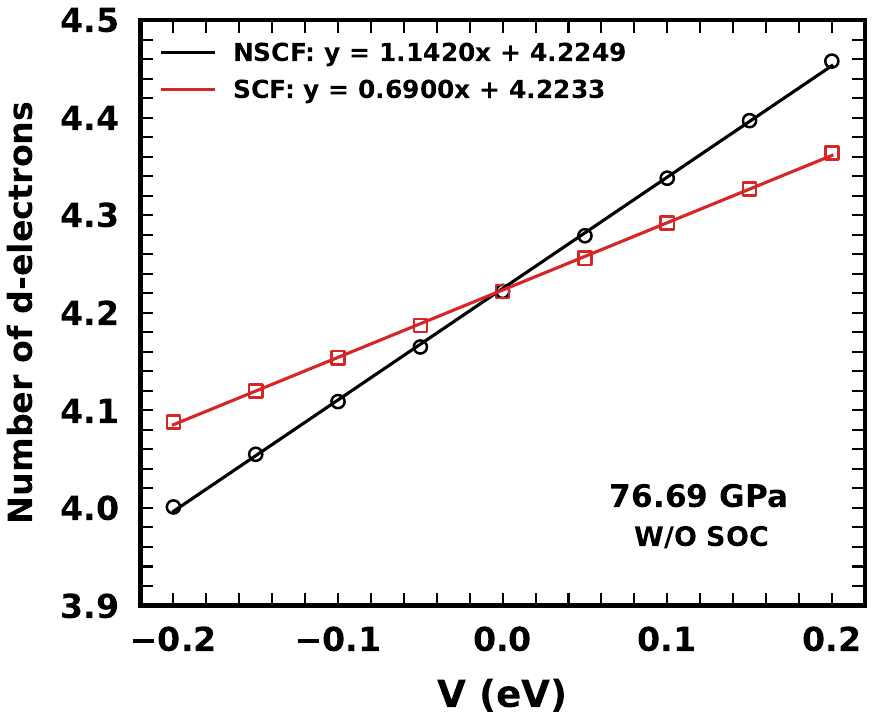}
\end{center}
\vspace*{-0.65cm}
\begin{center}
    \hspace*{0.7cm}
    \textbf{\small{(d)}}
    \hspace*{4.2cm}
    \textbf{\small{(e)}}
    \hspace*{4.3cm}
    \textbf{\small{(f)}}
\end{center}
\vspace*{-0.4935cm}
\caption{\label{fig:7} Linear fit of the number of d-electrons as a function of the additional potential V to find the response function in linear response ansatz method proposed by Cococcioni et al. \cite{cococcioni2005linear} on the site of Ta atom (Without SOC) at (a) 9.11 GPa, (b) 19.49 GPa, (c) 31.28 GPa, (d) 44.62 GPa, (e) 59.71 GPa and (f) 76.69 GPa}
\end{figure*}

\begin{figure*}[h!]
\begin{center}
    \includegraphics[width=0.27\linewidth]{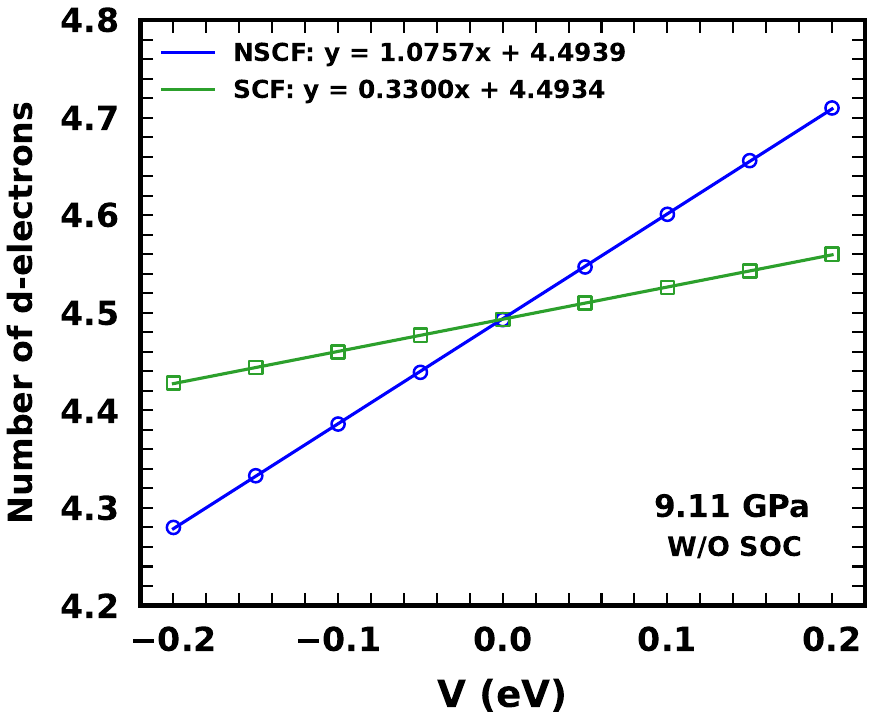}
    \includegraphics[width= 0.27\linewidth]{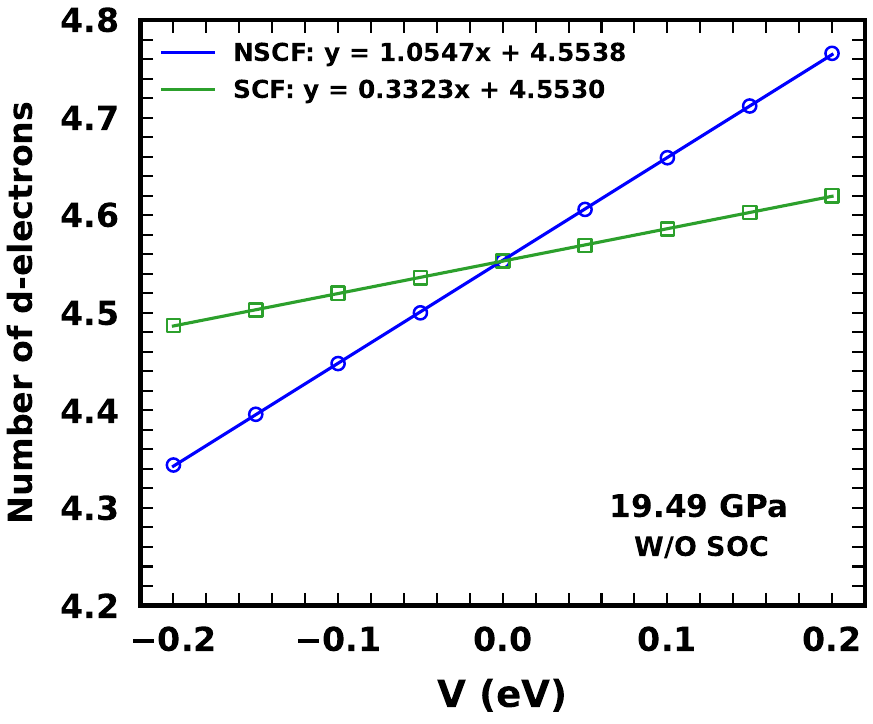}
    \includegraphics[width= 0.27\linewidth]{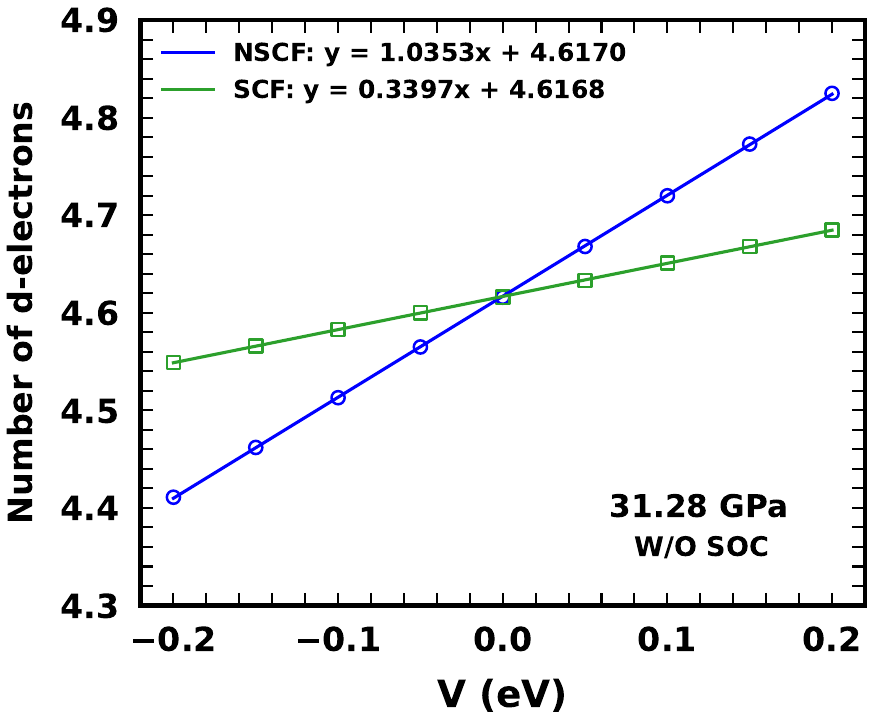}
\end{center}
\vspace*{-0.65cm}
\begin{center}
    \hspace*{0.7cm}
    \textbf{\small{(a)}}
    \hspace*{4.2cm}
    \textbf{\small{(b)}}
    \hspace*{4.3cm}
    \textbf{\small{(c)}}
\end{center}
\vspace*{-0.45cm}
\begin{center}
    \includegraphics[width=0.27\linewidth]{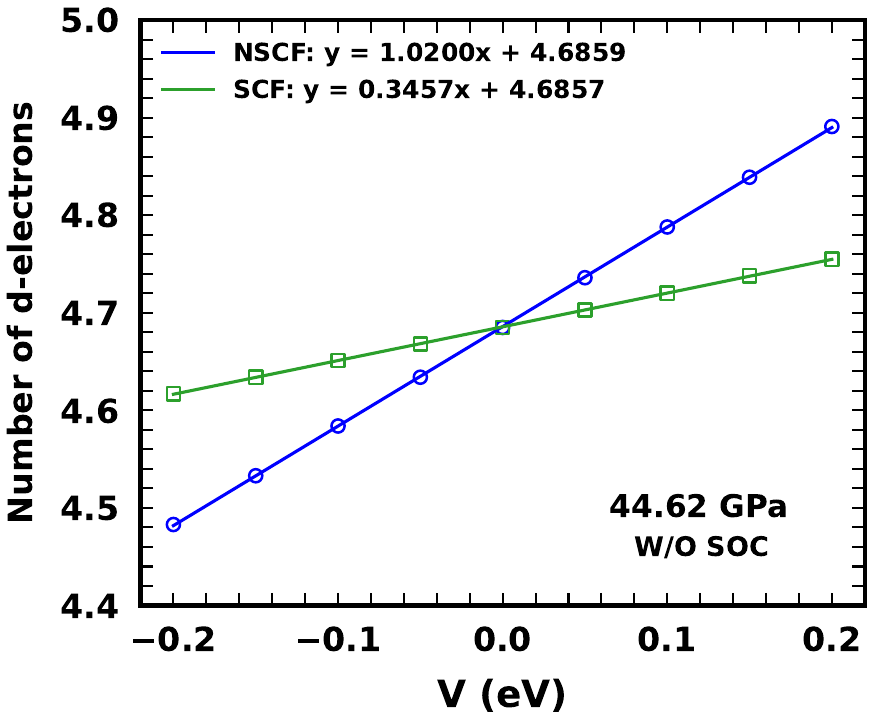}
    \includegraphics[width= 0.27\linewidth]{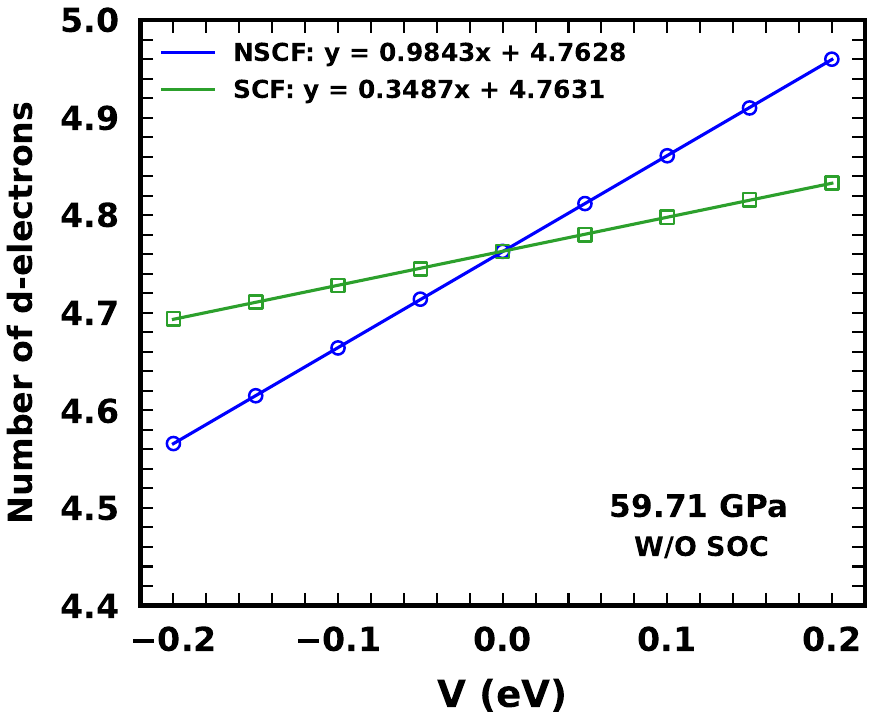}
    \includegraphics[width= 0.27\linewidth]{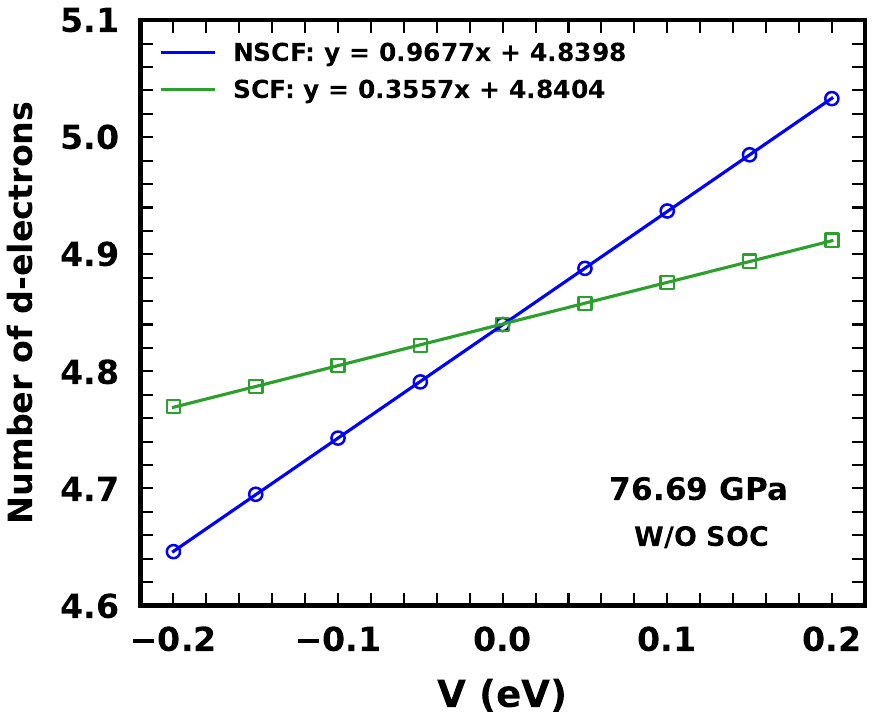}
\end{center}
\vspace*{-0.65cm}
\begin{center}
    \hspace*{0.7cm}
    \textbf{\small{(d)}}
    \hspace*{4.2cm}
    \textbf{\small{(e)}}
    \hspace*{4.3cm}
    \textbf{\small{(f)}}
\end{center}
\vspace*{-0.4935cm}
\caption{\label{fig:8} Linear fit of the number of d-electrons as a function of the additional potential V to find the response function in linear response ansatz method proposed by Cococcioni et al. \cite{cococcioni2005linear} on the site of Mo atom (Without SOC) at (a) 9.11 GPa, (b) 19.49 GPa, (c) 31.28 GPa, (d) 44.62 GPa, (e) 59.71 GPa and (f) 76.69 GPa}
\end{figure*}

In the next step, we have performed spin-polarized calculations including Hubbard U parameter. Since there is no experimental report about the value of U for Ta and Mo d-orbitals in $\mathrm{Ta(MoB)_2}$, we have calculated the values of U theoretically using the linear response ansatz method proposed by Cococcioni et al. \cite{cococcioni2005linear} implemented in VASP \cite{kresse1993ab,kresse1996efficiency}. In this purpose, we have made a 2 $\times$ 2 $\times$ 2 supercell of $\mathrm{Ta(MoB)_2}$. In the supercell, atoms 1-16 are Ta, atoms 17-48 are B and atoms 49 to 80 are Mo. For finding the value of U for Ta d-orbitals, Ta atoms are split into two groups: atom 1 and atoms 2-16. It breaks the symmetry of Ta sublattice and then we can treat that atom 1 differently from atom 2-16.

\begin{table}[h!]
\vspace{-10pt}
\caption{\label{tab:3}The values of Hubbard $\mathrm{U_{eff}}$ at different pressure}
\renewcommand{\arraystretch}{2.1}  
\centering
\vspace{10pt}
\begin{tabular}{|c|c|c|c|c|c|c|c|c|}
\hline
\multicolumn{2}{|c|}{Pressure (GPa)} & 0 & 9.11 & 19.49 & 31.28 & 44.62 & 59.61 & 76.69 \\
\hline
\multirow{2}{*}{$\mathrm{U_{eff}}$ (eV)} & Ta & 0.81 & 0.77 & 0.72 & 0.68 & 0.63 & 0.59 & 0.57 \\
\cline{2-9}
                   & Mo & 2.17 & 2.10 & 2.06 & 1.98 & 1.91 & 1.85 & 1.78 \\
\hline
\end{tabular}
\end{table}

\begin{table}[h!]
\vspace{-10pt}
\caption{\label{tab:4}The values of ground state energies at different pressures for various magnetic phases with Hubbard $\mathrm{U_{eff}}$ ($\mathrm{= U - J}$)}
\renewcommand{\arraystretch}{2.1}  
\centering
\vspace{10pt}
\begin{tabular}{|c|c|c|c|c|c|}
\hline
\multirow{2}{*}{Pressure (GPa)} & \multicolumn{5}{c|}{Energy (eV)} \\
\cline{2-6}
 & NM & FM & A-AFM & C-AFM & G-AFM \\
\hline
0.00   & -86.762 & -86.762 & -86.762 & -86.762 & -86.762 \\
\hline
9.11   & -87.064 & -87.064 & -87.064 & -87.064 & -87.064 \\
\hline
19.49  & -87.037 & -87.037 & -87.037 & -87.037 & -87.037 \\
\hline
31.28  & -86.954 & -86.954 & -86.954 & -86.954 & -86.954 \\
\hline
44.62  & -86.598 & -86.598 & -86.598 & -86.598 & -86.598 \\
\hline
59.61  & -85.898 & -85.898 & -85.898 & -85.898 & -85.898 \\
\hline
76.69  & -84.903 & -84.903 & -84.903 & -84.903 & -84.903 \\
\hline
\end{tabular}
\end{table}

To calculate the selfconsistent and non-selfconsistent response at ambient pressure, we have plotted the number of d-electrons of the atom 1 (Ta atom) with the additional spherical potential acting on the d-manifold on atom 1 in Figure \hyperref[fig:6]{6(a)},\hyperref[fig:6]{(c)}. After that, we have computed the value of Hubbard $\mathrm{U_{eff}}$ $\mathrm{(U-J)}$ taking the difference between inverse of selfconsistent response and inverse of non-selfconsistent response \cite{cococcioni2005linear}. The computed value of the Hubbard $\mathrm{U_{eff}}$ for Ta d-orbitals in $\mathrm{Ta(MoB)_2}$ at ambient pressure is 0.81 eV. We have checked this value in the absence of SOC (Figure \hyperref[fig:6]{6(c)}) and in the presence of SOC (Figure \hyperref[fig:6]{6(a)}). The values of $\mathrm{U_{eff}}$ for those two cases are nearly identical. So, SOC has no significant impact on the value of $\mathrm{U_{eff}}$ and therefore been neglected in further calculation at hydrostatic pressure. Similarly, the computed value of $\mathrm{U_{eff}}$ for Mo d-orbitals (Figure \hyperref[fig:6]{6(b)},\hyperref[fig:6]{(d)}) at ambient pressure is 2.17 eV. We have also checked whether there is any magnetic ordering dependence of $\mathrm{U_{eff}}$ at ambient pressure or not and found that $\mathrm{U_{eff}}$ does not change significantly for different magnetic orderings. Therefore, we have also neglected this for further high pressure calculation. In the same way, we have calculated the value of $\mathrm{U_{eff}}$ for Ta and Mo d-orbitals in $\mathrm{Ta(MoB)_2}$ at all pressure (Figure \hyperref[fig:7]{7} and \hyperref[fig:8]{8}) and attached those values in Table \hyperref[tab:3]{III}. The value of $\mathrm{U_{eff}}$ depends on the degree of electronic screening from the surrounding atoms. From Table \hyperref[tab:3]{III}, we observe that the value of $\mathrm{U_{eff}}$ is decreasing for Ta and Mo d-orbitals with pressure. Considering these computed values of $\mathrm{U_{eff}}$, we have calculated ground state energies of different magnetic orderings at each pressure and attached those values in Table \hyperref[tab:4]{IV}. It is evident that the energy values of various orderings are almost equivalent at any given pressure. In the output file (OUTCAR), no magnetic moments are induced in any Ta and Mo atoms. Here, SOC has no effect on this situation except making a slight reduction in energy value for all magnetic ordering. Therefore, although spin fluctuations are moderately strong in Ta and Mo atoms, spin-polarized states do not take advantage under pressure, as evident from our DFT+U studies.

\section{\label{sec:4}Superconductivity calculation\protect\\ }

\begin{figure*}[t]
\includegraphics[width=0.31\linewidth]{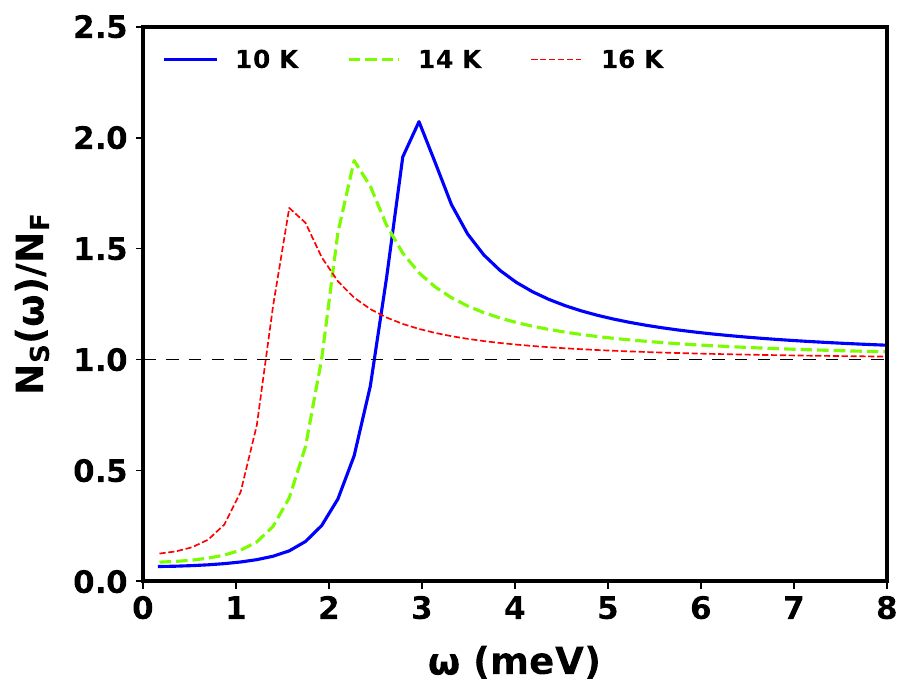}
\includegraphics[width=0.31\linewidth]{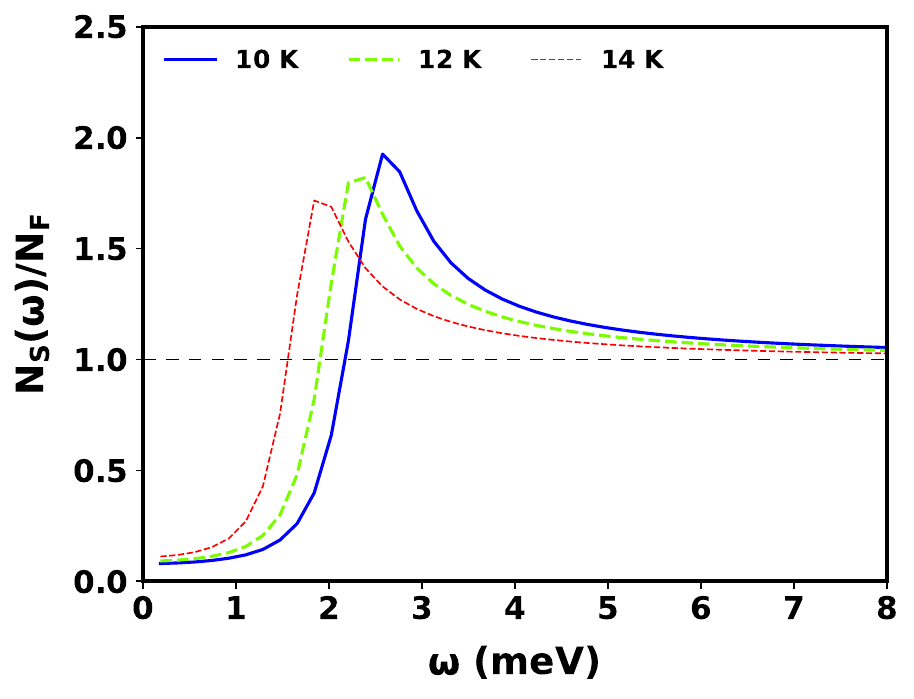}
\includegraphics[width=0.31\linewidth]{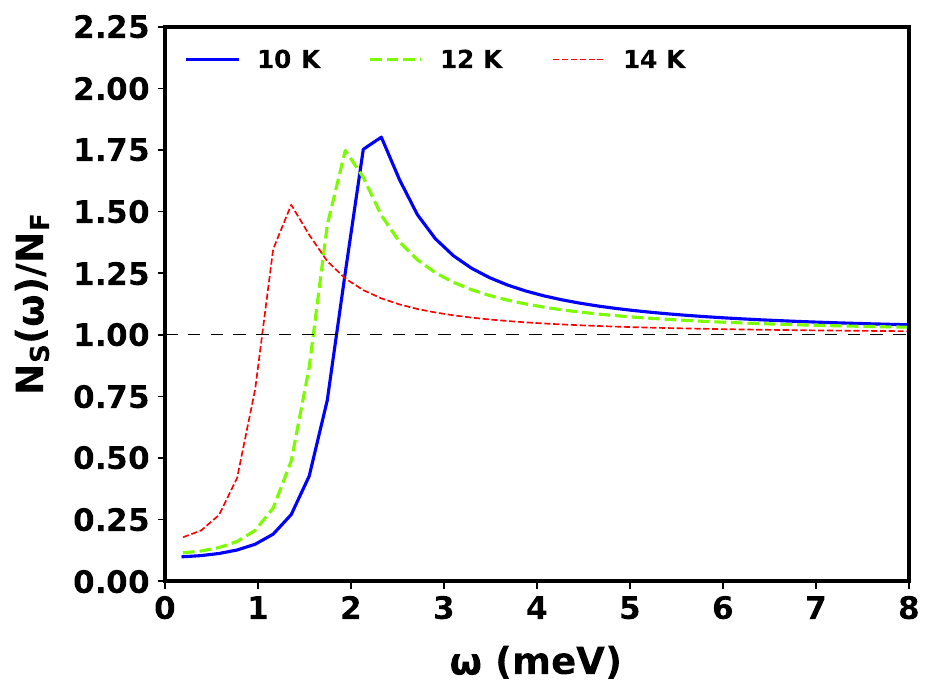}
\vspace*{-0.20cm}
    \begin{center}
        \hspace*{0.7cm}
        \textbf{\small{(a)}}
        \hspace*{4.9cm}
        \textbf{\small{(b)}}
        \hspace*{5.0cm}
        \textbf{\small{(c)}}
    \end{center}
\vspace*{-0.20cm}
\includegraphics[width=0.31\linewidth]{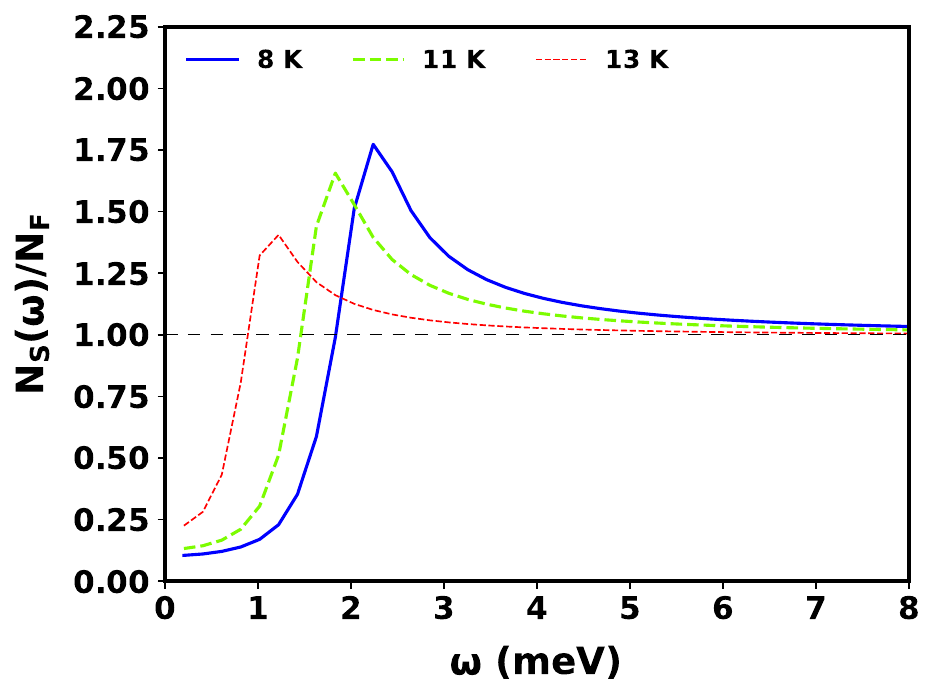}
\includegraphics[width=0.31\linewidth]{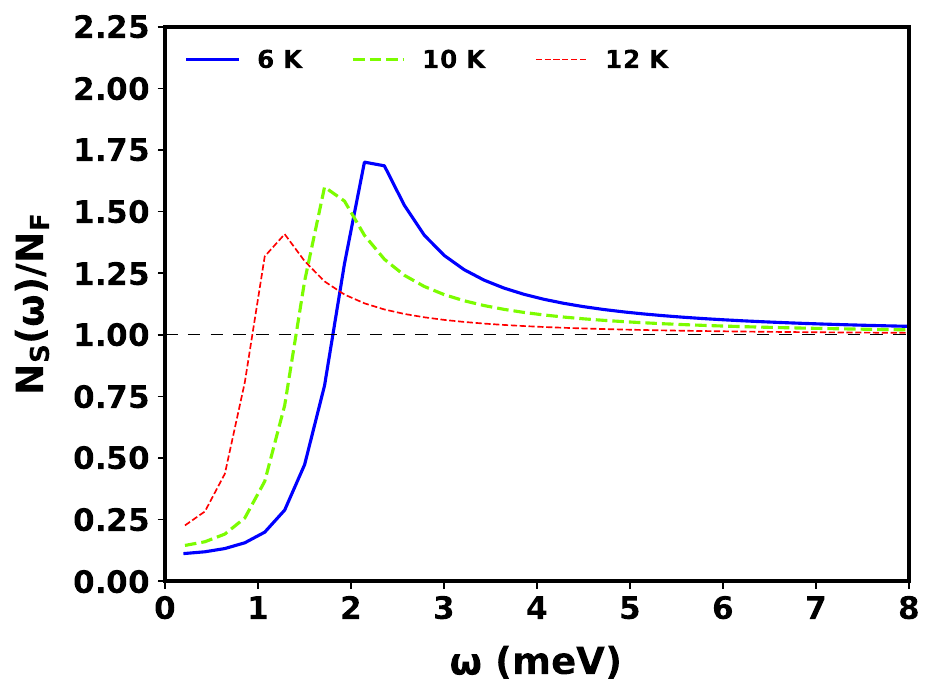}
\includegraphics[width=0.31\linewidth]{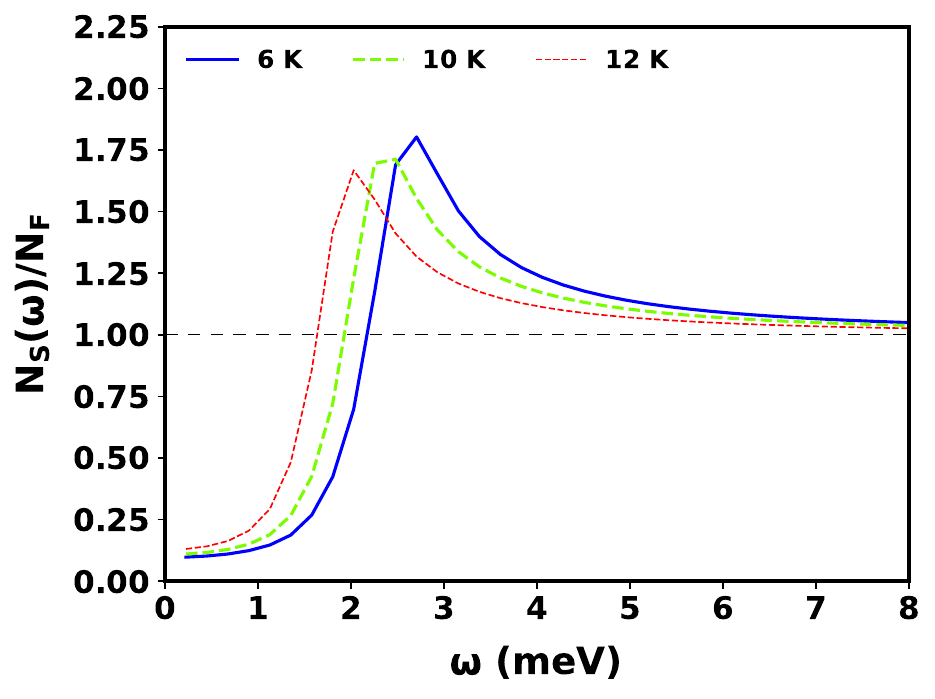}
\vspace*{-0.20cm}
    \begin{center}
        \hspace*{0.7cm}
        \textbf{\small{(d)}}
        \hspace*{4.9cm}
        \textbf{\small{(e)}}
        \hspace*{5.0cm}
        \textbf{\small{(f)}}
    \end{center}
\vspace{-10pt}
\caption{\label{fig:9} Quasiparticle density of states in the superconducting states with respect to normal states of $\mathrm{Ta(MoB)_2}$ calculated for (a) 9.11 GPa, (b) 19.49 GPa, (c) 31.28 GPa, (d) 44.62 GPa, (e) 59.71 GPa and (f) 76.69 GPa}
\end{figure*}

\begin{figure*}[h!]
\includegraphics[width=0.31\linewidth]{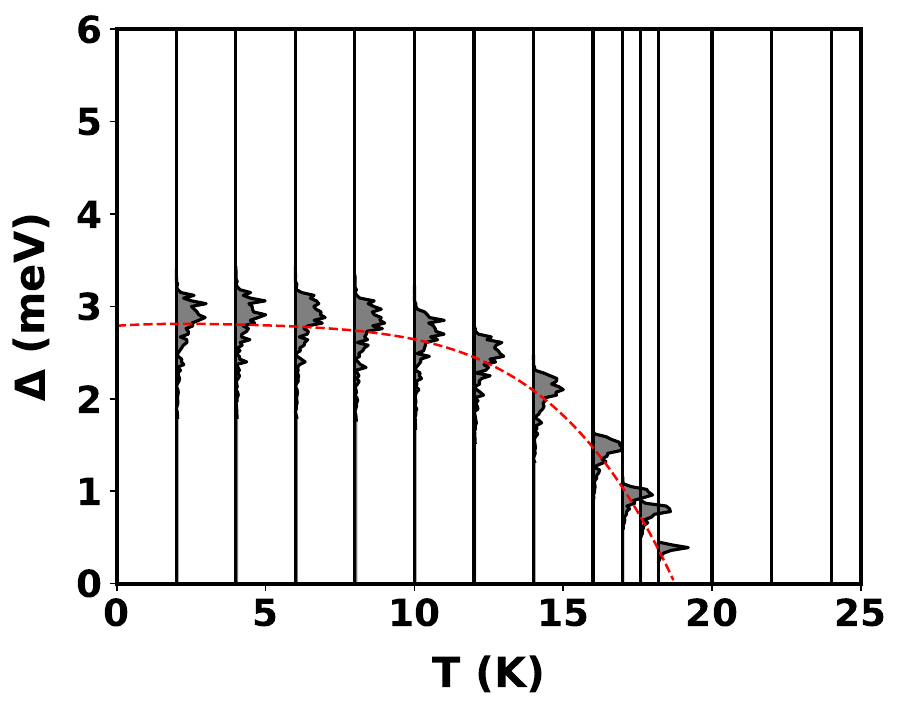}
\includegraphics[width=0.31\linewidth]{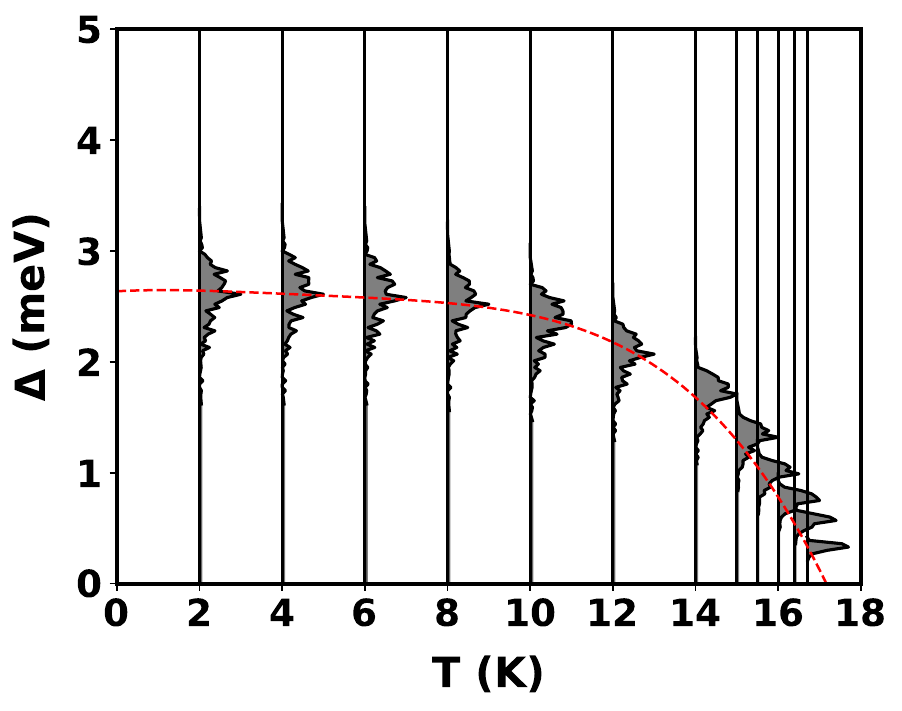}
\includegraphics[width=0.31\linewidth]{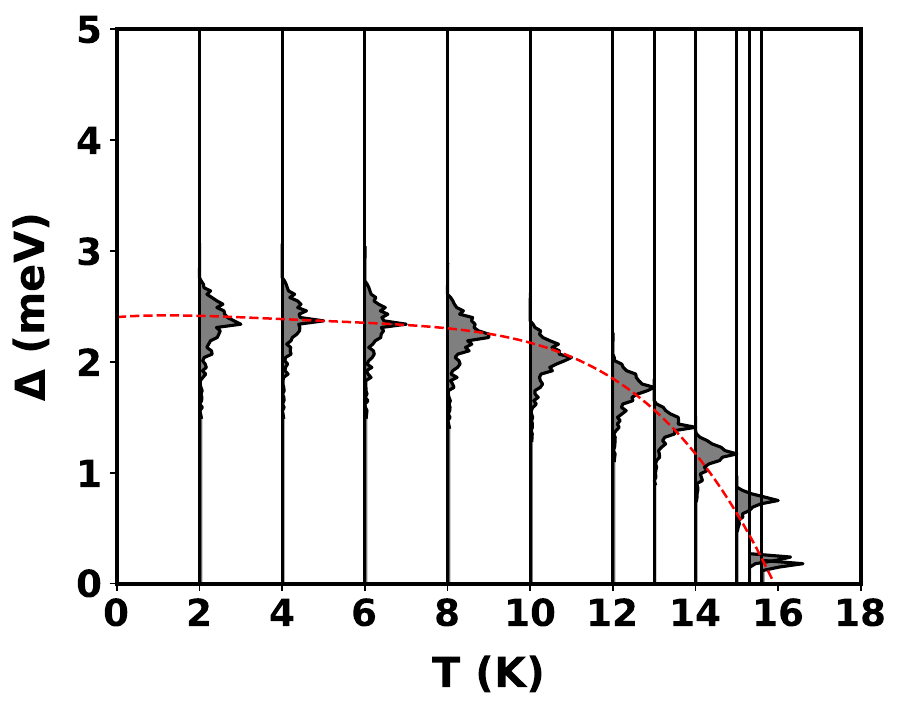}
\vspace*{-0.20cm}
    \begin{center}
        \hspace*{0.2cm}
        \textbf{\small{(a)}}
        \hspace*{5.0cm}
        \textbf{\small{(b)}}
        \hspace*{5.0cm}
        \textbf{\small{(c)}}
    \end{center}
\vspace*{-0.20cm}
\includegraphics[width=0.31\linewidth]{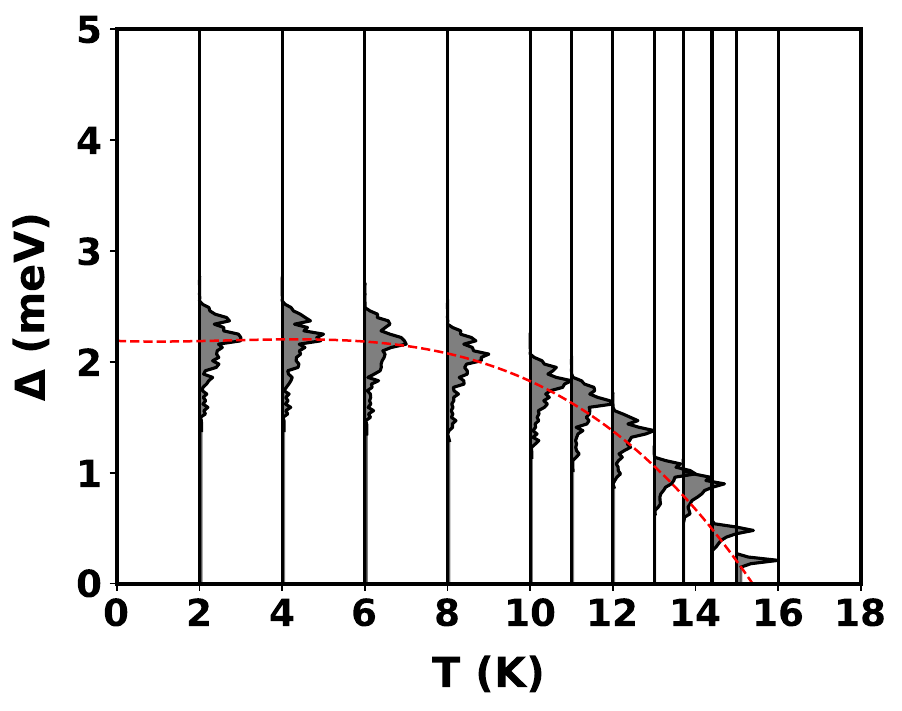}
\includegraphics[width=0.31\linewidth]{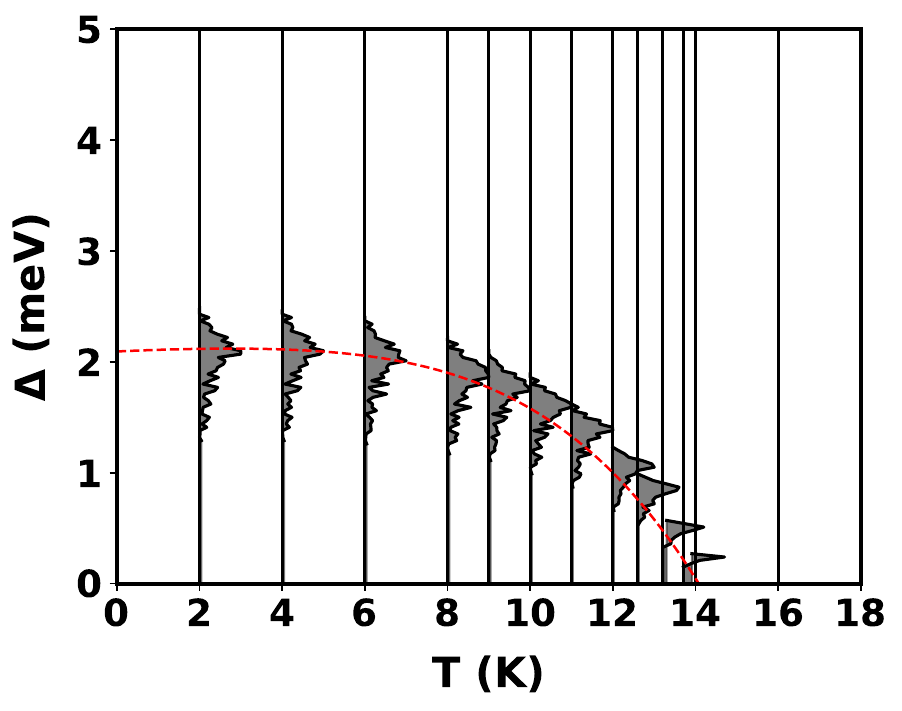}
\includegraphics[width=0.31\linewidth]{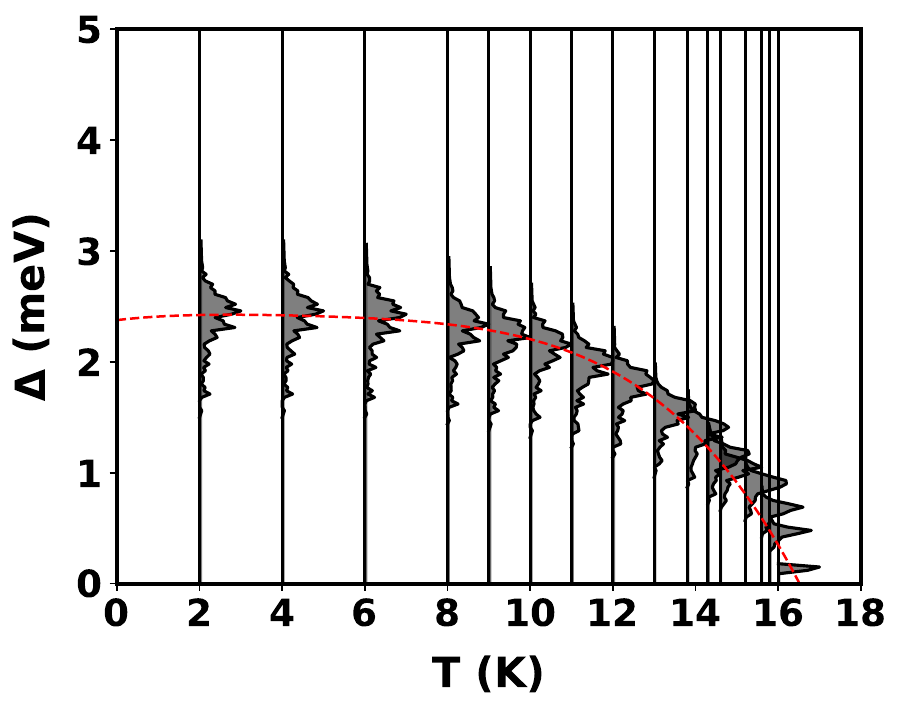}
\vspace*{-0.20cm}
    \begin{center}
        \hspace*{0.2cm}
        \textbf{\small{(d)}}
        \hspace*{5.0cm}
        \textbf{\small{(e)}}
        \hspace*{5.0cm}
        \textbf{\small{(f)}}
    \end{center}
\vspace{-10pt}
\caption{\label{fig:10} Anisotropic superconducting gap on the Fermi surface of $\mathrm{Ta(MoB)_2}$ calculated for (a) 9.11 GPa, (b) 19.49 GPa, (c) 31.28 GPa, (d) 44.62 GPa, (e) 59.71 GPa and (f) 76.69 GPa}
\end{figure*}

Allen-Dyne modified McMillan formula \cite{mcmillan1968transition,allen1975transition} and Migdal Eliashberg equations \cite{migdal1958interaction,eliashberg1960interactions} implemented in EPW code \cite{ponce2016epw,giustino2007electron,margine2013anisotropic} were employed to determine superconductivity. Figure \hyperref[fig:9]{9} depicts the ratio of the density of states in the superconducting state to the density of states in the normal state as a function of frequency. Here we have adjusted our Fermi level to a value of 1. The quasiparticle density of states in the superconducting state relative to the normal state diminishes with pressure up to 59.71 GPa and subsequently increases at 76.69 GPa. The anisotropic superconducting gap on the Fermi surface of $\mathrm{Ta(MoB)_2}$ is illustrated in Figure \hyperref[fig:10]{10} for each high pressure what we have applied. Figures \hyperref[fig:9]{9} and \hyperref[fig:10]{10} demonstrate that the superconducting temperature and superconducting gap diminish as the pressure increases up to 59.71GPa, and then increase again as the pressure increases from 59.71 GPa to 76.69 GPa. This electronic phase transition can be attributed to the improved nesting caused by Lifshitz transition.

\section{\label{sec:5}Convergence calculation\protect\\ }

\begin{figure*}[h]

\includegraphics[width=0.340\linewidth]{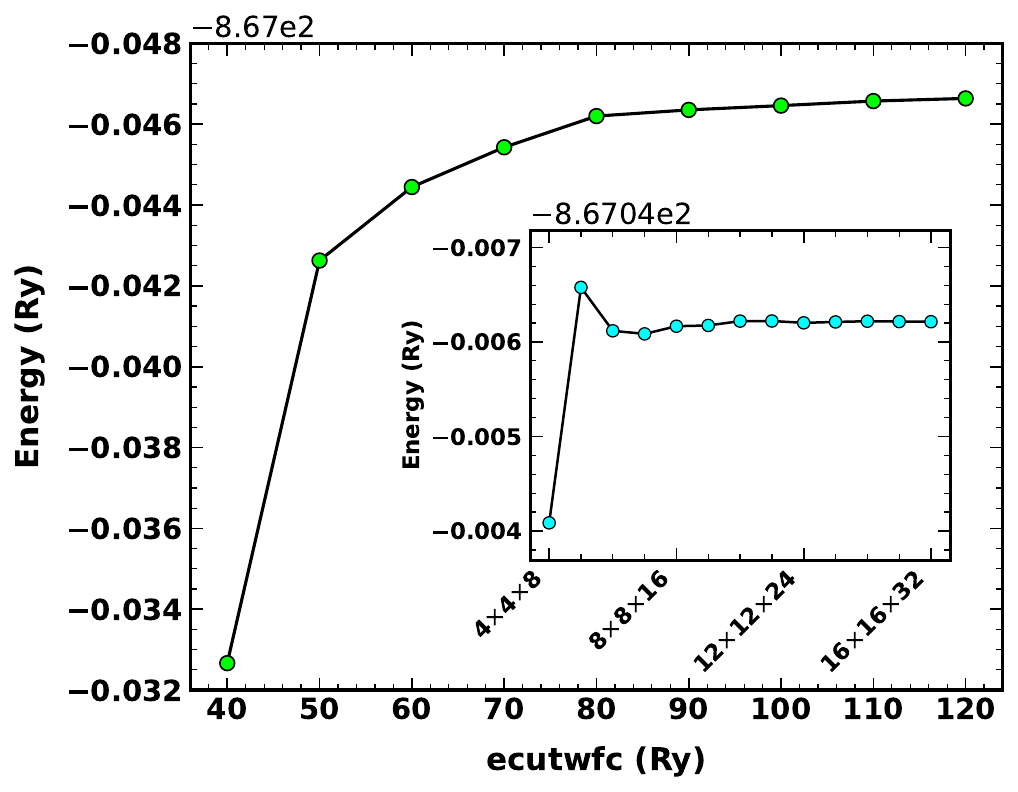}
\includegraphics[width=0.329\linewidth]{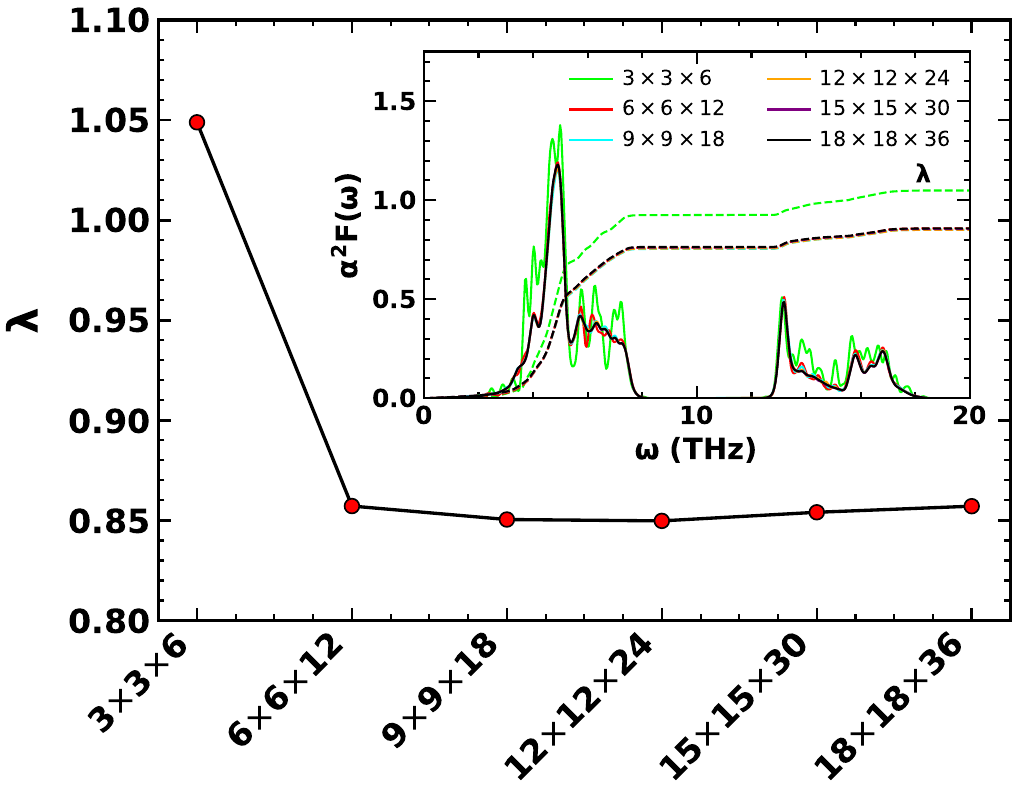}
    \begin{center}
        \hspace*{0.4cm}
        \textbf{\small{(a)}}
        \hspace*{5.1cm}
        \textbf{\small{(b)}}
    \end{center}
\includegraphics[width=0.320\linewidth]{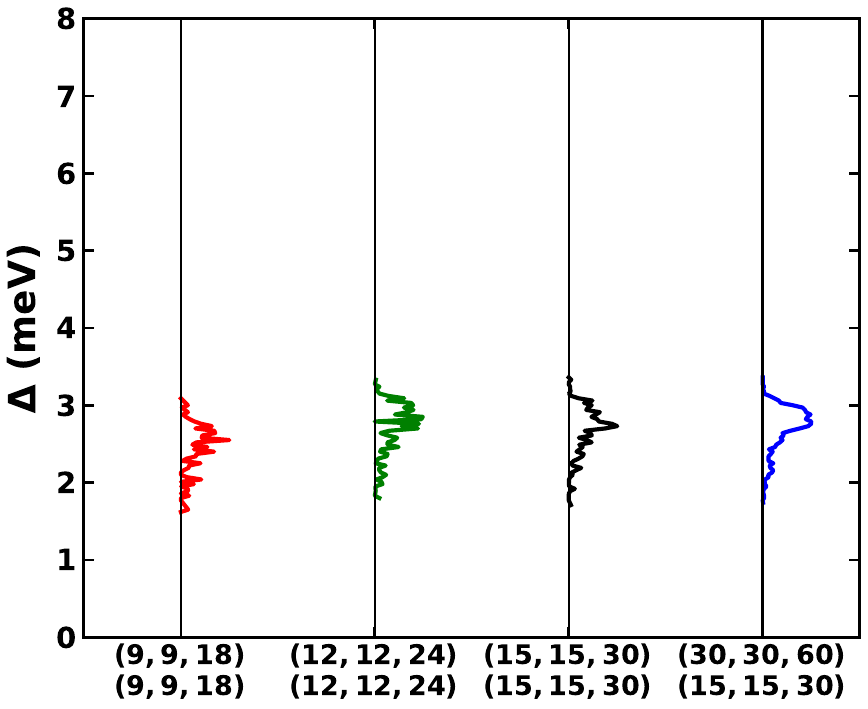}
\includegraphics[width=0.344\linewidth]{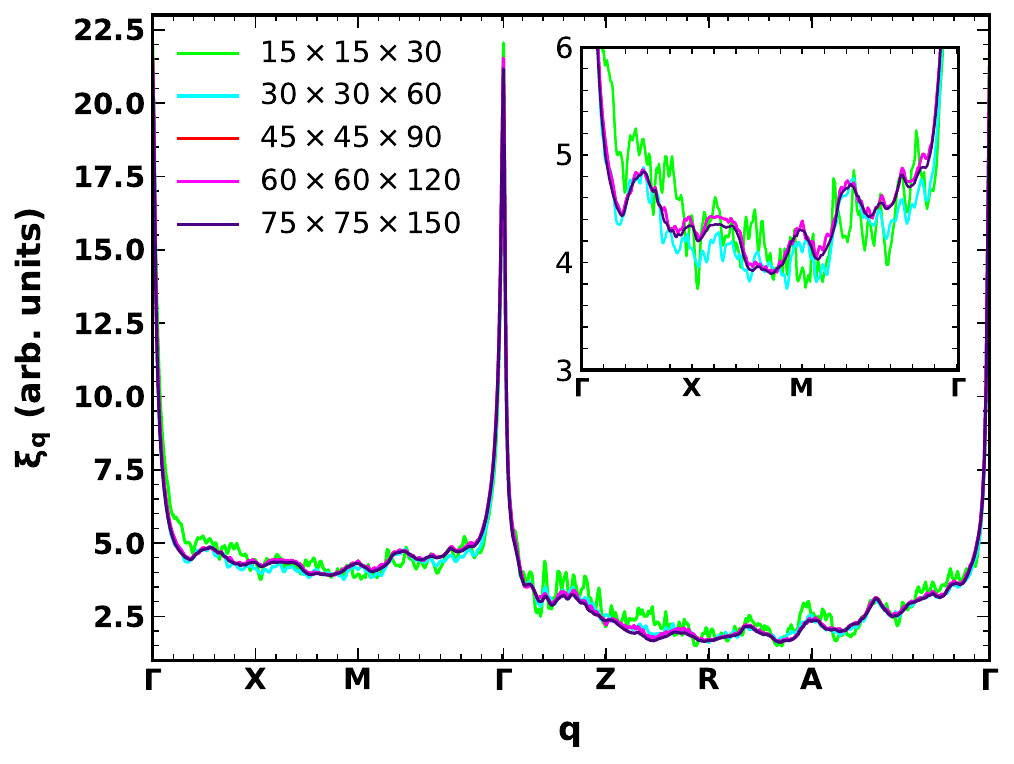}
    \begin{center}
        \hspace*{0.4cm}
        \textbf{\small{(c)}}
        \hspace*{5.1cm}
        \textbf{\small{(d)}}
    \end{center}
\vspace*{-0.20cm}
\caption{\label{fig:11} (a) Convergence test of energy in QE with respect to, Main plot: ecutwfc and Inset plot: k-points grid, (b) Main: Convergence test of EPC strength with respect to fine grid, Inset: Convergence test of Eliashberg spectral function with respect to fine grid in EPW, (c) Gap convergence with respect to different fine k and q grids at 10 K, (d) Main plot: Convergence test of nesting function calculated at 0.02 eV electronic smearing in EPW with respect to fine k-grid, Inset plot: zoomed version of main plot.}

\end{figure*}

We have plotted ground state energy in QE for different values of kinetic energy cut off in the main plot of Figure \hyperref[fig:11]{11(a)}. From here, we can see 80 Ry is one of the safest values of kinetic energy cut off. Using 80 Ry kinetic energy cut off, we have calculated ground state energy for different k-grids in the inset plot of Figure \hyperref[fig:11]{11(a)}. From this inset, we see that our calculated ground state properties in QE using $\mathrm{12\times12\times24}$ k-grid belongs to converged one.

Covergence test of EPC strength in main plot and Eliashberg spectral function in the inset plot of Figure \hyperref[fig:11]{11(b)} clearly indicate that a $\mathrm{15\times15\times30}$ fine k and q-grid are sufficient to obtain converged result. Anisotropic gap convergence test performed at 10 K in Figure \hyperref[fig:11]{11(c)} also confirms this. As Fermi nesting depends sensitively on the Fermi surface topology, we have performed convergence test of nesting function taking smaller electronic smearing (0.02 eV) for different fine k-grids in Figure \hyperref[fig:11]{11(d)}. Figure \hyperref[fig:11]{11(d)} confirms that well-converged Fermi nesting can be obtained for 0.02 eV electronic smearing and a $\mathrm{75\times75\times150}$ fine k-grid.

\nocite{*}
\bibliographystyle{unsrt}
\bibliography{apssamp_SM}